\documentclass[traditabstract]{aa}
\usepackage{graphicx}
\usepackage[varg]{txfonts}
\usepackage{color}
\usepackage{natbib}
\usepackage{epsfig}
\usepackage{float}

\usepackage[usenames,dvipsnames,svgnames,table]{xcolor}
\usepackage[colorlinks=true,
            linkcolor=blue,
            urlcolor=blue,
            citecolor=blue]{hyperref}

\titlerunning{Observations of a Flux Rope Formation}
\authorrunning{Kumar et al.}

\begin{document}

\title{Multiwavelength Observations of a Flux Rope Formation by Series of Magnetic Reconnection in the Chromosphere}

\author{Pankaj Kumar\inst{1,2}, Vasyl Yurchyshyn\inst{3,1}, Kyung-Suk Cho\inst{1,5}, Haimin Wang\inst{4,3}}
\institute{Korea Astronomy and Space Science Institute (KASI), Daejeon, 305-348, Republic of Korea\\
\email{pankaj@kasi.re.kr}
\and
Heliophysics Science Division, NASA Goddard Space Flight Center, Greenbelt, MD 20771, USA
\and
Big Bear Solar Observatory, New Jersey Institute of Technology, Big Bear City, CA 92314, USA
\and
Space Weather Research Laboratory, New Jersey Institute of Technology, University Heights, Newark, NJ 07102-1982, USA
\and
University of Science and Technology, Daejeon 305-348, Republic of Korea
}
%\email{mano@ncra.tifr.res.in}}
%*****************************************************************************
\abstract
{Using high resolution observations from the 1.6 m New Solar Telescope (NST) operating at the Big Bear Solar Observatory (BBSO), we report direct evidence of merging/reconnection of cool H$\alpha$ loops in the chromosphere during two homologous flares (B- and C-class) caused by a shear motion at the footpoints of two loops. The reconnection between these loops caused the formation of an unstable flux rope which showed counterclockwise rotation. The flux rope could not reach the height of torus instability and failed to form a coronal mass ejection. The HMI magnetograms revealed rotation of the negative/positive (N1/P2) polarity sunspots in the opposite directions, which increased the right and left-handed twist in the magnetic structures rooted at N1/P2. Rapid photospheric flux cancellation (duration$\sim$20-30 min, rate$\approx$3.44$\times$10$^{20}$ Mx h$^{-1}$) was observed during and even after the first B6.0 flare and continued until the end of the second C2.3 flare. The RHESSI X-ray sources were located at the site of the loop's coalescence. To the best of our knowledge, such a clear interaction of chromospheric loops along with rapid flux cancellation has not been reported before. These high resolution observations suggest the formation of a small flux rope by a series of magnetic reconnection within chromospheric loops associated with very rapid flux cancellation.}
 
%*****************************************************************************
\maketitle
\section{Introduction}
Flux ropes contain helical twisted field lines wrapped around a central axis and are considered an integral part of a coronal mass ejection (CME)    \citep{rust1996,chen2011}. The flux ropes are recognized as a magnetic cloud in the interplanetary medium which generally exhibit strong magnetic field, low plasma beta, and smooth rotation of magnetic field vector (B$_\theta$) at 1 AU \citep{burlaga1982,marubashi1986,bothmer1998,vasyl2001}. These flux ropes (with a strong southward magnetic field component) can produce severe geomagnetic storms by interacting with Earth's magnetosphere. Therefore, the study of the flux rope formation/eruption is important for the space weather point of view \citep{srivastava2004,gopalswamy2005,kumar2011}.

The formation of erupting flux ropes on the Sun is not very well understood. The presence of a flux rope before or during an eruption is a crucial condition for CME initiation models. Some CME initiation models (e.g., emerging flux model: \citealt{chen2000}, MHD kink instability models: \citealt{fan2004,kliem2004,torok2005}) assume a pre-existing twisted flux rope before the eruption,
%that emerges from below the photosphere \citep{rust1994,mckaig2001}
 whereas in other models (e.g., tether cutting model: \citealt{moore2001}, breakout models: \citealt{antiochos1998,karpen2012}) the flux rope is formed (during the eruption) in the corona by a series of magnetic reconnection within an arcade of loops. The forward S-or inverse S shaped sigmoids are generally observed in the X-ray images of active regions (ARs), which may be the manifestation of kink unstable flux ropes \citep{canfield1999,rust1996,kliem2004,gibson2006}. According to the flux rope model of the prominence/filament \citep{van1989}, a series of magnetic reconnections between the sheared arcade of loops at the polarity inversion line (PIL) driven by photospheric converging motions leads to the formation of a helical flux rope before it erupts. Other models create a flux rope during the eruption process utilizing  multiple reconnections of the arcade loops during a two-ribbon flare leaving behind a post-eruptive arcade \citep{gosling1995,longcope2007}. 
 An alternative assumption is that the flux rope is emerged from below the photosphere and was already twisted in the convection zone \citep{rust1994,mckaig2001}.

The three part structure of a CME includes a frontal loop/leading edge, a dark cavity, and a bright core region (filament/prominence material). The cavity region of a CME is assumed to be a flux rope \citep{bak2013}. According to the flux rope model \citep{priest1989} of a solar prominence, the filament cool material is supported in the lower portion of the helical field lines. The circular features within the CMEs observed by coronagraphs are considered to be a helical flux rope structure \citep{vourlidas2013,vourlidas2014}. The interplanetary scintillation images (density) also show the evolution of the large-scale CME flux ropes in the interplanetary medium (50-250 Rs) \citep{manoharan2010}. 

There are a number of observational reports on the activation of a helical flux rope associated with kink instability. For example, helical field lines with 3-4 turns have been reported during a kink unstable prominence eruption \citep{gary2004,kumar2012}.
Also, using multiwavelength observations (Hinode/SOT, TRACE), \citet{srivastava2010} and \citet{kumar2010b} provided evidence of a helical kinked flux tube  (3-4 turns) that triggered multiple B- and M-class flares associated with a failed eruption. \citet{Kumar2011a} reported a failed flux rope eruption observed not only in the STEREO and TRACE EUV cool channels but also in the XRT hot channel, which suggest that multi-temperature plasma may be contained within a flux rope. Recently, SDO/AIA observations revealed high temperature flux ropes visible in 131 and 94 \AA~ channels \citep{cheng2011,zhang2012,kumar2013b}. However, the above reports did not discuss the formation mechanism of the flux ropes mostly due to lack of photospheric magnetograms and/or high-resolution chromospheric observations.

Using high resolution Hinode/SOT images, \citet{Okamoto2008} presented evidence for the emergence of a helical flux rope from below a pre-existing prominence. 
\citet{green2009} and \citet{green2011} suggested the formation of a S-shaped flux rope via flux cancellation at the PIL. \citet{rui2010} showed evidence of tether-cutting reconnection between two opposite J-shaped coronal loops that formed a continuous S-shaped flux rope. Recently, \citet{kumar2014} reported reconnection signatures above a small kinked filament and formation of a twisted hot flux rope observed in the AIA 131, 94 \AA, and XRT images during magnetic reconnection. These observations show the formation of coronal flux ropes.

There are several mechanisms capable of triggering and driving the flux rope eruption. The flux rope eruption can be triggered either by a reconnection process (e.g., associated with tether-cutting or an emerging flux) or by an MHD instability (i.e., kink instability if the twist angle $\geq$2.49$\pi$, \citealt{hood1981,einaudi1983}),which may cause it to attain a certain height within an AR. Further, the successful eruption largely depends on the onset of the torus instability \citep{kliem2006,olmedo2010,aulanier2010}, which is defined by the rate of decrease of the overlying magnetic field with height (i.e, decay index$\geq$1.5 \citep{kliem2006}).
   
The high resolution observations from the NST are extremely useful to investigate the issues related to the flux rope formation/eruption from the photosphere to chromosphere.
A recent study utilizing  NST data along with a NLFFF modeling revealed a pre-existing small flux rope in the H$\alpha$ chromosphere that became unstable most likely due to the onset of a kink instability \citep{wang2015}. Alternatively, \citet{kumar2015fr} provided evidence  of reconnection (inflows) between cool chromospheric sheared loops and the associated appearance/formation of a S-shaped twisted flux rope presumably as a result of magnetic reconnection. In addition, \citet{vasyl2015} reported  an in-situ formation of a flux rope by multiple reconnections starting from the low chromosphere to the corona. The flux rope was observed as a circular shape (axial view) in the hot AIA channels (131 and 94 \AA) and LASCO C2 coronagraph images. 

In this paper, we mainly focus on the interaction/reconnection between  H$\alpha$ loops, the flux rope formation, and its dynamics in the chromosphere and corona during two small homologous flares (B6.0 and C2.3) that occurred in AR NOAA 12353 on 23 May 2015. During both flare, we detected the presence of oscillatory reconnection in X-ray/EUV channels. A small flux rope was observed in high-resolution NST H$\alpha$ images allowing us to resolve the dynamics of small-scale chromospheric loops not otherwise detectable in the AIA data. We utilized SDO/AIA images and HMI magnetograms to investigate the structure and dynamics of the large-scale coronal magnetic field configuration (e.g., loops connectivity) and the associated photospheric field changes before and during the flare. RHESSI hard X-ray images are used to locate the particle transportation/precipitation site during the flares.  
In Section~2, we present the observations and the results. In the last section, we summarize and discuss our findings.
%%%%%%%%%%%%%%%%%%%%%%%%%%%%%%%%%%%%%%%%%%%%%%%%%%%%%%%%%%%%%%%%%%%%%%%%%%%
%%%%%%%%%%%%%%%%%%%%%%%%%%%%%%%%%%%%%%%%%%%
%%%%%%%%%%%%%%%%%%%%%%%%%%%%%%%%%%%%%%%%%%%%%%%%%%%%%%%%%%%%%%%%%%%%%%%%%%%%%%%%%%%%%%%%%%%
%------------------------------------------------------------------------------------ 
\begin{figure}
\centering{
\includegraphics[width=9cm]{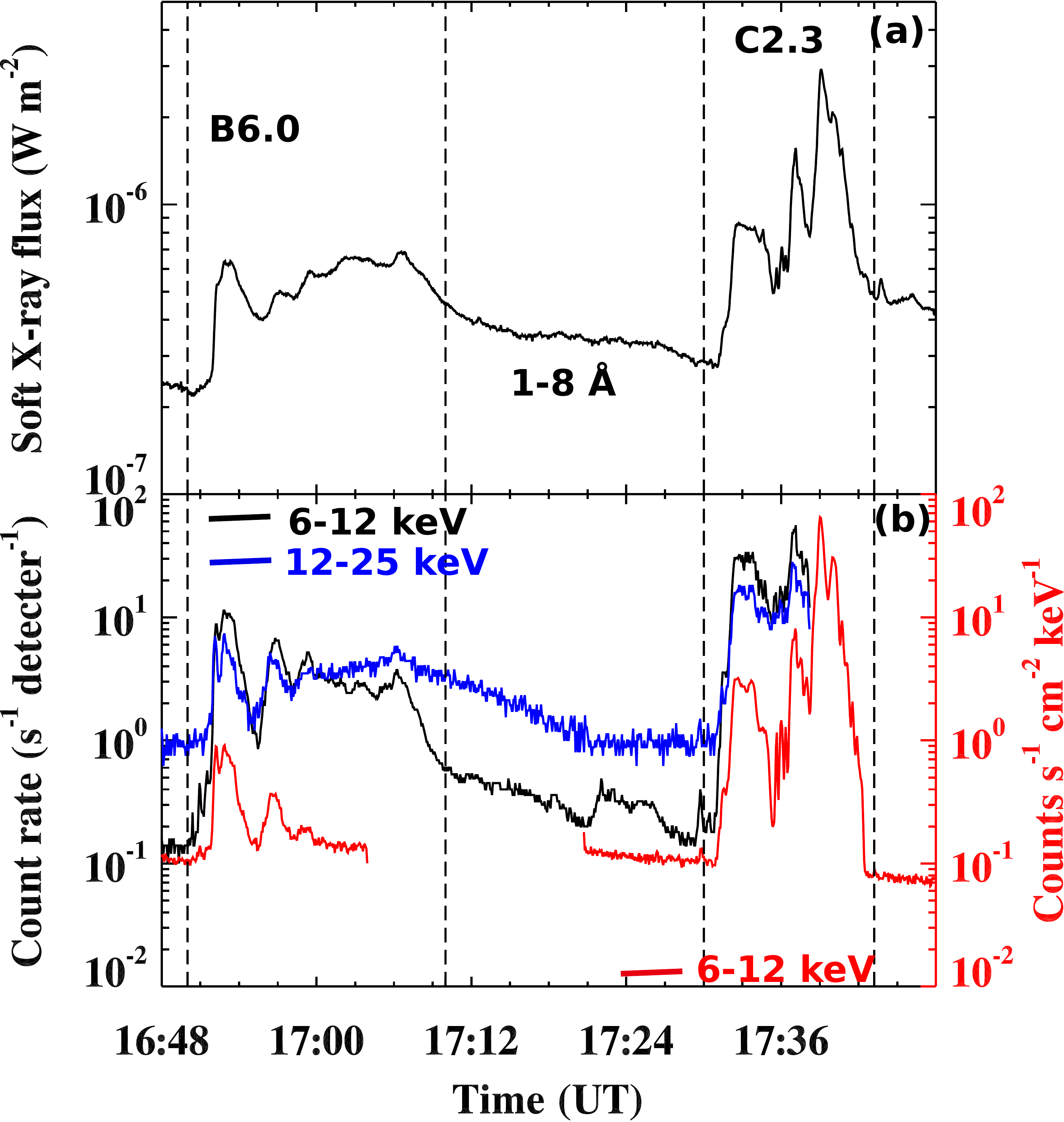}
}
\caption{(a) GOES soft X-ray flux profile in 1-8 \AA~ channel. (b) RHESSI X-ray flux in the 6-12 keV (black) and 12-25 (blue) keV channels. The Fermi GBM X-ray flux in the 6-12 keV (red) is also included. The vertical dashed lines indicate the interval of the B6.0 and C2.3 flare. }
\label{flux}
\end{figure}

%%%%%%%%%%%%%%%%%%%%%%%%%%%%%%%%%%%%%%%%%%%%%%%%%%%%%%%%%%%%%%%%%%%%%%%%%%%%%%%%%%%%%%%%%%%
%------------------------------------------------------------------------------------ 
\begin{figure*}
\centering{
\includegraphics[width=8cm]{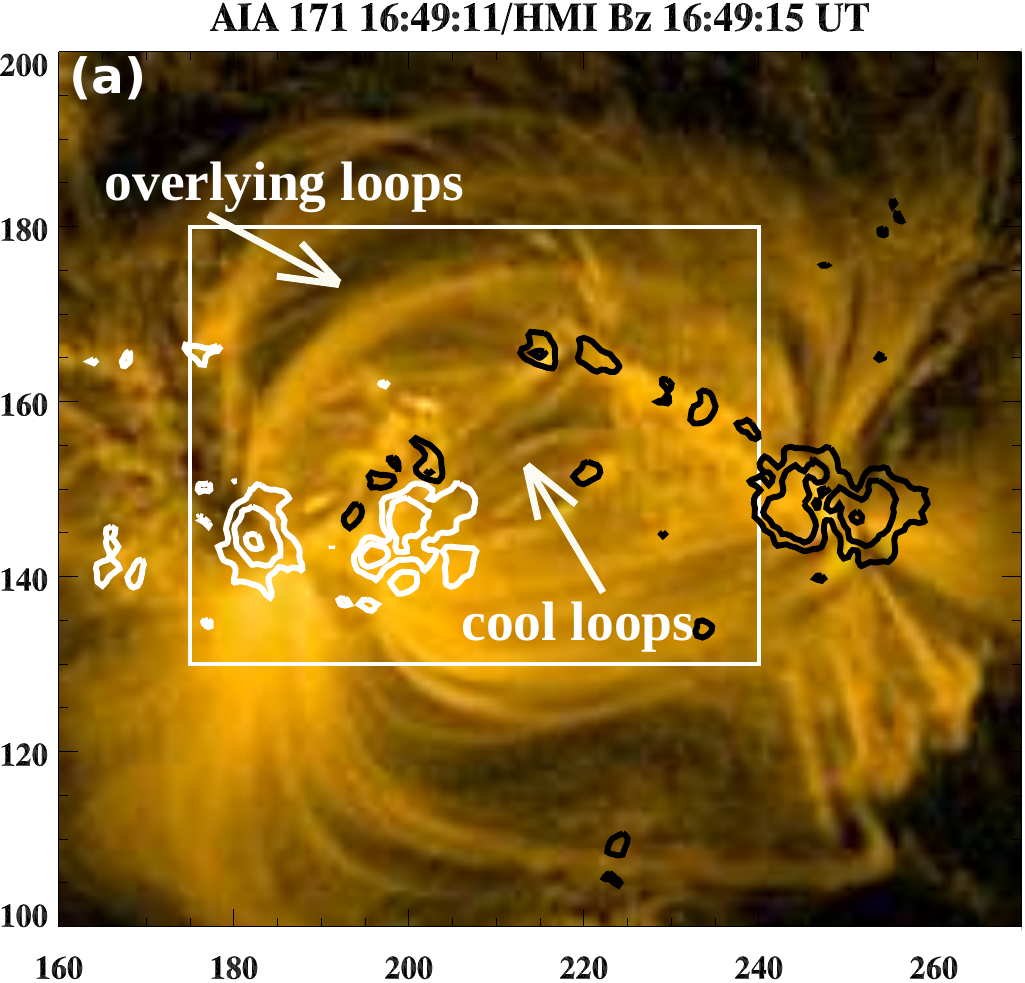}
\includegraphics[width=8cm]{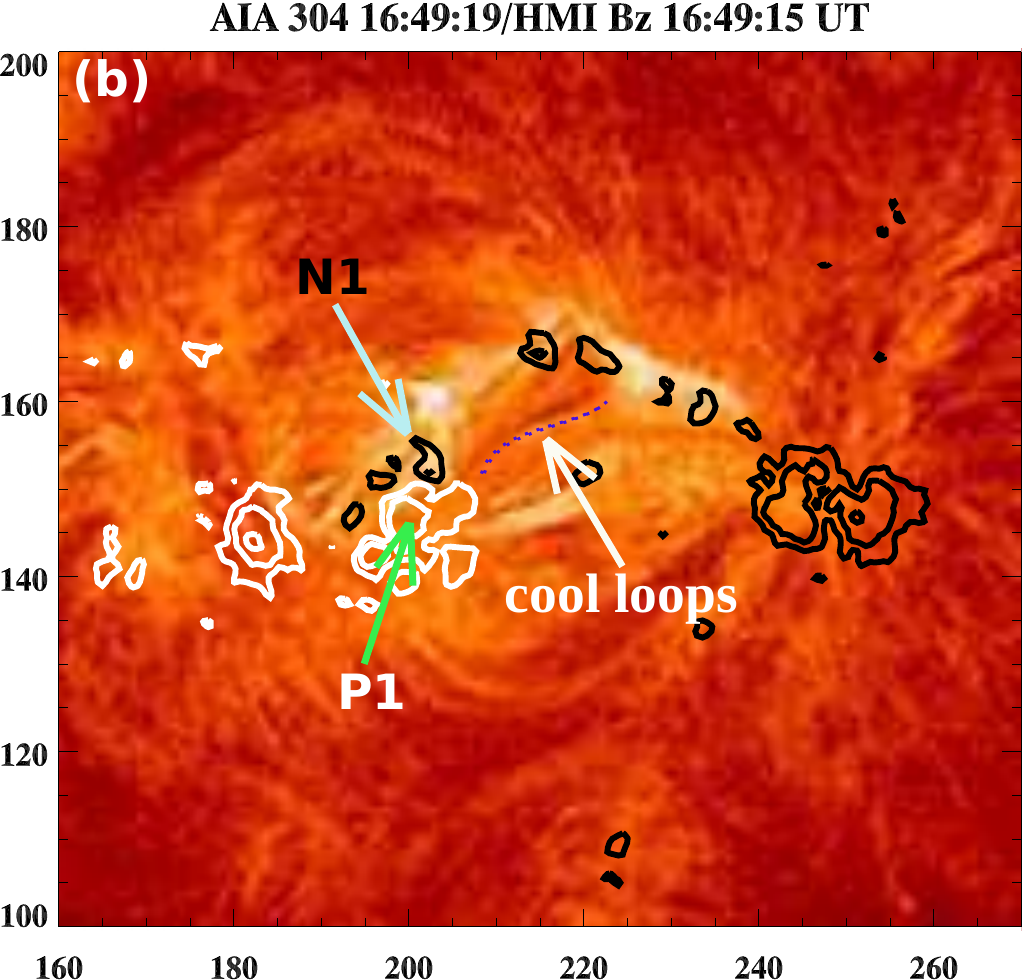}

\includegraphics[width=5.3cm]{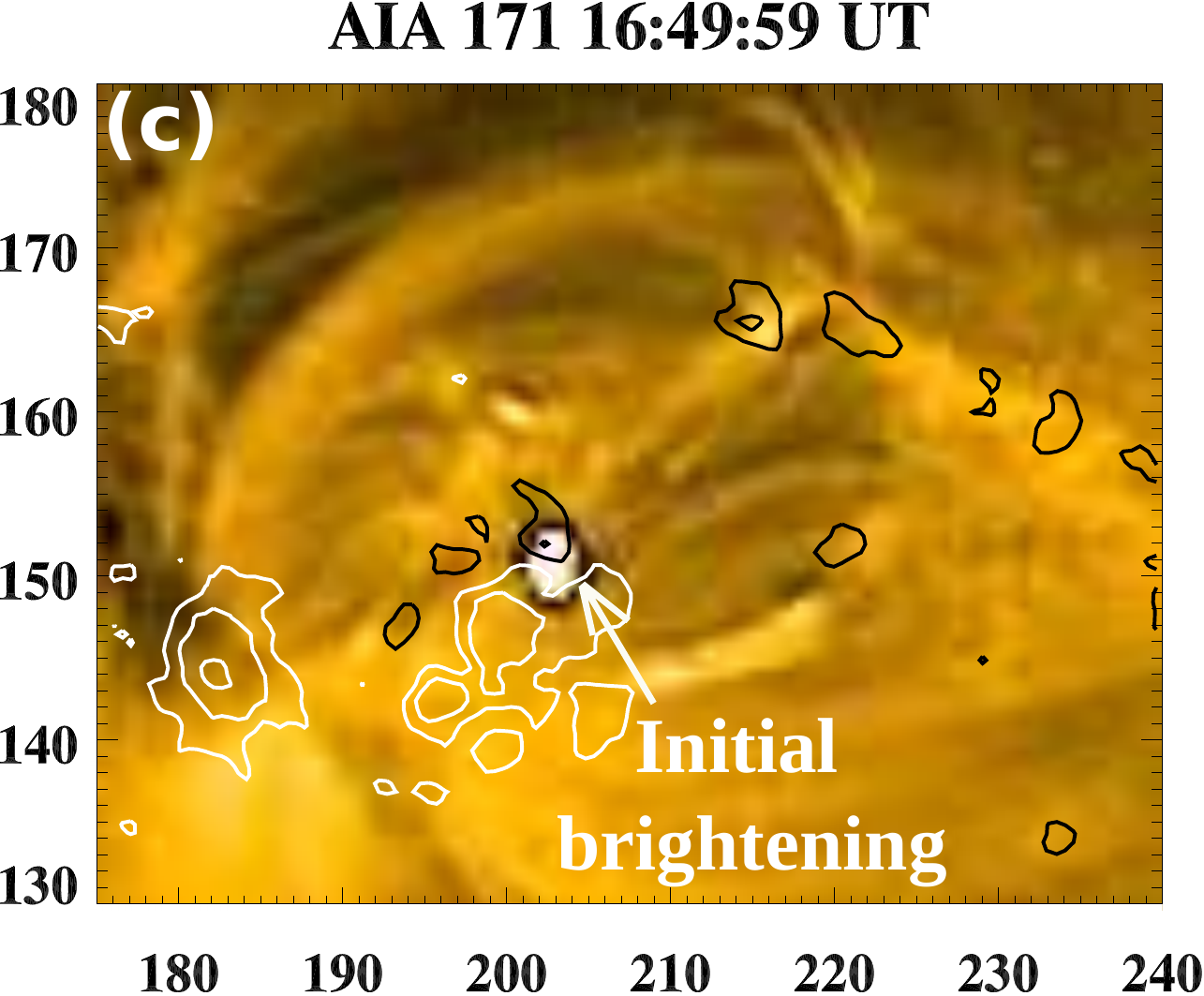}
\includegraphics[width=5.3cm]{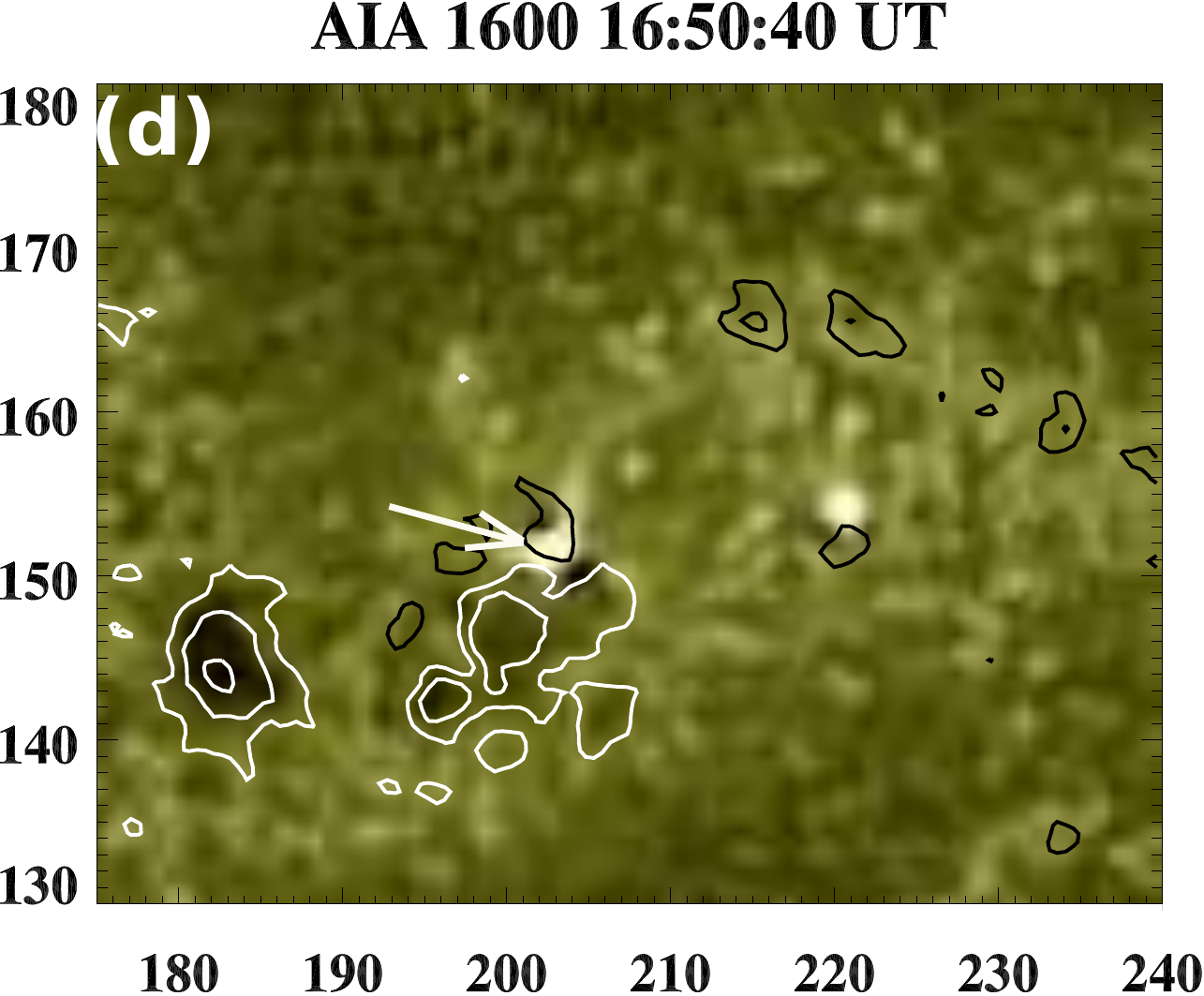}
\includegraphics[width=5.3cm]{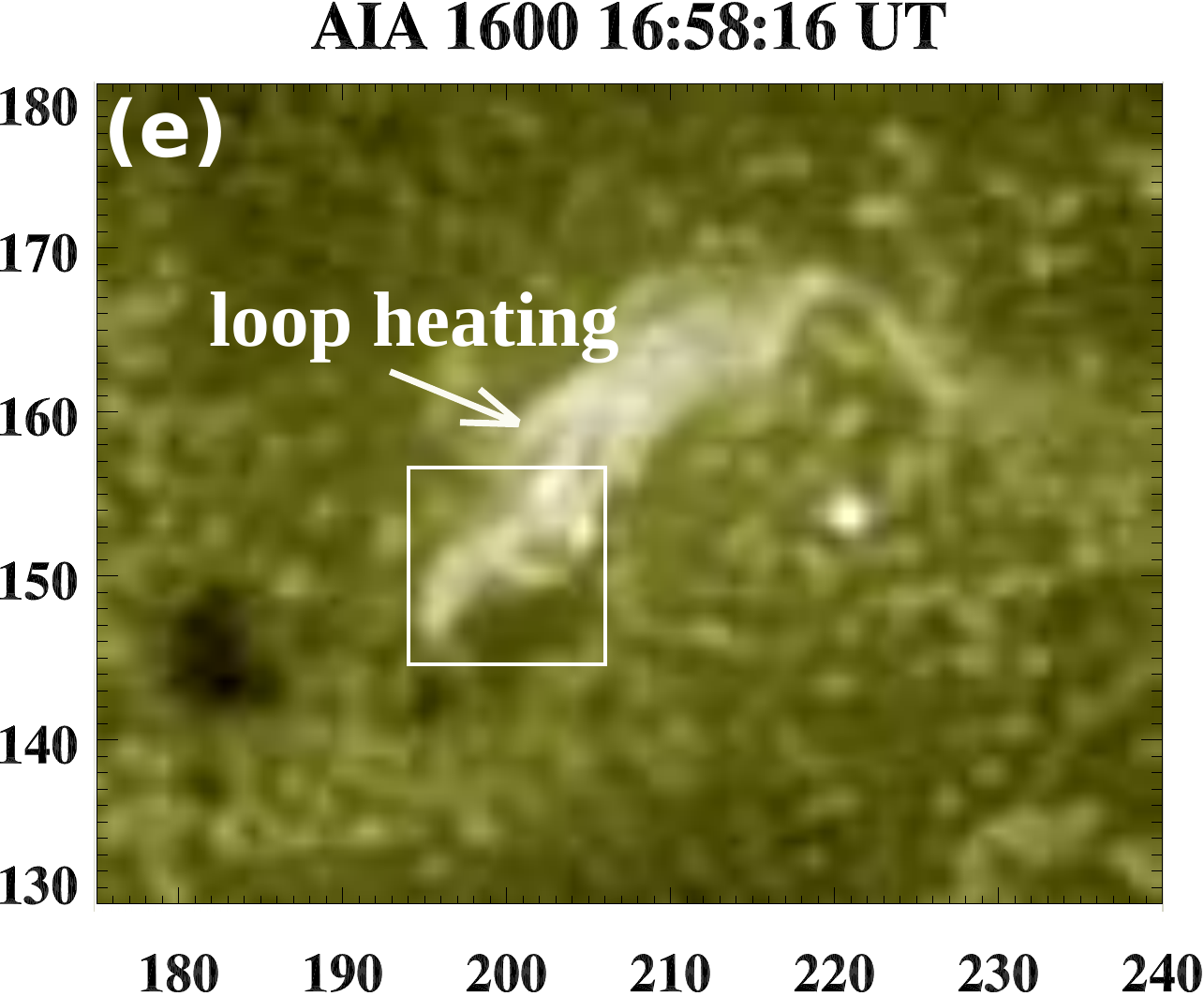}

\includegraphics[width=5.3cm]{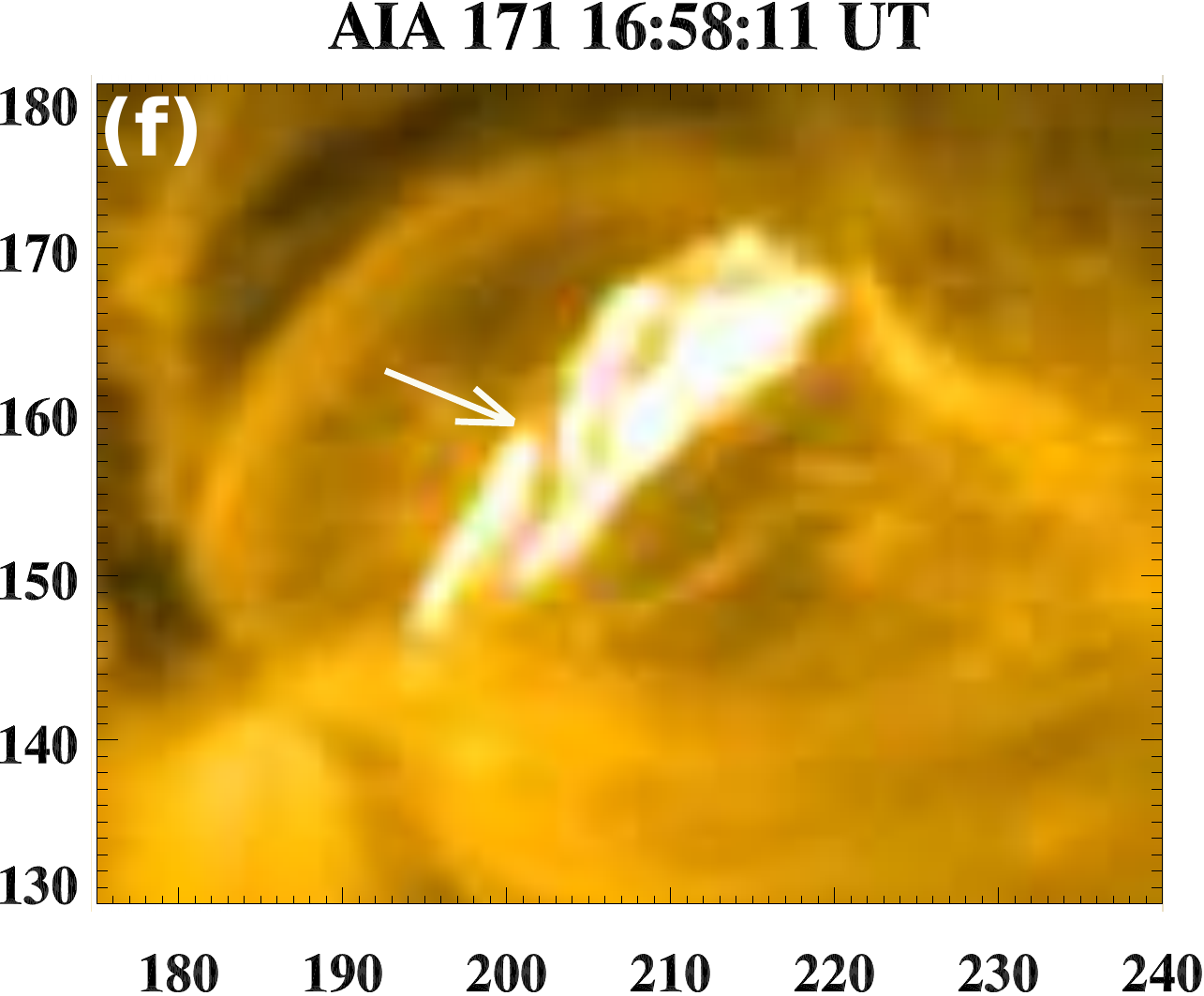}
\includegraphics[width=5.3cm]{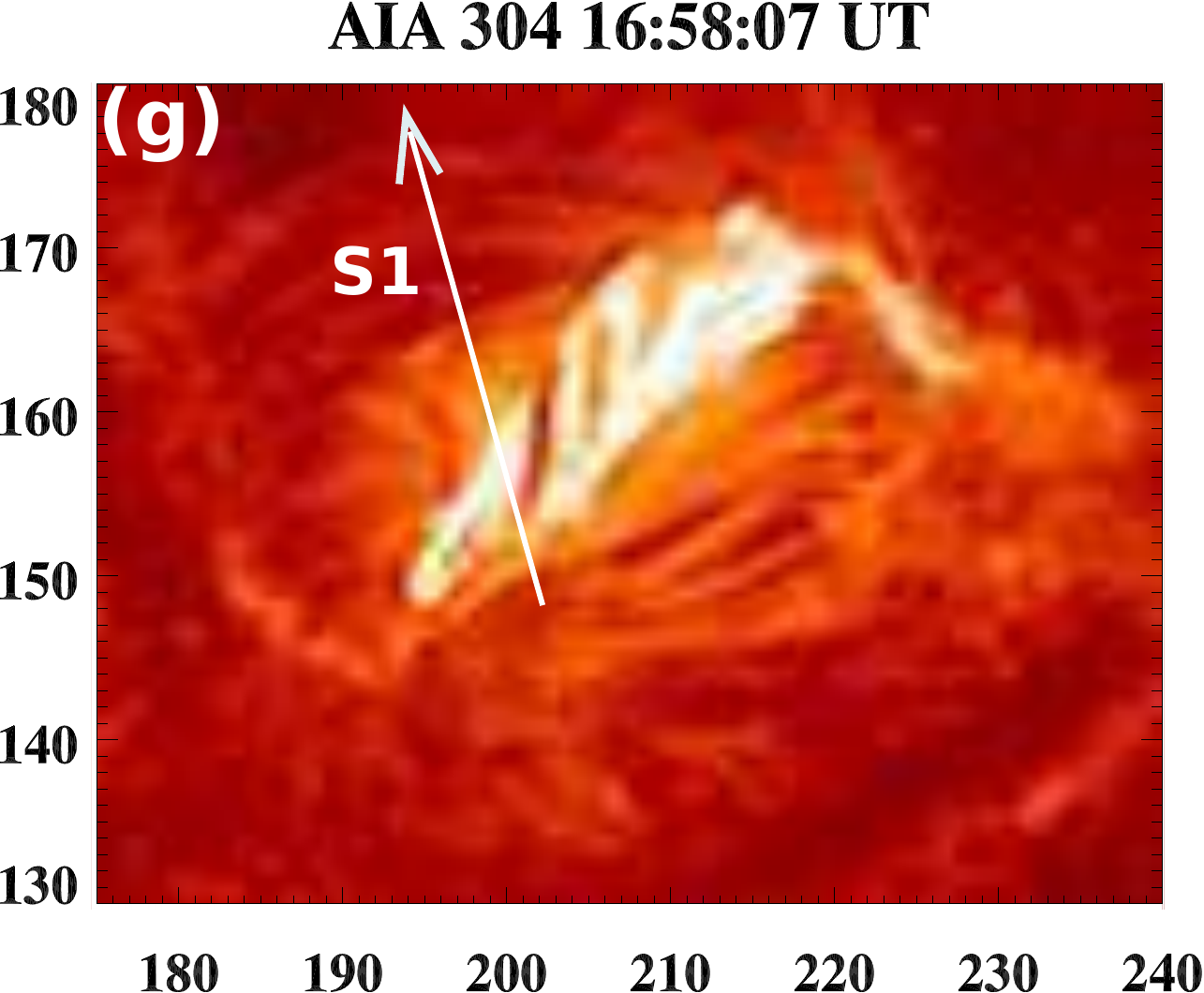}
\includegraphics[width=5.3cm]{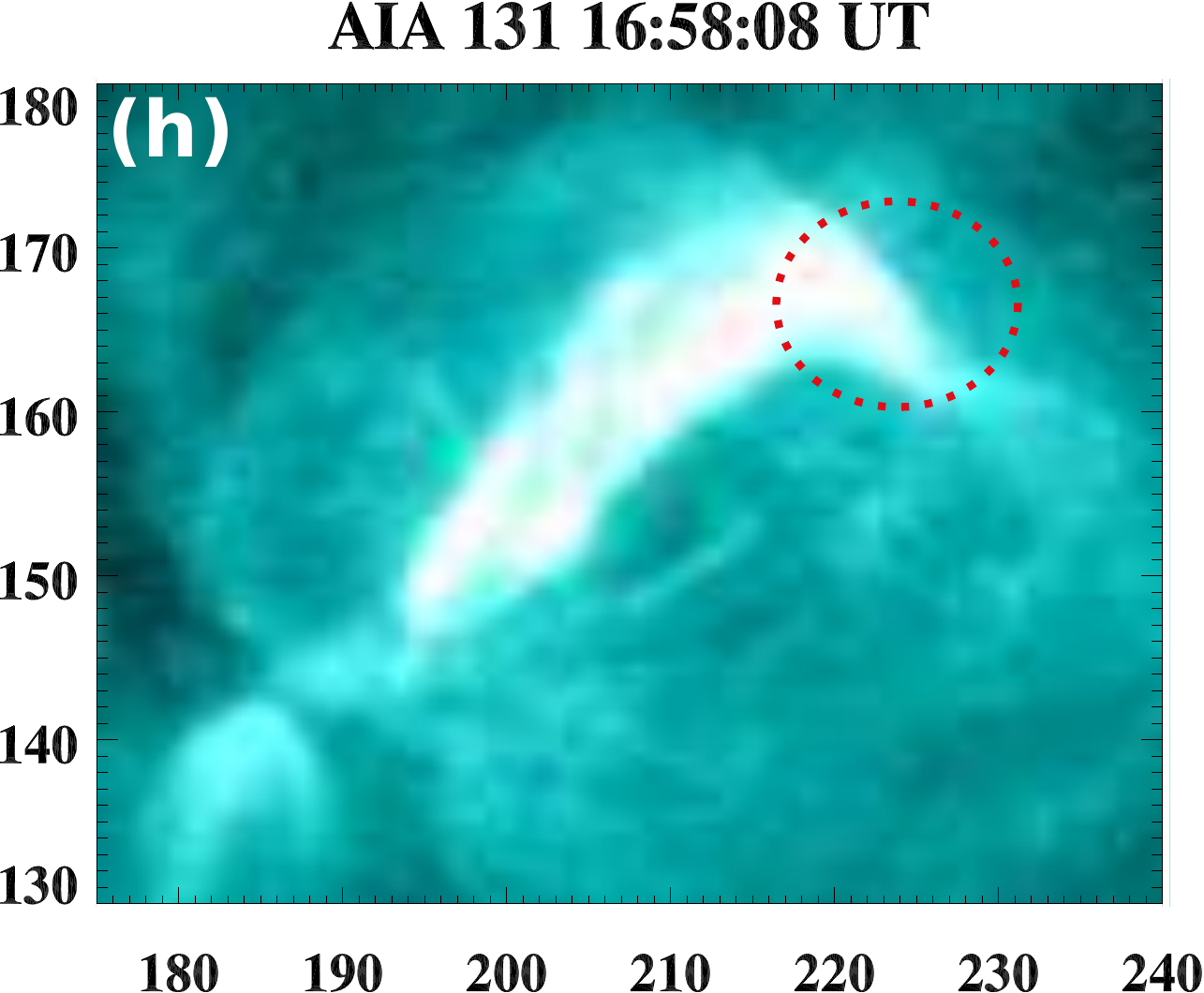}
}
\caption{(a,b) AIA 171 and 304 \AA~ images before the onset of the first flare (B6.0). These images are overlaid by the HMI magnetogram contours of positive (white) and negative (black) polarities. The contour levels are $\pm$500, $\pm$1000, $\pm$1500 Gauss. The rectangular (white) box indicates the size of the middle/bottom panels. P1 and N1 are positive and negative polarity sunspots. (c,d) AIA 171 and 1600 \AA~ images showing the preflare brightening between N1 and P1. (e-h) The heated loop was observed in the cool (304, 171, and 1600 \AA) and hot (AIA 131 \AA) channels. S1 indicates the slice cut (over the footpoint of the loop) used to create the stack plot. The arrow-head shows the direction of the slice. The x- and y axes are labeled in arcsecs. The evolution of this flare is shown in an animation (i.e., AIA 304, 171, and 131 \AA, bottom panels) available in the online edition.}
\label{fl1}
\end{figure*}

%%%%%%%%%%%%%%%%%%%%%%%%%%%%%%%%%%%%%%%%%%%%%%%%%%%%%%%%%%%%%%%%%%%%%%%%%%%%%%%%%%%%%%%%%%%
\section{OBSERVATIONS AND RESULTS}
The NST images used in this study were captured with the help of the 308 sub-aperture adaptive optics (AO-308) system. We used a series of narrow-band H$\alpha$ (6563 \AA) images taken at $\pm$0.8 \AA~, $\pm$0.4 \AA~, and 0.0 \AA~ from the line center with the NST's Visible Imaging Spectrometer (VIS, pixel size of 0.029$\arcsec$). The VIS combines a 5 \AA~ interference filter with a Fabry-P\'erot etalon to produce a resulting bandpass of 0.07 \AA~ over a 70$\arcsec$$\times$70$\arcsec$ field of view. A Series of broadband (10 \AA) images of the photosphere were acquired with a TiO filter (7057 \AA, pixel scale of 0.0375$\arcsec$) to study the evolution of the fine photospheric structures associated with flux emergence and cancellation. 

The Atmospheric Image Assembly (AIA; \citealt{lemen2012}) onboard the Solar Dynamics Observatory 
(SDO; \citealt{pesnell2012}) acquires full-disk images of the Sun (field of view $\sim$1.3 R$_\odot$) with a spatial resolution of 1.5$\arcsec$ 
(0.6$\arcsec$  pixel$^{-1}$) and a cadence of 12 sec in 10 extreme ultraviolet (EUV) and UV channels. This study utilizes 171~\AA\ \ion{Fe}{IX}, $T\approx$0.7 MK), 131 \AA~ (\ion{Fe}{VIII}, \ion{Fe}{XXI}, \ion{Fe}{XXIII}, i.e., 0.4, 10, 16 MK), 304 \AA~(\ion{He}{II}, T$\approx$0.05 MK) and 1600~\AA\ (\ion{C}{IV} + continuum, $T\approx$0.1 MK \& 5000 K) images. We also used Heliospheric and Magnetic Imager (HMI) magnetograms \citep{schou2012} to investigate the magnetic field evolution before and during the flares. 

We also used the Reuven Ramaty High Energy Solar Spectroscopic Imager (RHESSI; \citealt{lin2002}) data to investigate the particle acceleration/precipitation sites during the flares. The CLEAN algorithm was used for the image reconstruction with the integration time of 30 s.

%%%%%%%%%%%%%%%%%%%%%%%%%%%%%%%%%%%%%%%%%%%%%%

%------------------------------------------------------------------------------------ 
\begin{figure*}
\centering{
\includegraphics[width=7cm]{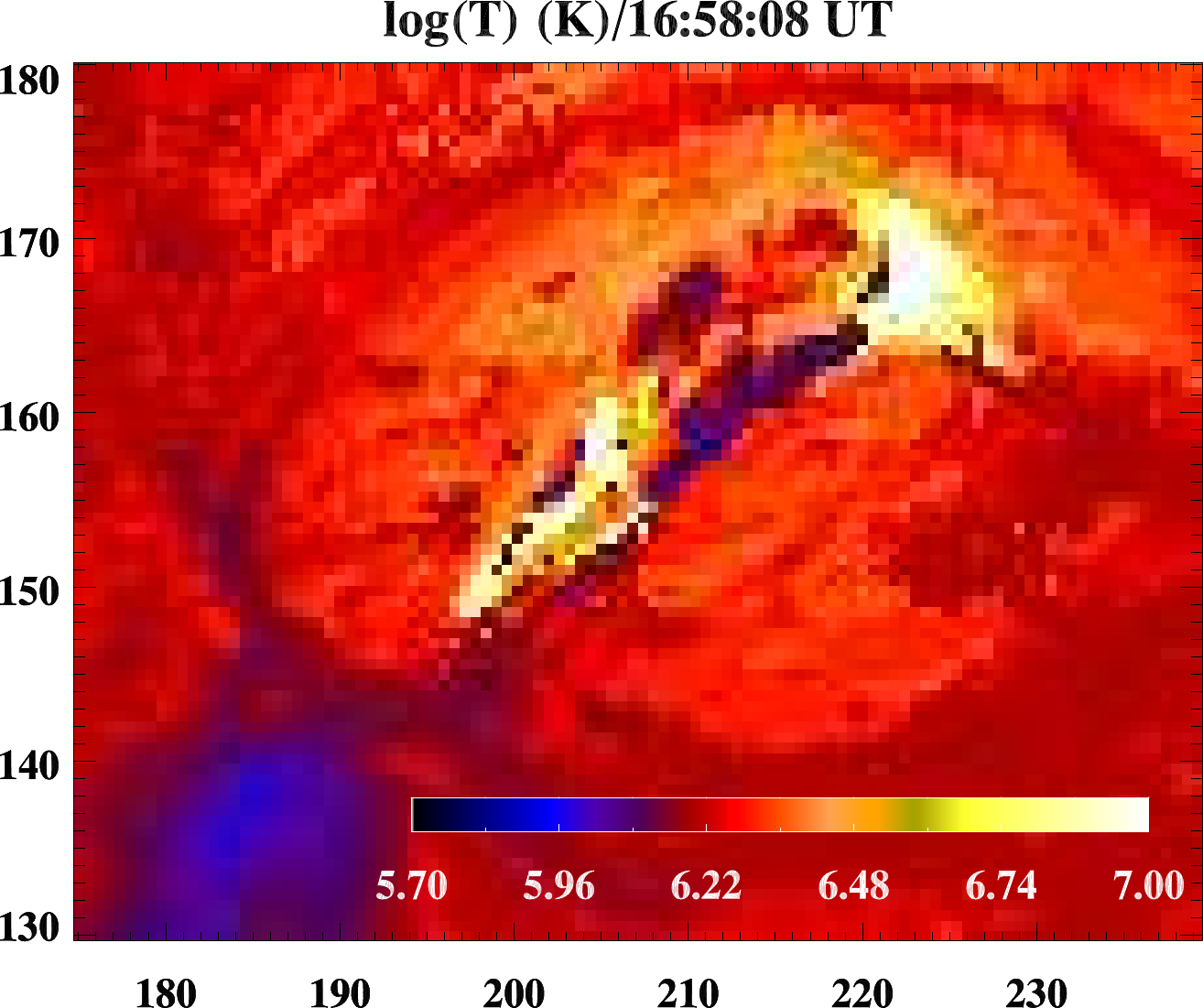}
\includegraphics[width=7cm]{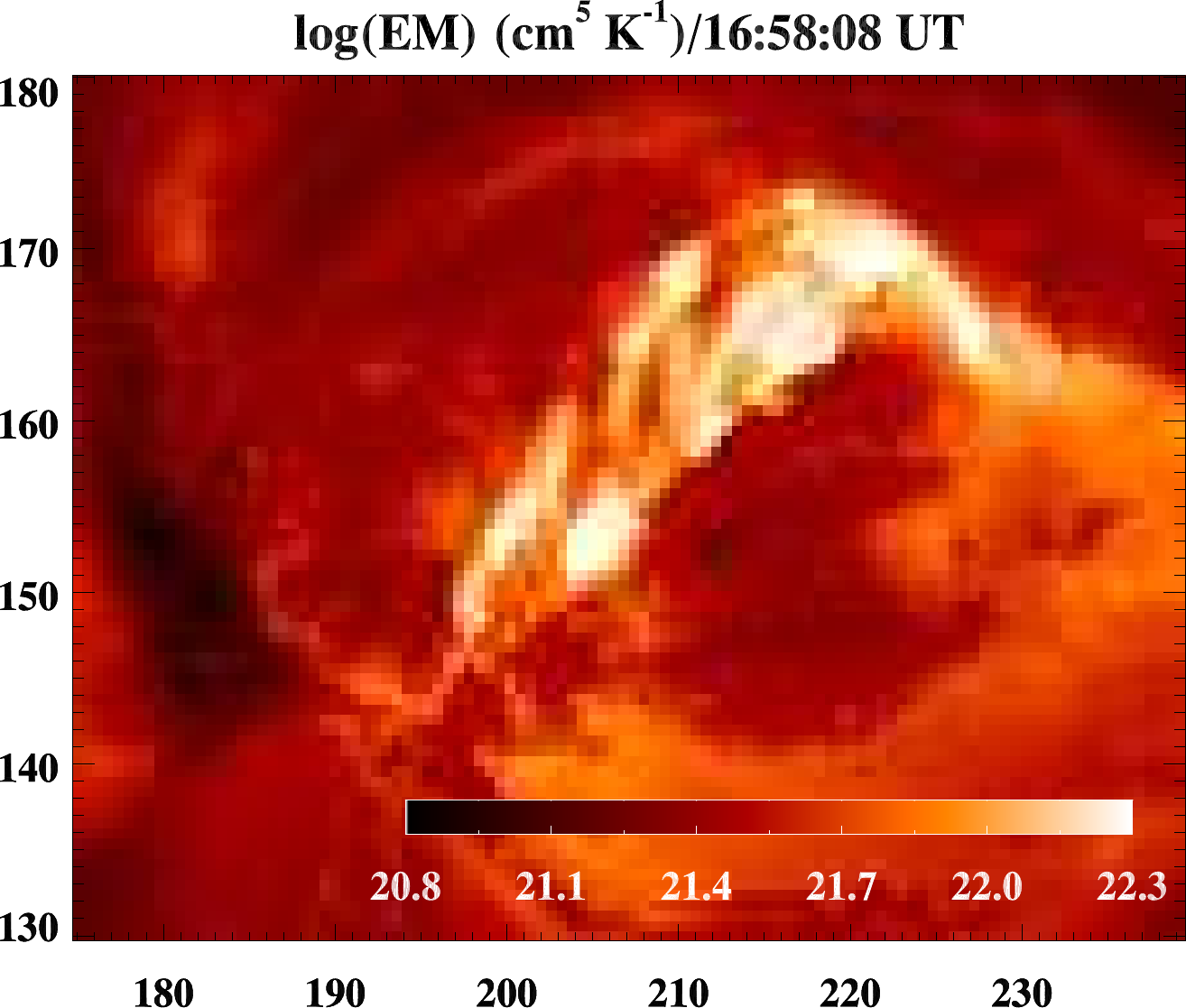}
}
\caption{Peak temperature and emission measure of the loop during the flare maximum (16:58:08 UT) derived from the near simultaneous AIA images in six channels. The loop contains multi-temperature plasma.}
\label{dem}
\end{figure*}
%%%%%%%%%%%%%%%%%%%%%%%%%%%%%%%%%%%%%%%%%%%%%%%%%%%%%%%%%%%%%%%%%%%%%%%%%%%%%%%%%%%%
%%%%%%%%%%%%%%%%%%%%%%%%%%%%%%%%%%%%%%%%%%%%%%%%%%%%%%%%%%%%%%%%%%%%%%%%%%%%%%%%%%%%%%%%%%%
%------------------------------------------------------------------------------------ 
\begin{figure*}
\centering{
\includegraphics[width=5.3cm]{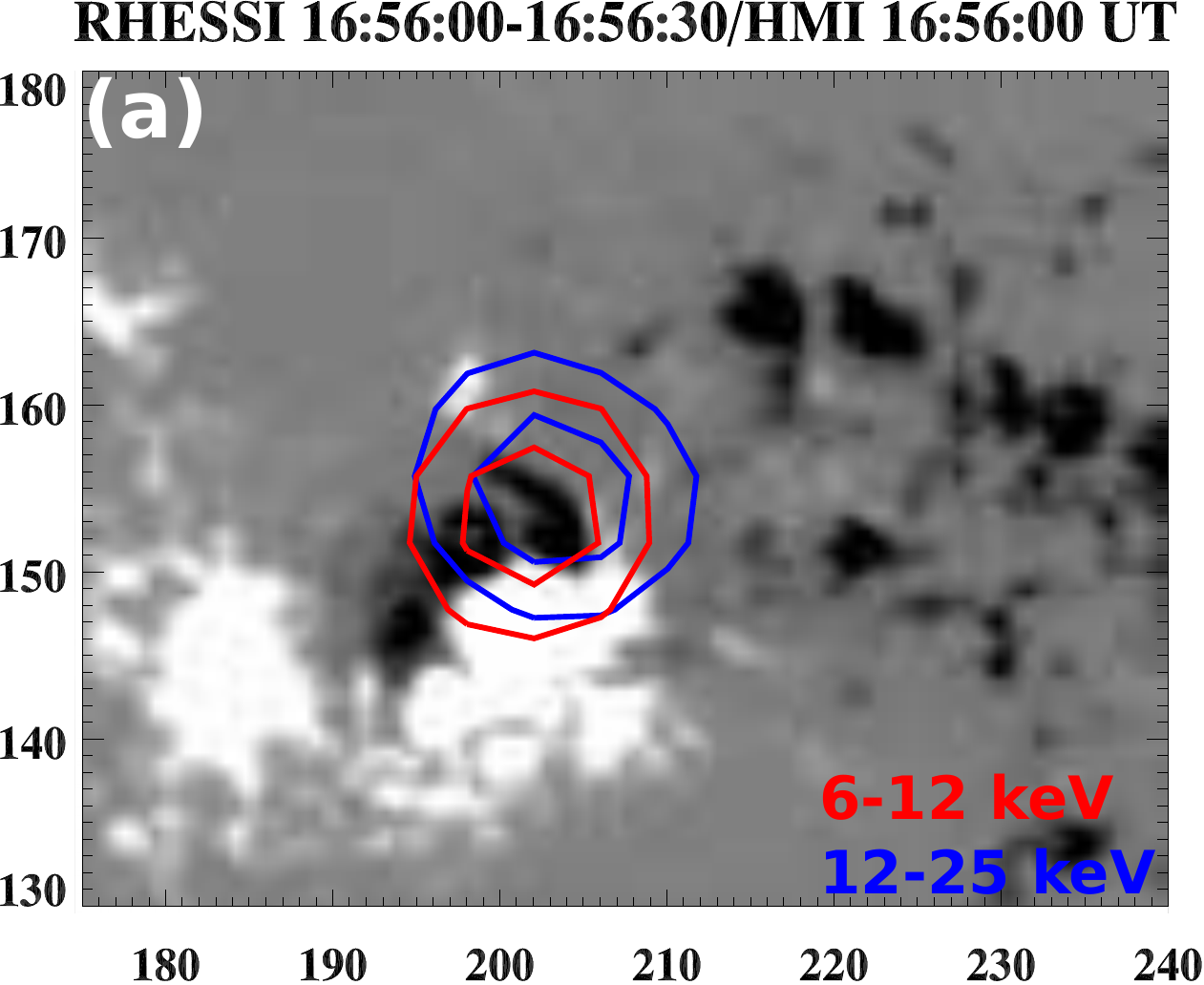}
\includegraphics[width=5.3cm]{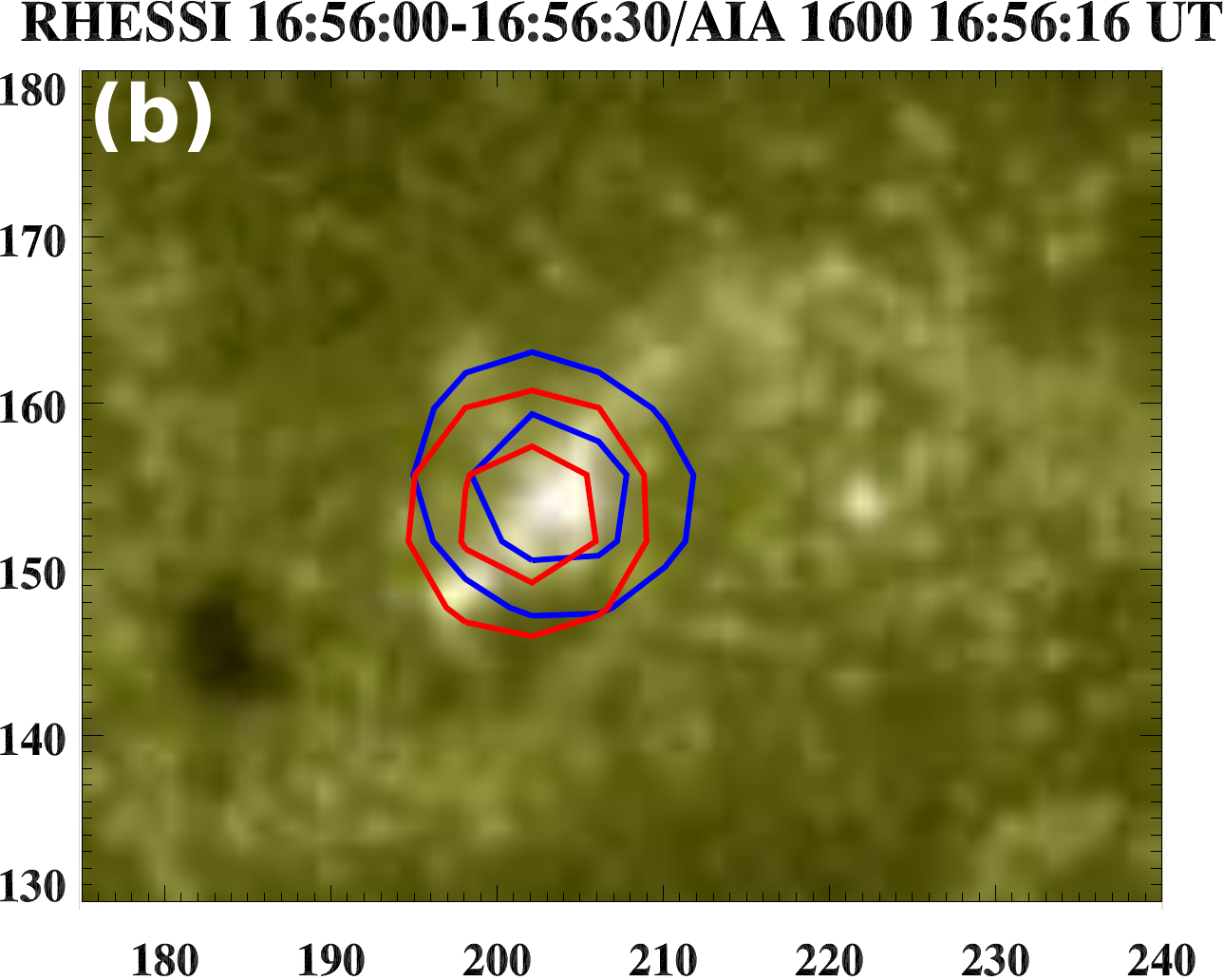}
\includegraphics[width=5.3cm]{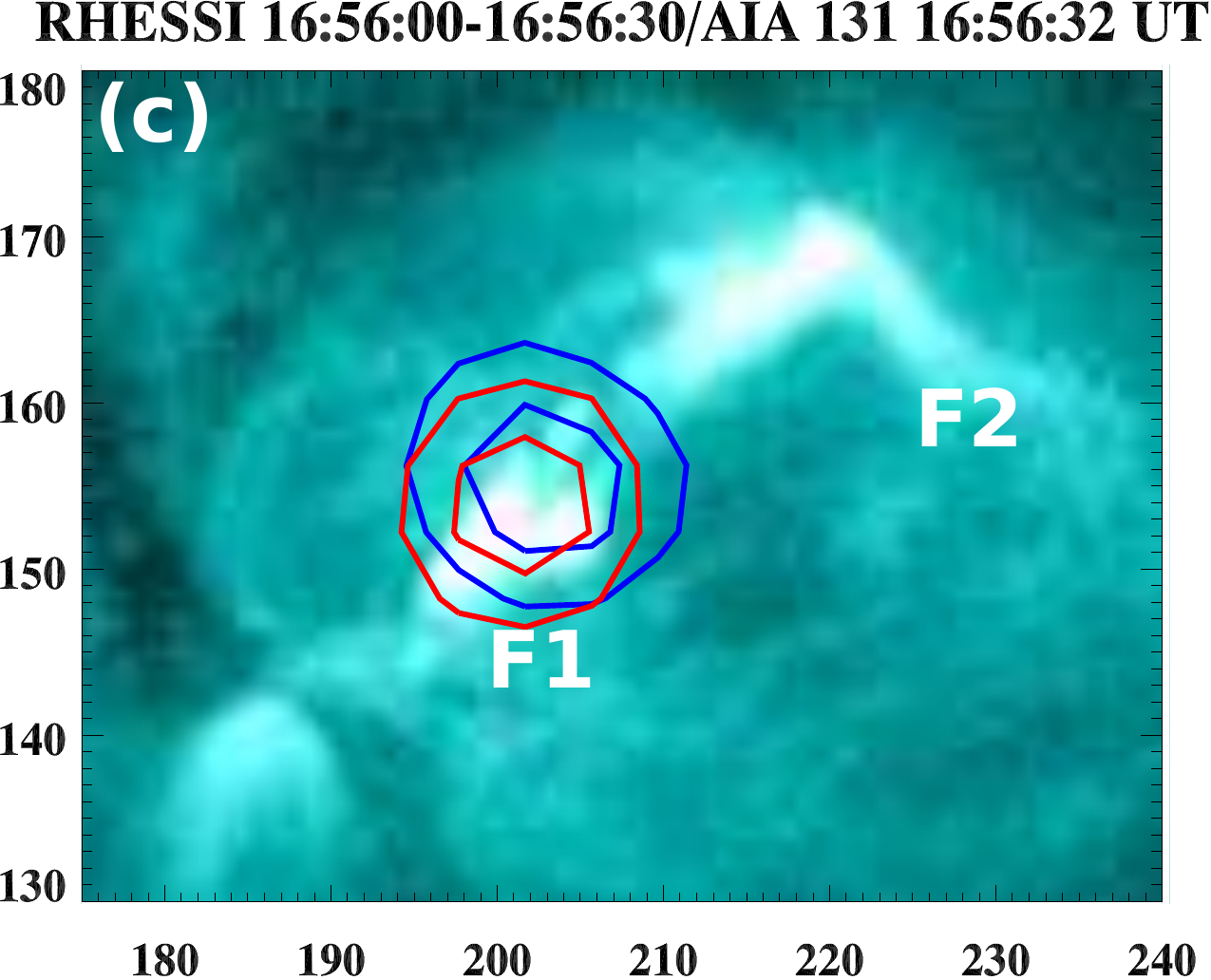}
}
\caption{RHESSI X-ray image contours (red: 6-12 keV, blue: 12-25 keV) overlaid on the HMI magnetogram, AIA 1600, and 131 \AA~ images during the B6.0 flare. The contour levels are 85$\%$ and 95$\%$ of the peak intensity. F1 and F2 are the footpoints of the heated loop. The x- and y axes are labeled in arcsecs.}
\label{hessi1}
\end{figure*}
%%%%%%%%%%%%%%%%%%%%%%%%%%%%%%%%%%%%%%%%%%%%%%%%%%%%%%%%%%%%%%%%%%%%%%%%%%%%%%%%%%%%
%------------------------------------------------------------------------------------ 
\begin{figure}[p]
\centering{
\includegraphics[width=8.5cm]{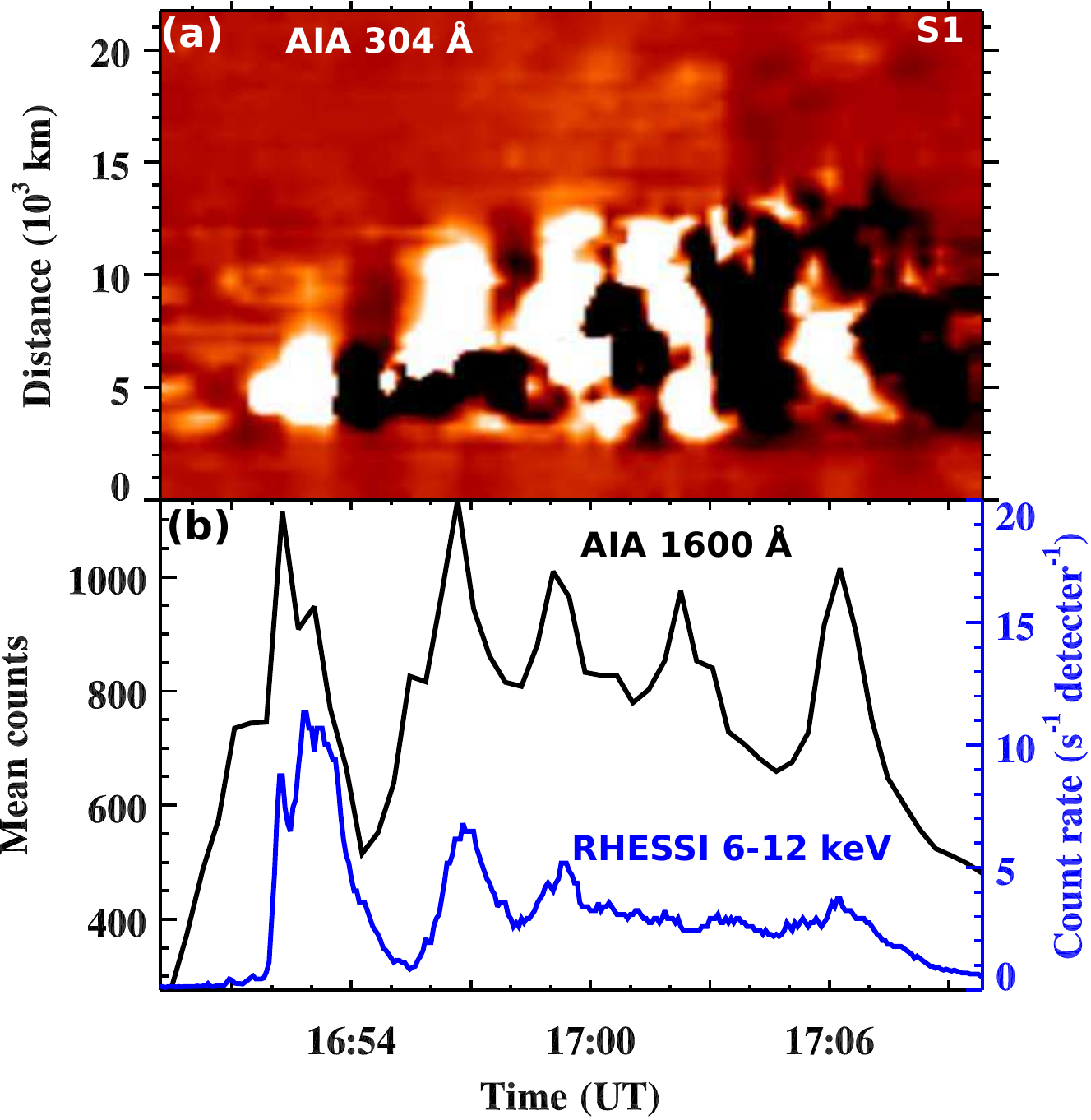}
}
\caption{(a) Stack plot along slice S1 (marked in Figure \ref{fl1}(g)) using AIA 304 \AA~ running difference images. (b) AIA 1600 \AA~ mean counts extracted within a box region shown in Figure \ref{fl1}(e). The blue curve is the RHESSI X-ray flux in 6-12 keV channel.}
\label{stack1}
\end{figure}
%%%%%%%%%%%%%%%%%%%%%%%%%%%%%%%%%%%%%%%%%%%%%%%%%%%%%%%%%%%%%%%%%%%%%%%%%%%%%%%%%%%%

%\subsection{Periodic reconnection}
AR NOAA 12353 of $\beta\gamma$ magnetic configuration was located near the disk center (N07W17) on 23 May 2015. We focus on two homologous flares that occurred at the same site within the AR and the associated reconnection events that formed a twisted flux rope. Figure \ref{flux}(a) displays the GOES soft X-ray flux profile obtained in the 1-8 \AA~ channel. The bottom panel of Figure \ref{flux}(b) shows RHESSI X-ray flux in the 6-12 keV (black) and 12-25 keV (blue) channels. The RHESSI did not observe the third peak of the second flare, therefore we overplotted Fermi GBM (Gamma Ray Burst Monitor; \citealt{meegan2009}) X-ray flux profile (6-12 keV, red color) in the same panel. The first B6.0 flare started at $\sim$16:50 UT, peaked at $\sim$16:53 UT, and ended at $\sim$17:10 UT. Note that there was another short duration B-class flare during 16:50-16:55 UT interval that occurred in an AR NOAA 12349 (S21W34). That flare contaminated the first peak of the B6.0 X-ray burst. The best way to exclude the contribution from the other AR is to plot the EUV flux profile of the studied AR in order to show the periodic behaviour of the first B6.0 flare (please see next section). The second C2.3 flare started at $\sim$17:30 UT, peaked at $\sim$17:39 UT, and ended at $\sim$17:45 UT. Interestingly, in both soft (6-12 keV) and hard (12-25 keV) X-ray time profiles we detected oscillatory behaviour during both flares.

%------------------------------------------------------------------------------------ 
\begin{figure*}
\centering{
\includegraphics[width=8.0cm]{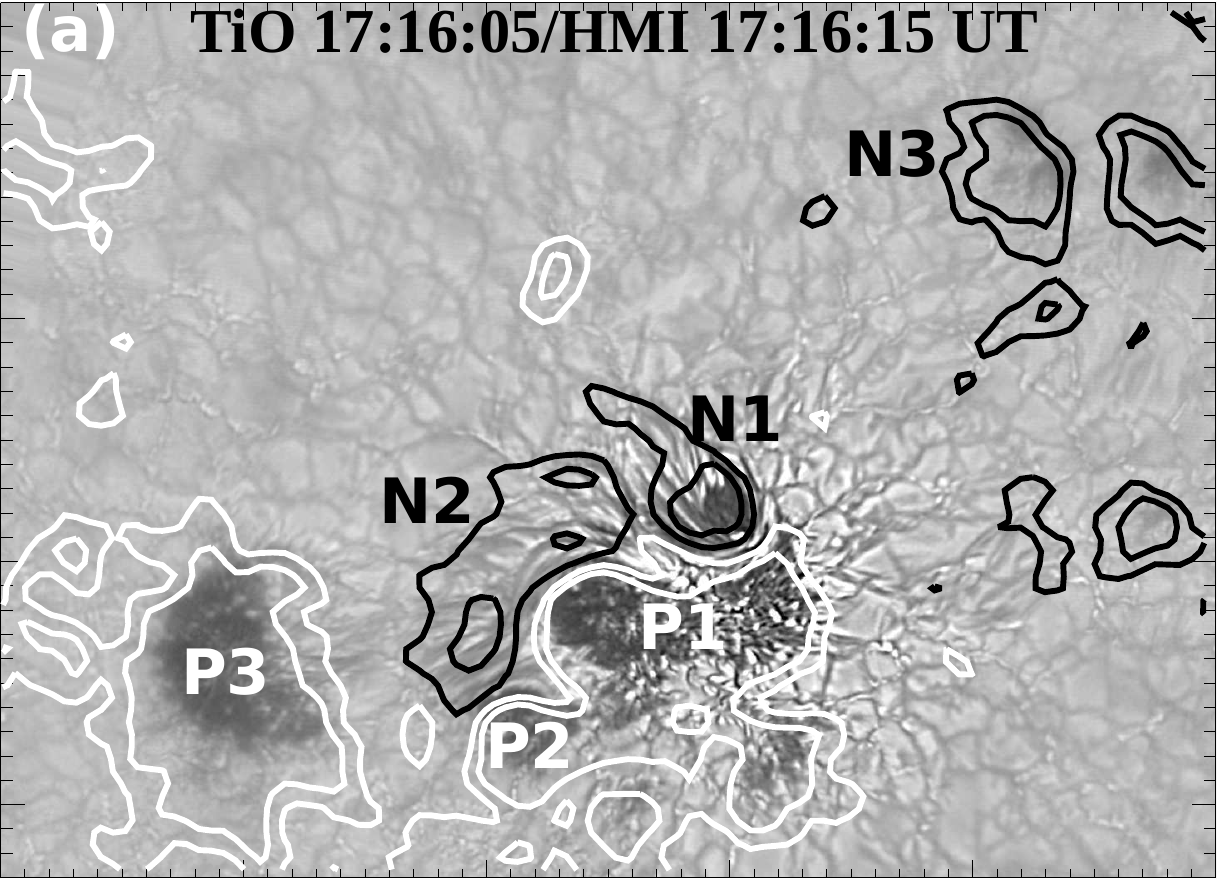}
\includegraphics[width=8.1cm]{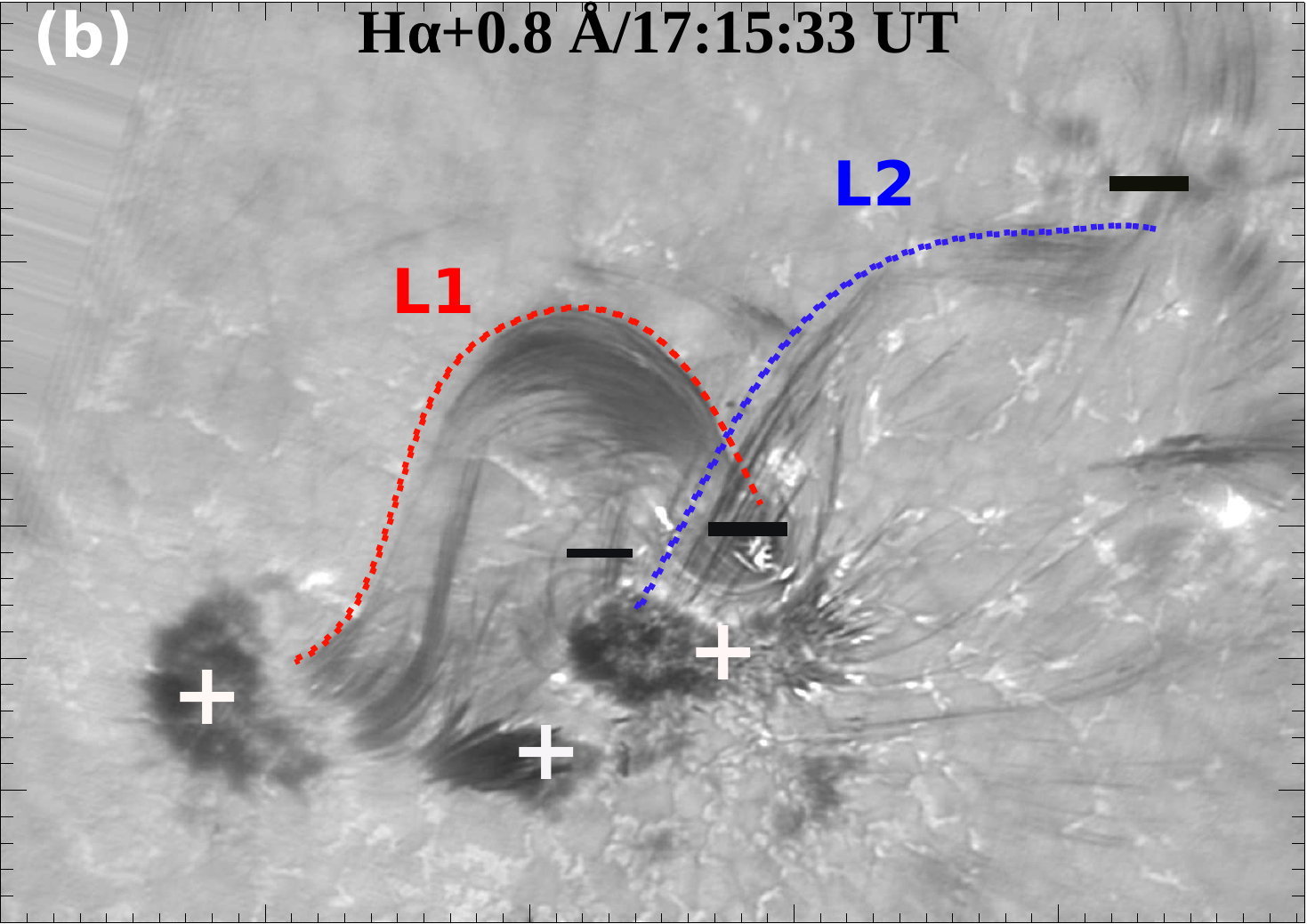}
\includegraphics[width=8.0cm]{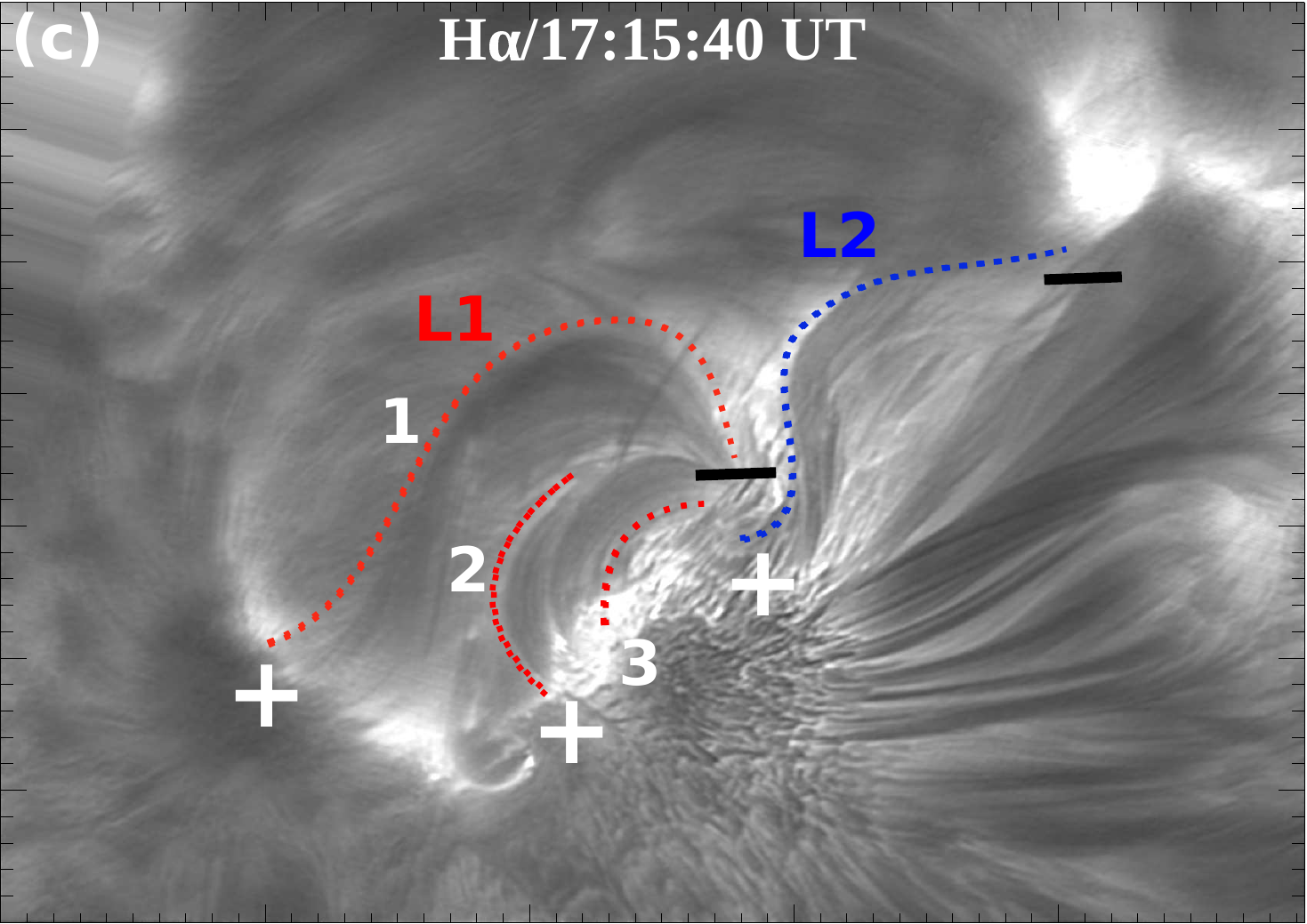}
\includegraphics[width=8.1cm]{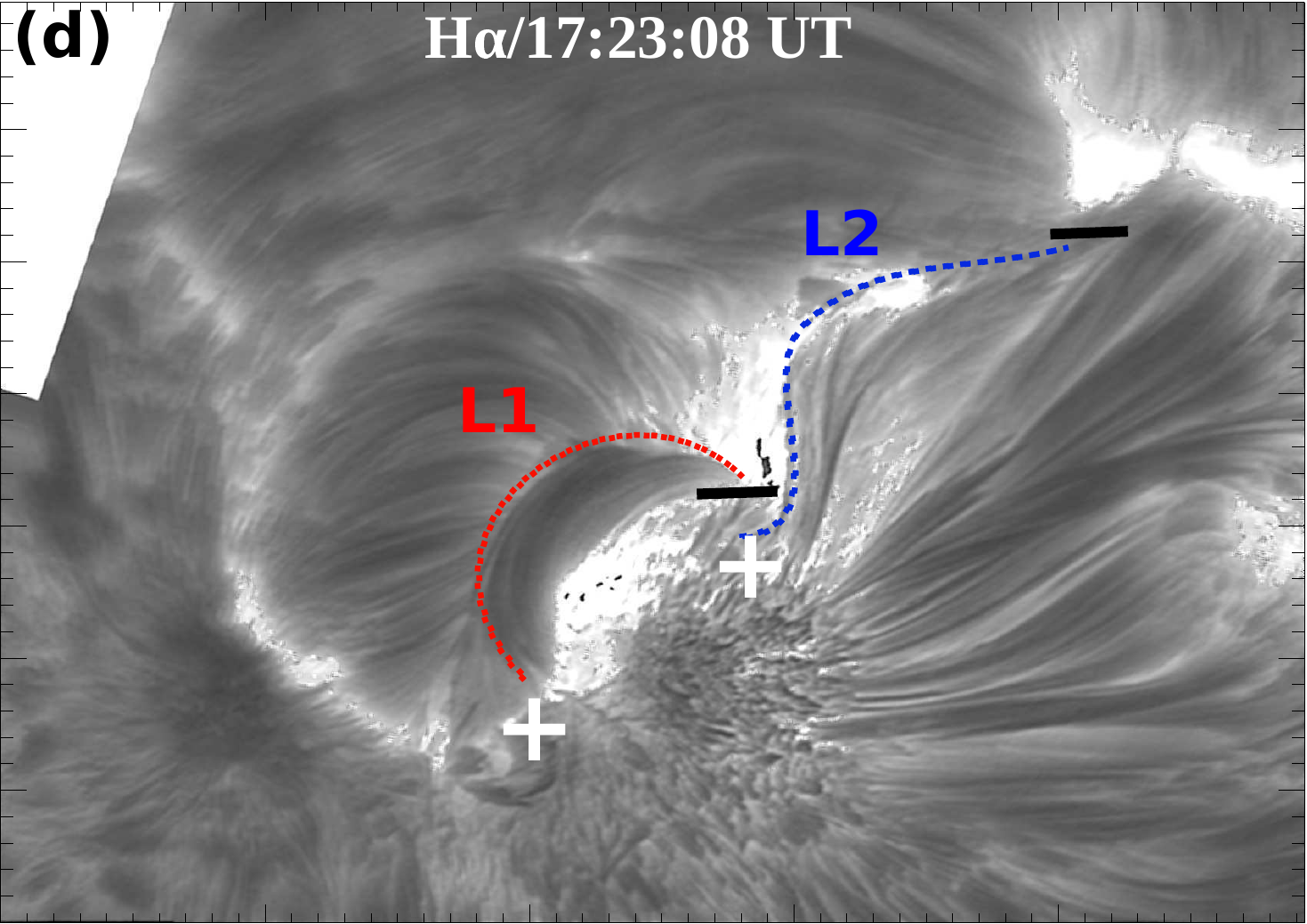}
}
\caption{(a) NST TiO (7057 \AA~) image overlaid by HMI magnetogram contours of positive (white) and negative (black) polarities. The contour levels are $\pm$200 and $\pm$500 Gauss. (b-d) NST H$\alpha$+0.8 \AA~ and H$\alpha$ line center images showing two loop system (L1 and L2) and their connectivity (marked by + and - signs). The size of each image (a-d) is $\sim$50\arcsec$\times$35$\arcsec$. Each division on the x- and y axes is equal to 1$\arcsec$. }
\label{nst1}
\end{figure*}

%------------------------------------------------------------------------------------ 
%%%%%%%%%%%%%%%%%%%%%%%%%%%%%%%%%%%%%%%%%%%%%%%%%%%%%%%%%%%%%%%%%%%%%%%%%%%%%%%%%%%%
%%%%%%%%%%%%%%%%%%%%%%%%%%%%%%%%%%%%%%%%%%%%%%%%%%%%%%%%%
\subsection{The First B6.0 flare}
Figure \ref{fl1}(a,b) displays the AIA 171 and 304 \AA~ images of the AR  overplotted  with the HMI magnetogram contours of positive (white) and negative (black) polarities. In the AIA 171 \AA~ image, there are a set of overlying coronal loops (above the flare site) connecting opposite polarities of the AR. The AIA 304 \AA~ image shows the connectivity of the dark/cool chromospheric/transition region loops (indicated by arrows). 
We do not see any pre-existing filament lying along the polarity inversion line (PIL). Therefore, the dark structures are mentioned as a system of cool loops connecting opposite polarity fields. Cool loops are also seen at this location in the high-resolution NST H$\alpha$ images (Figure \ref{nst1}). Figure \ref{fl1}(c-h) shows the zoomed in area outlined by the rectangular box in the panel (a)) as it appears in the AIA 171, 1600, 304, and 131 \AA~ channels. The initial brightening occurred at $\sim$16:49-16:50 UT between N1 and P1 and a loop-like structure appeared at $\sim$16:58 UT in cool (1600, 171, and 304 \AA) and hot (131 \AA) AIA channels, which suggests a wide range of plasma temperatures ($\sim$0.1-10 MK) within the loop. The bottom panels show few dark threads (marked by arrow) in the loop, which indicates the existence of cool chromospheric plasma within the loop. 

To determine the plasma temperature (T) and emission measure (EM) of the loop, we created the peak temperature and EM maps using an automatic differential emission measure (DEM) code developed by \citet{asc2013}. This code utilizes six channels of near simultaneous AIA images (171, 304, 193, 211, 131, 94 \AA) to derive the peak T and EM. We used six-channel AIA images taken at $\sim$16:58:08 UT. The peak T and EM map (Figure \ref{dem}) confirm that the bright loop contains a multi-temperature plasma. We observed $\sim$10 MK plasma near both the footpoints of the loop, while the middle section of the loop contained cool plasma ($\leq$1 MK). AIA 304 and 171 \AA~ images did not show the hot portion ($\sim$10 MK) at the end of the loop, whereas the 131 \AA~ shows it (see Figure \ref{fl1}(f,g,h)). This indirectly suggests the existence of $\sim$10 MK plasma at the opposite end of the heated loop (marked by a circle in Figure \ref{fl1}(h)). The EM map also reveals high EM near the ends of the loop and low EM in the middle.

To investigate the particle acceleration/precipitation sites during the flare, we overlaid RHESSI X-ray 6-12 keV (red) and 12-25 keV (blue)  contours (16:56:00-16:56:30 UT) and on the top of an HMI magnetogram, AIA 1600, and 131 \AA~ images (Figure \ref{hessi1}). We reconstructed RHESSI X-ray images in the 6-12 keV and 12-25 keV using the CLEAN algorithm with 30-s integration time. The contour levels are at 85$\%$ and 95$\%$ of the peak intensity. Interestingly, the X-ray sources were co-spatial with sunspots N1 and P1, where the initial brightening in the AIA channels was detected. The AIA 131 and 1600 \AA~ images show the sources location at the footpoint (F1) of the heated loop. We do not see any X-ray source at the opposite footpoint (F2) of the loop. The energy release site is basically between N1 and P1. 

The initial brightening and the location of the X-ray sources at the footpoint F1 suggest that the magnetic reconnection most likely occurred at the same location between the opposite polarity sunspots (N1 and P1). It is likely that a portion of the accelerated electrons produced during the reconnection (above F1) were injected into the loop and then precipitated at the remote footpoint F2, which caused chromospheric evaporation and heating of the footpoints of the loop up to $\sim$10 MK temperatures.

%------------------------------------------------------------------------------------ 
\begin{figure*}
\centering{
\includegraphics[width=8cm]{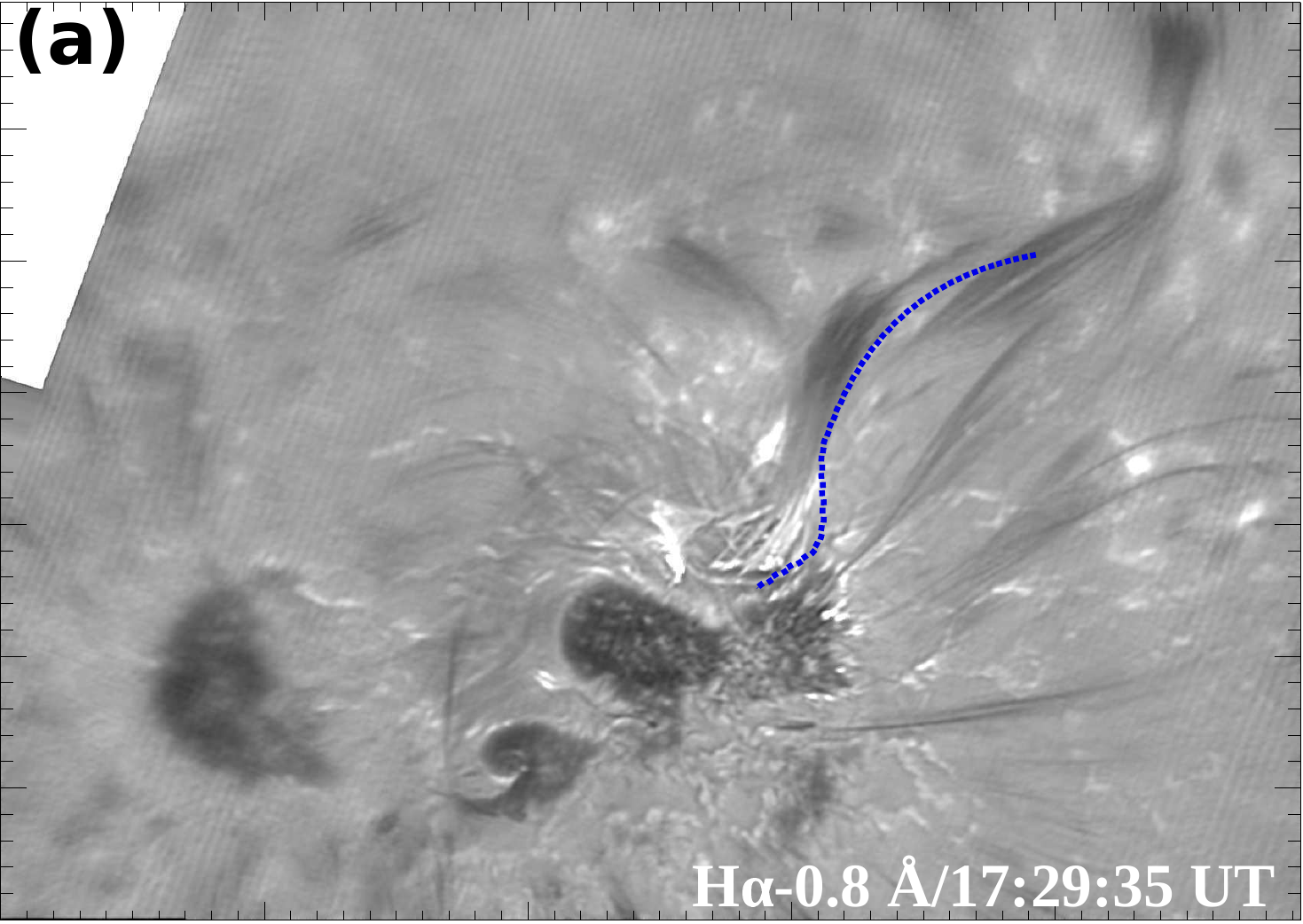}
\includegraphics[width=8cm]{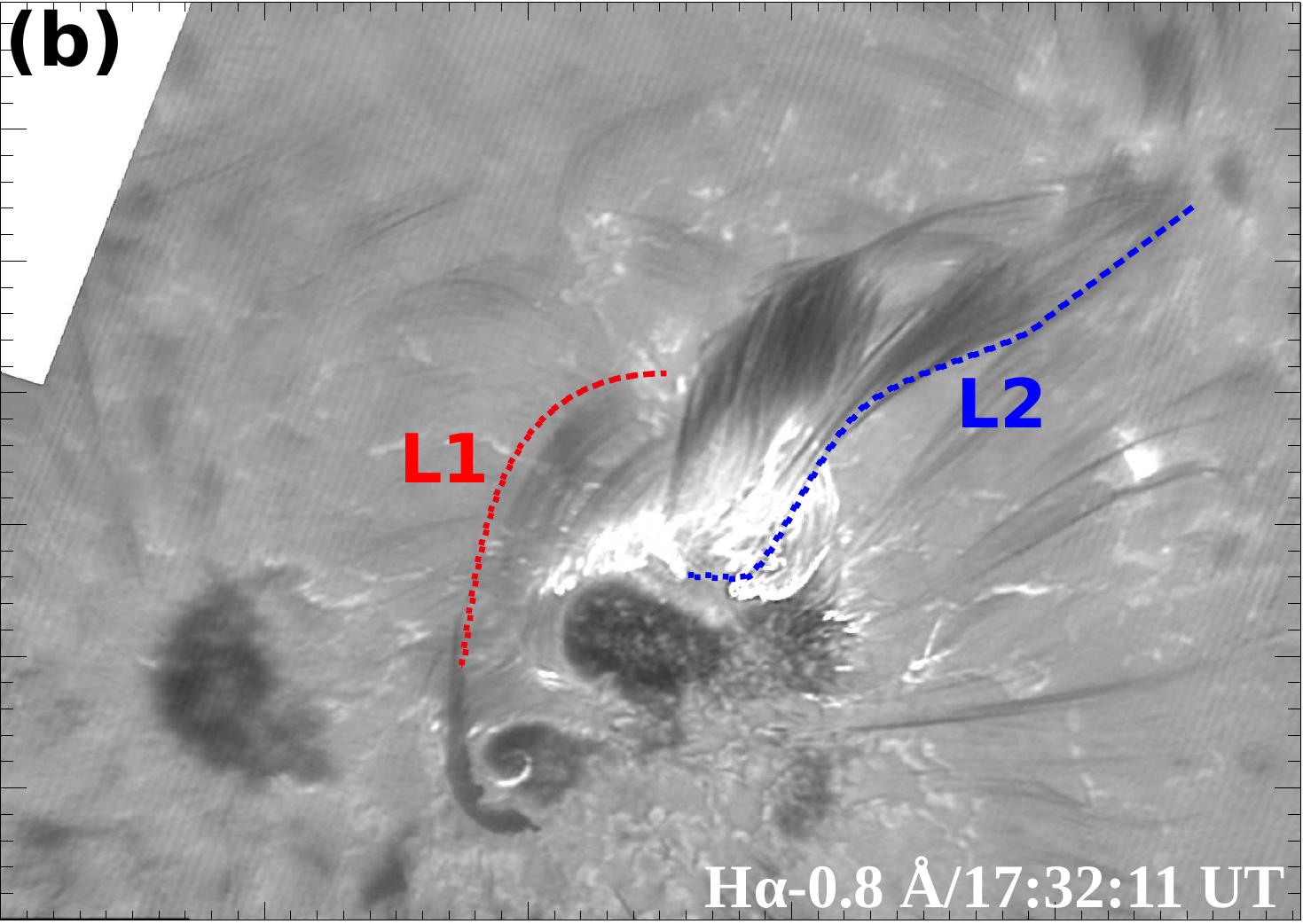}

\includegraphics[width=8cm]{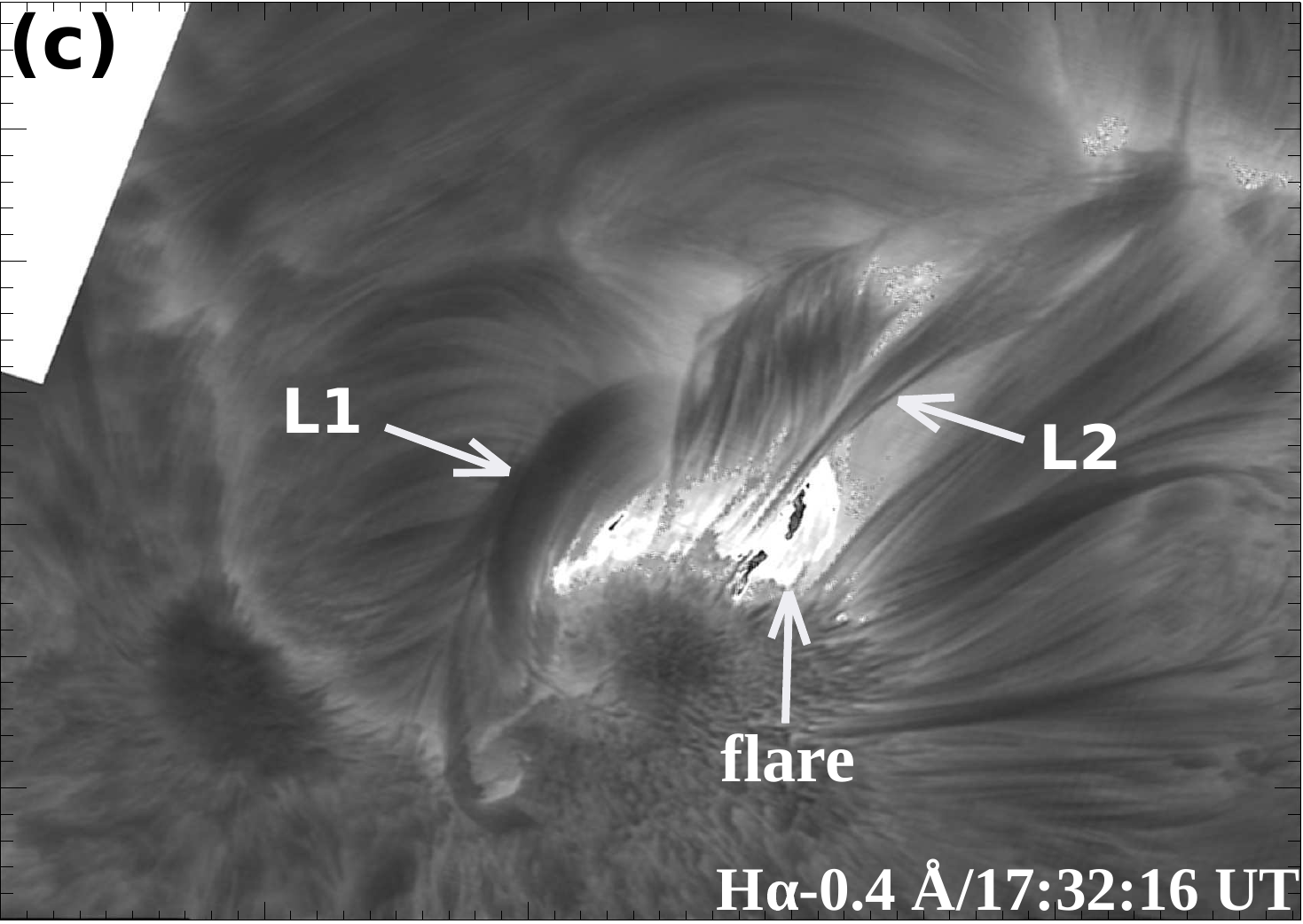}
\includegraphics[width=8cm]{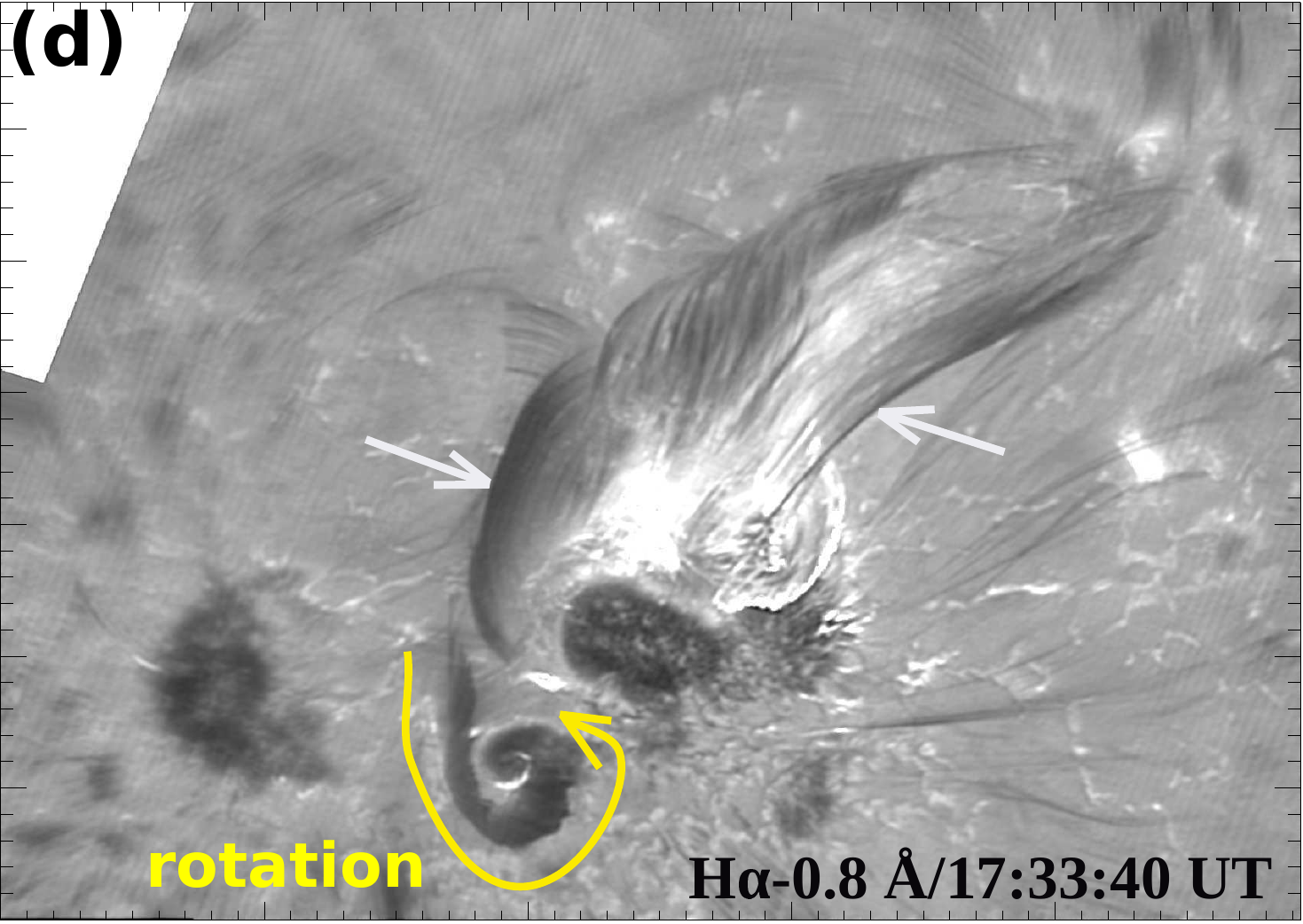}

\includegraphics[width=8cm]{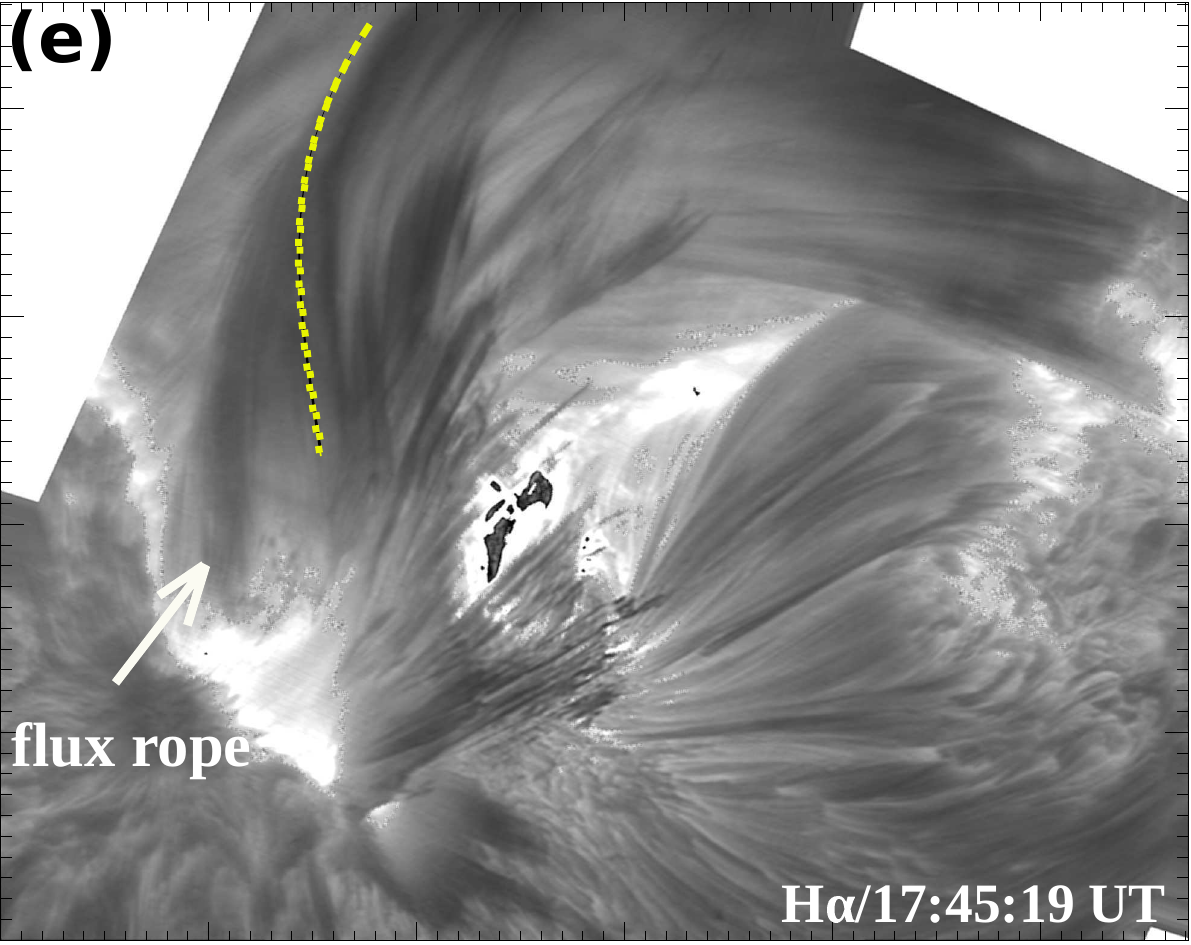}
\includegraphics[width=8cm]{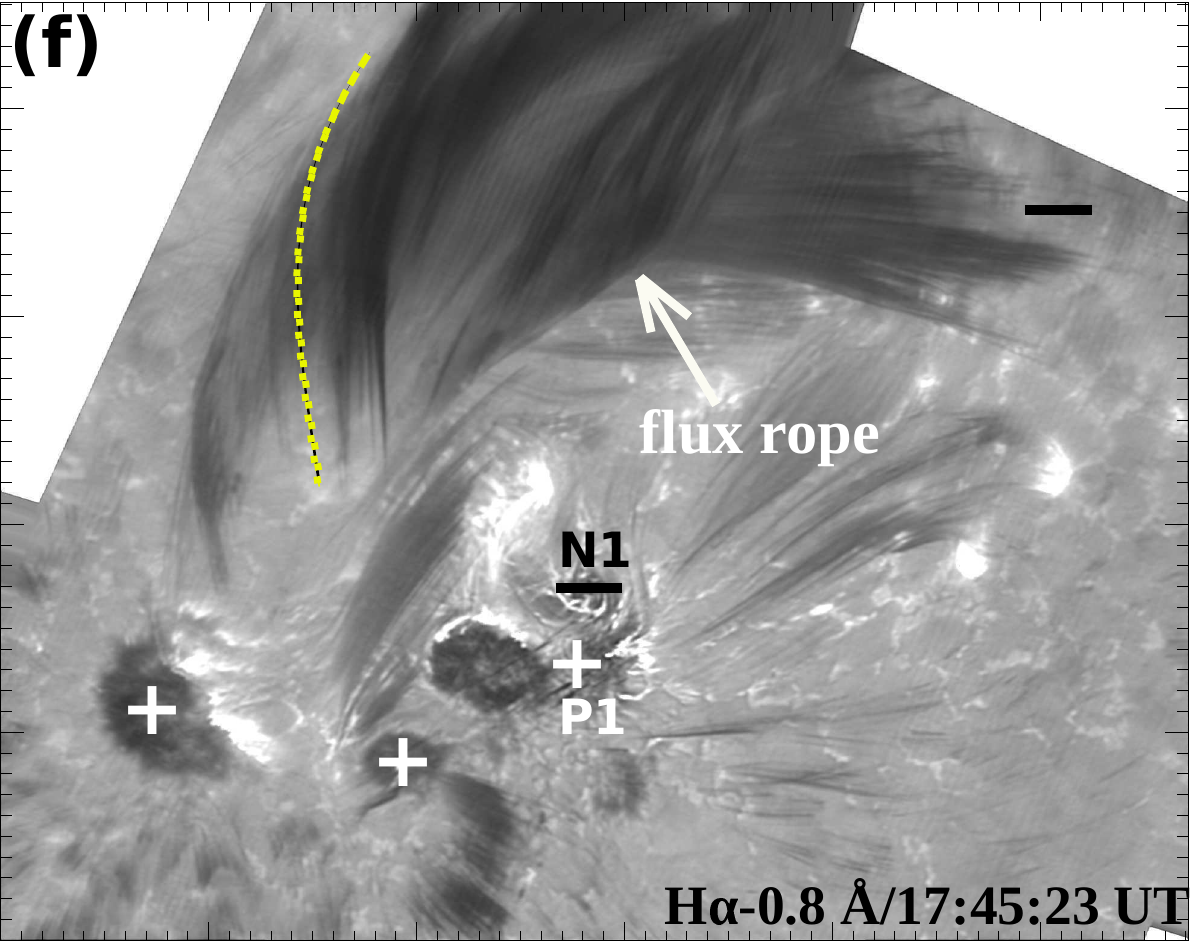}
}
\caption{(a-d) NST H$\alpha$-0.8 \AA, H$\alpha$-0.4 \AA, and H$\alpha$ line center images showing the interaction of two loop system (L1 and L2) and associated C-class flare. The size of each image (a-d) is $\sim$50\arcsec$\times$35$\arcsec$. (e-f) NST  H$\alpha$ line center and H$\alpha$-0.8 \AA~ images showing the flux rope produced by the interaction of cool loops. The size of each image (e-f) is $\sim$57\arcsec$\times$45$\arcsec$. Each division on the x- and y axes is equal to 1$\arcsec$. The coalescence of loops (L1 and L2) and formation of a resulting flux rope is shown in the H$\alpha$ line center (panel(e)) and H$\alpha$-0.8 \AA~ (panels(a,b,d,f)) animations available in the online edition.}
\label{nst2}
\end{figure*}

%------------------------------------------------------------------------------------ 
%%%%%%%%%%%%%%%%%%%%%%%%%%%%%%%%%%%%%%%%%%%%%%%%%

%------------------------------------------------------------------------------------ 
\begin{figure*}
\centering{
\includegraphics[width=5.2cm]{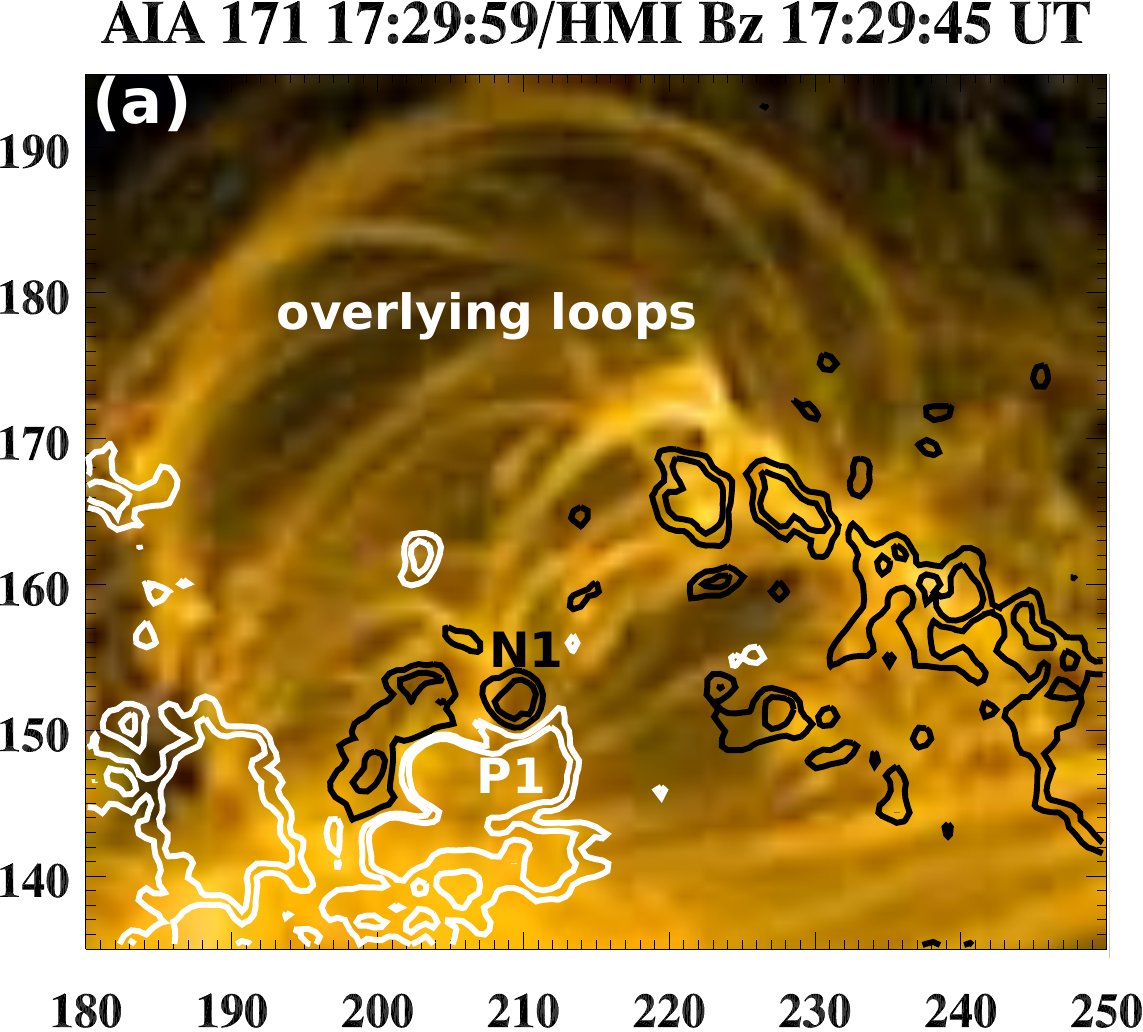}
\includegraphics[width=5.4cm]{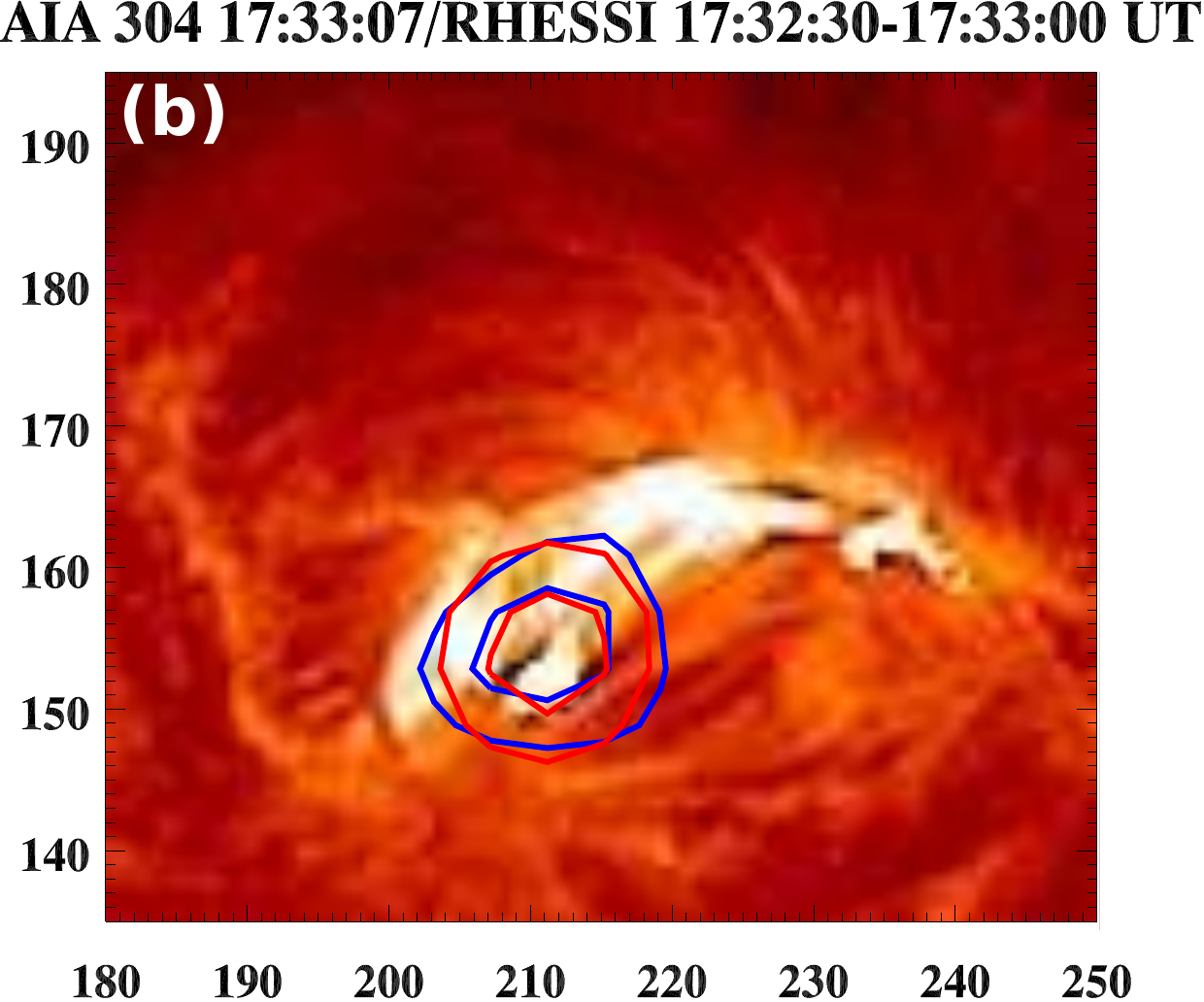}
\includegraphics[width=5.4cm]{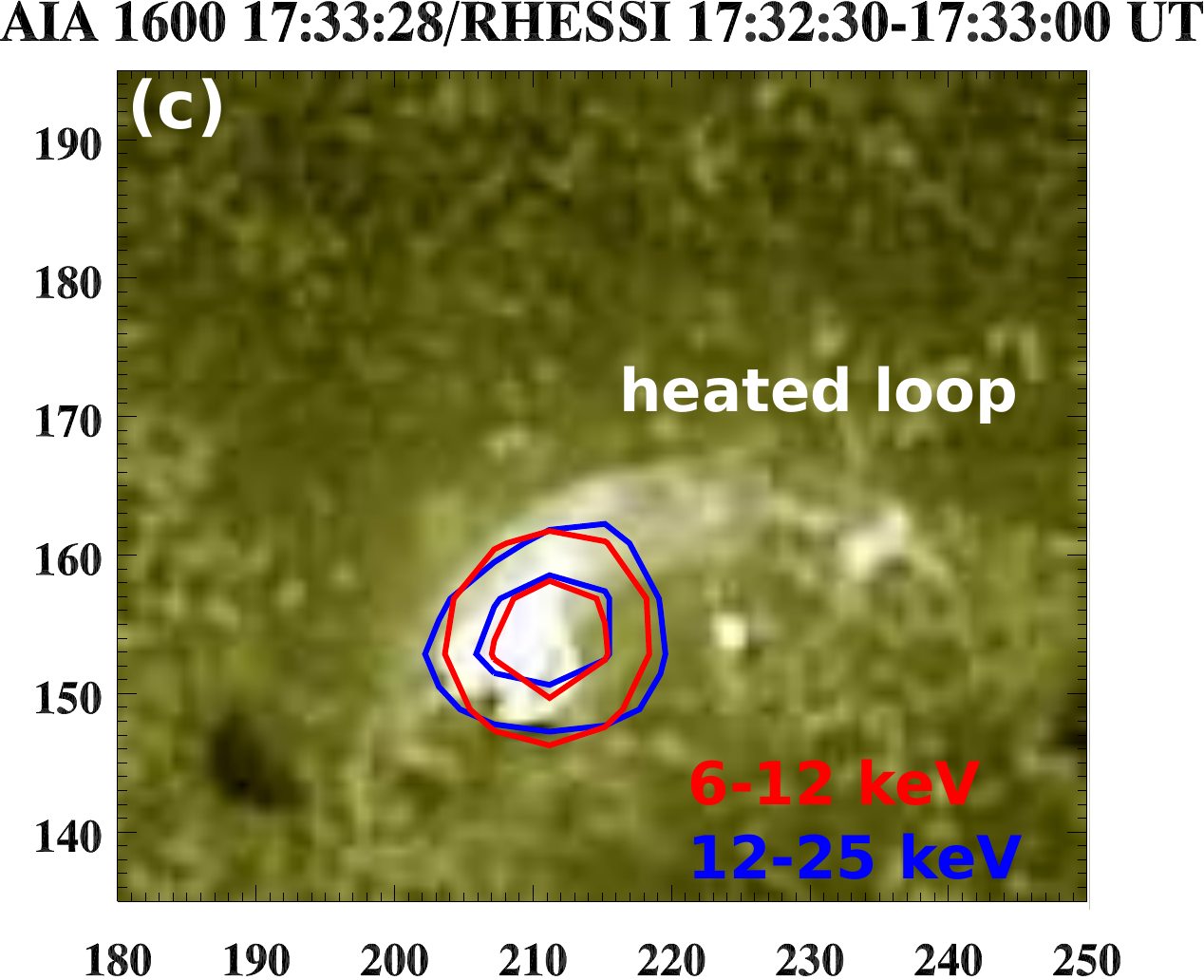}

\includegraphics[width=5.2cm]{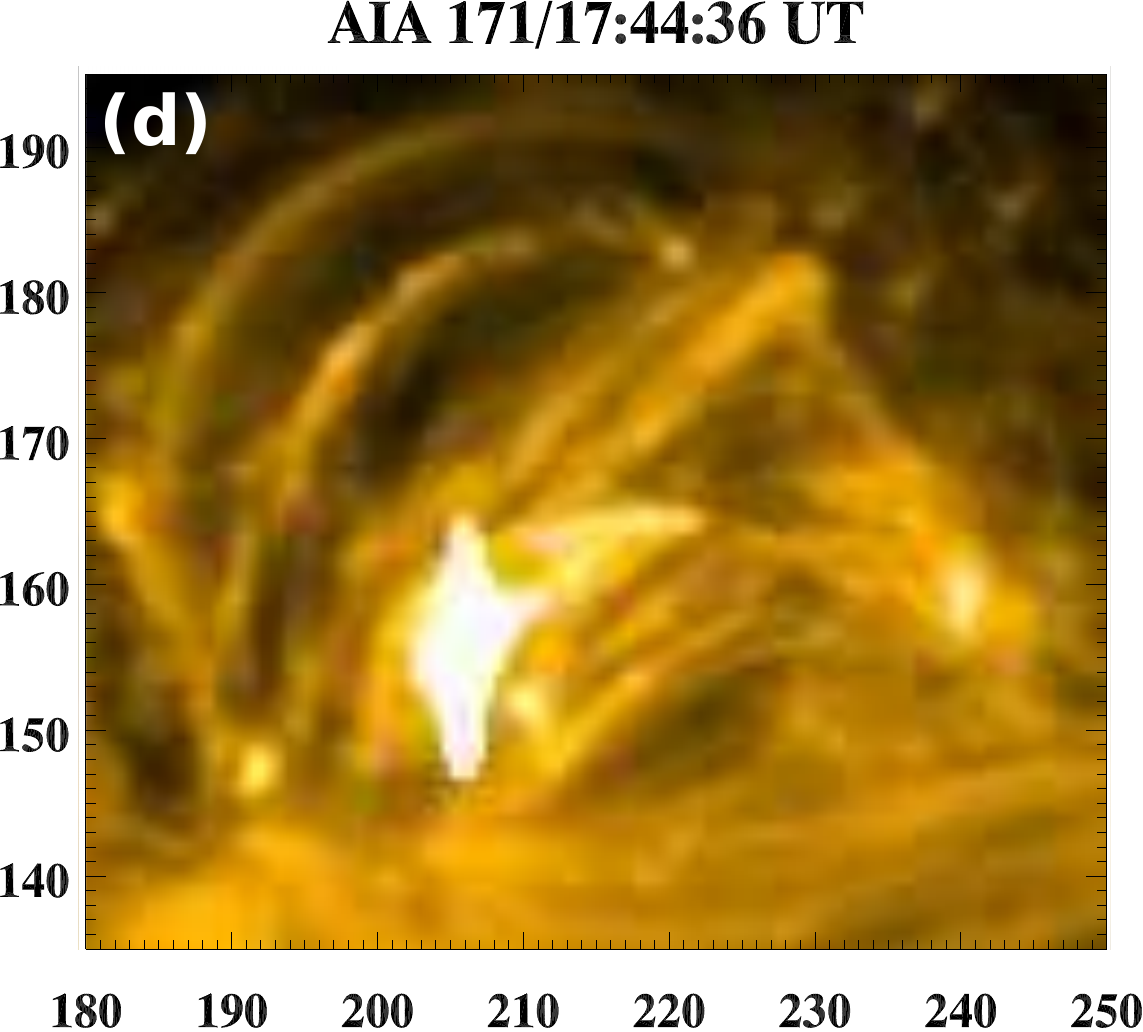}
\includegraphics[width=5.3cm]{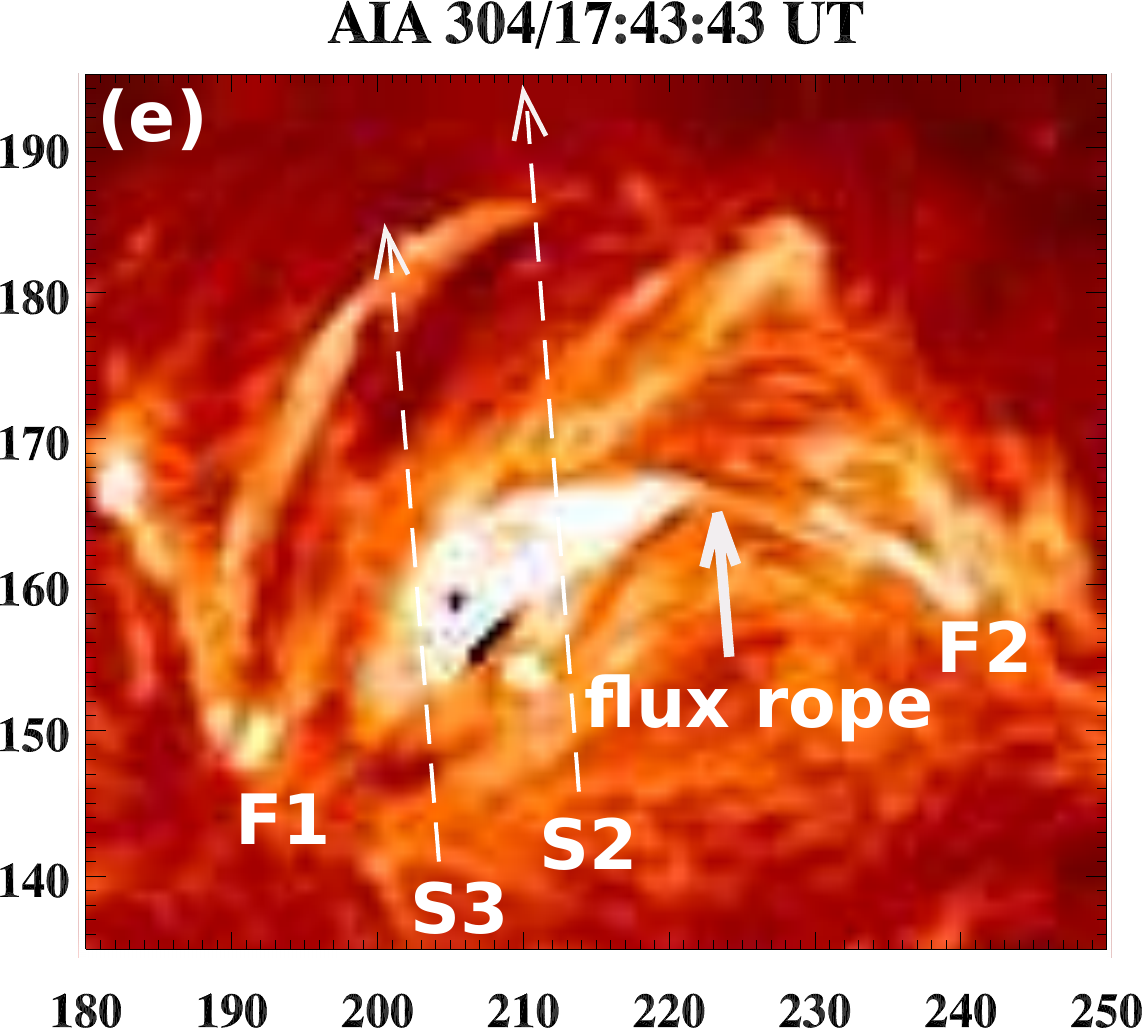}
\includegraphics[width=5.3cm]{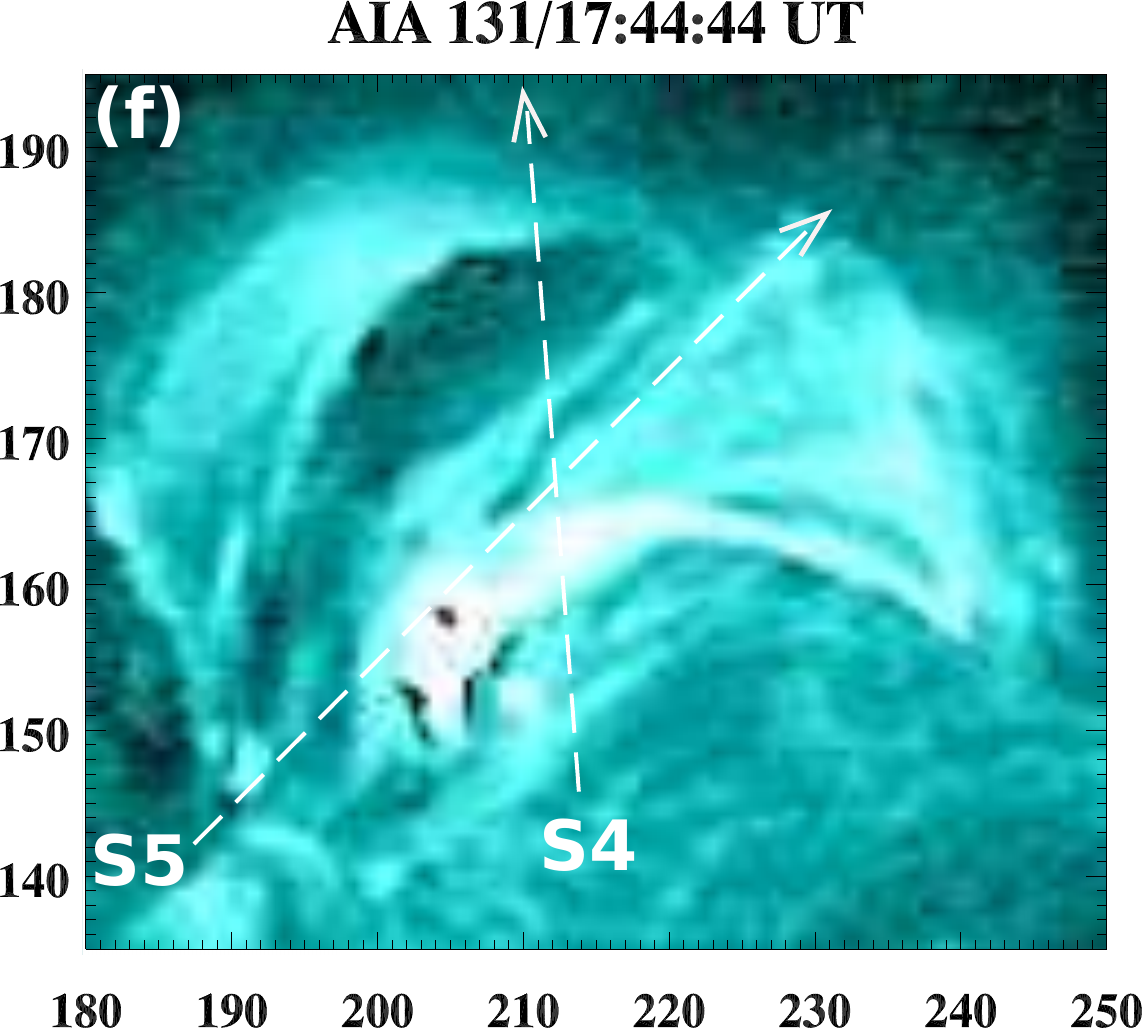}
}
\caption{(a) AIA 171 \AA~ image with HMI magnetogram contours of positive (white) and negative (black) polarities. The contour levels are $\pm$200 and $\pm$500 Gauss. (b-c) RHESSI X-ray contours (red: 6-12 keV, blue: 12-25 keV) overlaid on the AIA 304 and 1600 \AA~ images. The contour levels are 85$\%$ and 95$\%$ of the peak intensity. (d-f) Observation of the flux rope in different AIA channels (304, 171, and 131 \AA). S2, S3, S4, and S5 are the slices used to create the stack plots. F1 and F2 are the new footpoints of the flux rope. The oscillatory reconnection and associated flux rope formation with counterclockwise rotation is shown by a composite movie (AIA 171, 304, and 131 \AA, bottom panels) available in the online edition.}
\label{fr} 
\end{figure*}

%------------------------------------------------------------------------------------

%%%%%%%%%%%%%%%%%%%%%%%%%%%%%%%%%%%%%%%%%%%%%%%%%%%%%%%%%%%%%%%%%%%%%%%%%%%%%%%%%%%%
\subsection{Oscillatory behaviour in X-ray and EUV flux profiles}
The periodic variations of the X-ray flux in 6-12 keV energy range during the B6.0 flare indicate periodic release of energy with five distinct cycles.
To determine the source of the oscillation, we created a time-distance plot (Figure \ref{stack1} (a)) using the AIA 304 \AA~ images (running-difference) along the slice S1 shown in Figure \ref{fl1}(g). The AIA 304 \AA~ channel images show the chromosphere and the transition region. The location of the slice was chosen to be just above the sunspots P1. The slice crosses the footpoint F1 and the initial brightening site. The stack plot shows repeated brightenings occurring along S1 with total five clear brightening features. These periodic brightenings are the result of repetitive magnetic reconnection between N1 and P1. These periodic disturbances can be interpreted by reconnection generated outflows.

To check the upper photospheric (T=5000 K)/transition region (T=0.1 MK) responses of oscillatory energy release, we also selected a small region (Figure \ref{fl1}(e)) that includes the sunspot P1 and N1 in AIA 1600 \AA~ intensity images and extracted the mean counts within it. Figure \ref{stack1}(b) shows five AIA 1600 \AA~ peaks during the flare. Note that the AIA 1600 \AA~ plot shows the intensity oscillation (during the flare) over the sunspot region. These intensity oscillations are likely to be the result of oscillatory reconnection at F1 during the B6.0 flare.

As we mentioned earlier, during the first X-ray peak of the B6.0 flare (16:51-16:54 UT), there was a B-class flare that occurred in a different AR. The time-distance plot shows that the first burst in the 6-12 keV flux has as well contribution from the studied B6.0 flare. Also, we can see simultaneous $\sim$3 minute oscillation in the AIA 1600 \AA~ and 304 \AA~ channels. The periodic brightenings above spots P1 and N1 strongly indicate the occurrence of repeated reconnections at the F1 footpoint of the heated loop. We also plotted the RHESSI 6-12 keV flux (blue curve) along with the AIA 1600 mean counts, which shows a good correlation between X-ray and EUV periodic energy release.

%%%%%%%%%%%%%%%%%%%%%%%%%%%%%%%%%%%%%%%%%%%%%%%%%%%%%%%%%%%%%%%%%%%%%%%%%%%%%%%%%%%%%%%
%------------------------------------------------------------------------------------ 
\begin{figure}
\centering{
\includegraphics[width=8.5cm]{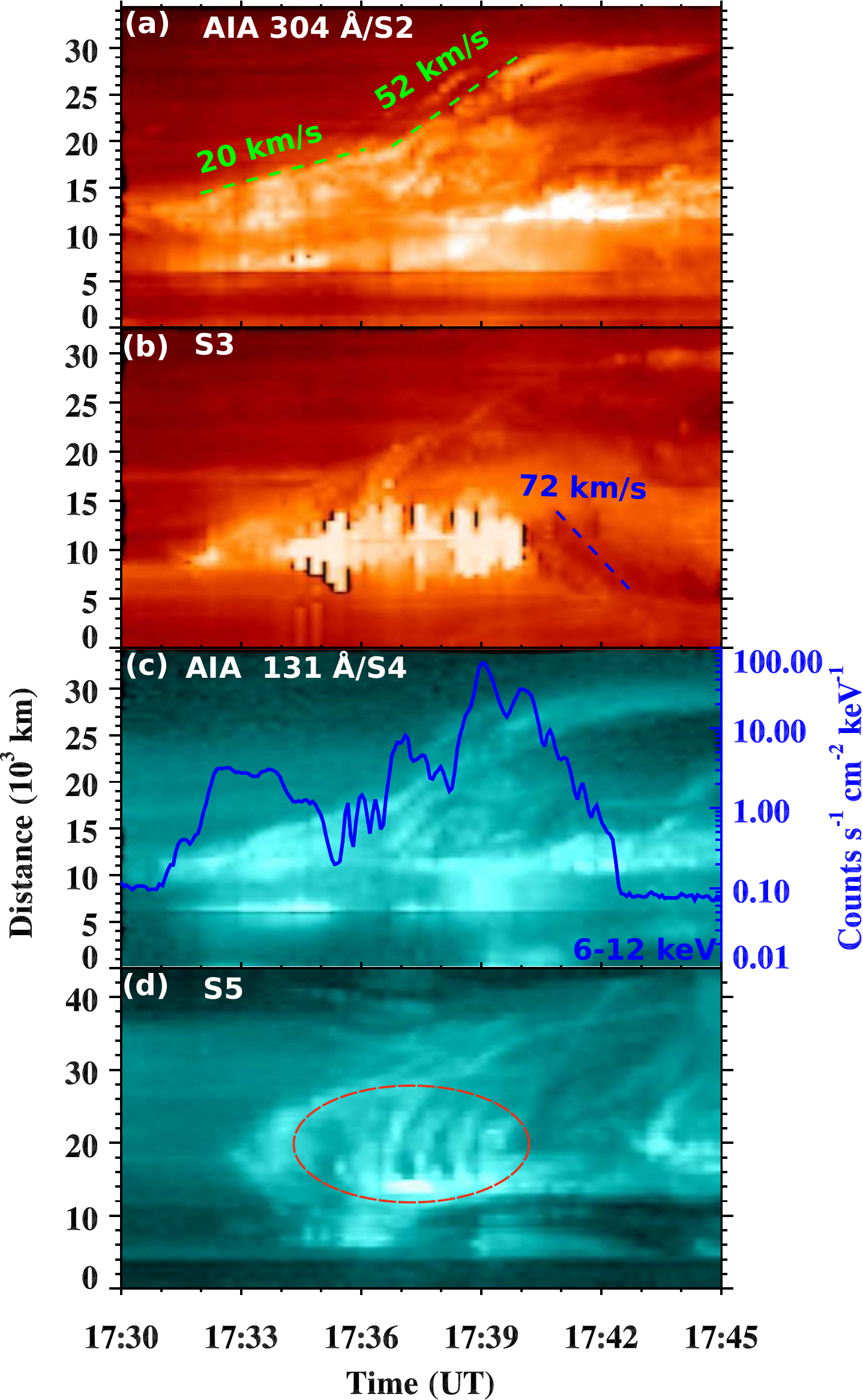}
}
\caption{Stack plots along the slices S2, S3, S4, and S5 using AIA 304 and 131 images. The Fermi GBM X-ray flux in 6-12 keV channel is also included in panel (c). The periodic rise of multiple reconnected loops is shown within a dashed ellipse (panel (d)). The co-temporal X-ray bursts (panel (c)) and strong chromospheric brightening in the AIA 304 \AA~ channel (panel (b)) indicate the particle acceleration/precipitation during the periodic rise of reconnecting loops.}
\label{stack2}
\end{figure}
%%%%%%%%%%%%%%%%%%%%%%%%%%%%%%%%%%%%%%%%%%%%%%%%%%%%%%%%%%%%%%%%%%%%%%%%%%%%%%%%%%%%
\subsection{Second C2.3 flare: merging of cool loops in the chromosphere and the flux rope formation}
To investigate the characteristics of the magnetic structure at the flare site, we utilized high resolution NST TiO and VIS images taken before and during the C2.3 flare. The first B6.0 flare was not observed by the NST. The AIA resolution (0.6$\arcsec$ per pixel) is not sufficient  to resolve the smallest-scale chromospheric fine structures, therefore, the NST observations (0.029$\arcsec$ per pixel) are extremely useful for that purpose. In Figure \ref{nst1}(a), we show a  photospheric TiO image before the flare onset (17:16:05 UT) overlaid with HMI magnetic field contours of the positive (white) and negative (black) polarities. The positive and negative polarity sunspots are indicated by P1, P2, P3, N1, N2, and N3. The initial AIA flare brightening (see Figure \ref{fl1}(d)) was observed between N1 and P1. The negative polarity sunspot, N1, had an  elongated shape. The H$\alpha$ red wing image (H$\alpha$+0.8 \AA, \ref{nst1}(b)) shows two system of loops (17:15:33 UT) at the flare site: loop L1 (red) connecting P3 and N1, and loop L2 (blue) spanning between P1 and N3. Both loops are highlighted with red and blue dotted lines, and their footpoints polarities are indicated  by ``+'' and ``-'' signs. The co-temporal H$\alpha$ line center image (\ref{nst1}(c)) of the same region shows multiple chromospheric loops marked by 1, 2, and 3. Note that the loop 1 is the same as loop L1 whereas loops 2 and 3 connect different element of positive flux (P2) in the photosphere. However, loop L2 has a single structure. After some time (17:23:08 UT, \ref{nst1}(d)), we observed the disappearance of the higher loop, 1, and merging of loops 2 and 3 into a single structure marked as  L1. The eastern footpoint of the merged loop L1 in Figure \ref{nst1}(d) is now connected to P2, which is different from L1 as shown in \ref{nst1}(b). So finally we have two loop system in the chromosphere (L1 and L2) before the onset of the C2.3 flare. 

Figure \ref{nst2} displays a series of H$\alpha$ line center, H$\alpha$-0.4 \AA~ and H$\alpha$-0.8 \AA~ flare images. During the initiation of the C2.3 flare, the  initial brightening occurred at 17:29:35 UT at the footpoints of the penumbral filaments between P1 and N1 along with the activation of a cool loop L2 (blue dotted line). This is the similar loop L2 as shown in Figure \ref{nst1}. The loop was present before the trigger of the C2.3 flare. The right footpoint of the loop L2 was anchored in the negative polarity sunspot N3, while its left footpoint was anchored in a positive polarity fields between N1 and P1.
 
A clear interaction between two loops L2 (blue) and L1 (red) can be observed starting at 17:32:11 UT as evidenced by a chromospheric flare brightening seen between the interacting loops (Figure \ref{nst2}(b,c)). In addition, we also noted the counterclockwise rotation of the chromospheric loop structures above the positive polarity sunspot P2 (marked by a yellow arrow in Figure \ref{nst2}(d)). The viewing direction is toward the positive footpoint of the flux rope. In addition, we also observe the counterclockwise rotation of P2 in NST TiO (photospheric) movie. Therefore, it is likely that the rotation of P2 can increase the twist of the left-handed flux rope.

The loop L2 seems to display left-handed twisted threads, whereas loop L1 was smaller and seemed to have a potential field structure. After merging of loops L1 and L2, we see a combined twisted flux rope structure (at $\sim$17:45 UT). The interaction between L1 and L2 that continued for about 15 min to produce a single merged flux rope with rotation in the counter-clockwise direction. The direction of the loop threads (marked by yellow dotted line) suggests that the flux rope has as well the left-handed twist. The counterclockwise rotation corresponds to an increasing the left-handed twist in the rope. The high resolution NST images allowed us to observe for the first time details of a loop-loop interaction in the chromosphere. 

In Figure \ref{fr}, we show the C2.3 flare brightening and the flux rope structure in the AIA channels. The AIA 171 \AA~ image (overlaid by HMI magnetogram contours) just before the flare onset (at 17:29:59 UT) shows the connectivity of the overlying loops at the flare site (Figure \ref{fr}(a)). It is obvious that there were several overlying loops at different heights. At least, two sets of loops were clearly observed, (i.e., low lying and higher loops). AIA 304 and 1600 \AA~ images at $\sim$17:33 UT clearly revealed the appearance of a  hot loop with the onset of the impulsive brightening at the footpoint F1 (Figure \ref{fr}(b,c)). The enlarged view of the interacting loops L1 and L2 has been already discussed with the help of the NST H$\alpha$ images. The RHESSI X-ray image show a single X-ray sources (red: 6-12 keV, blue: 12-25 keV) located over the left footpoint of the loop L2, exactly between the interacting loops L1 and L2 (Figure \ref{fr} (b,c)). The location of the sources suggest the particle acceleration/precipitation site to be between N1 and P1 sunspots (i.e., as a result of reconnection between L1 and L2). The RHESSI X-ray movie does not show any shift of the X-ray source during the C2.3 flare energy release, suggesting a stationary particle precipitation site.

The AIA composite movie clearly shows the dynamics of the flux rope with its counterclockwise rotation. We see a resulting merged flux rope structure during 17:43-17:44 UT (Figure \ref{fr}(d-f)). The left-handed strands are observed in the flux rope structure (see AIA 304 \AA~ image). The flux rope could not escape from the overlying arcade loops (i.e., no removal of the overlying field) and the flux rope plasma drained back to the solar surface after the counterclockwise rotation.

To study the dynamics/kinematics of the flux rope along with the signatures of magnetic reconnection, we analyse AIA 304, 171, 131, and 1600 \AA~  images during the C2.3 flare (17:30-17:45 UT). Figure \ref{stack2} displays the stack plots created along the slices S2, S3, S4, and S5 marked on the AIA 304 and 131 \AA~ images. The slices S2 and S4 are across the flux rope and showing the two rise phases, (i) slow rise with $\sim$20 km s$^{-1}$ during 17:30-17:36 UT and (ii) a fast-rise with 52 km s$^{-1}$ during 17:36-17:41 UT. The flux rope showed rotation in its positive footpoint area and could not go further after reaching a projected height of $\sim$30 Mm. We included Fermi GBM 6-12 keV X-ray flux in order to compare the flux rope dynamics along with the particle acceleration (X-ray bursts). We see the first energy-release/burst (17:30-17:35 UT) during the interaction of L1 and L2 in the chromosphere as seen in the NST H$\alpha$ images. During the interaction time, resulting flux rope rises slowly with $\sim$20 km s$^{-1}$. Later, we see multiple X-ray bursts with fast-rise of the rope and strong chromospheric brightenings below the flux rope during 17:36-17:41 UT. Interestingly, we see multiple loops/brightening patches (in the AIA 131 \AA~ channel, marked within ellipse) moving above the reconnection site (i.e., below the flux rope) suggesting the occurrence of oscillatory reconnection generated outflows. The fast-rise of the flux rope correlates with the quasi-periodic variation in the X-ray flux (17:36-17:41 UT), which suggests the quasi-periodic reconnection among the loops during the expansion of the the flux rope. 
We also see the counterclockwise motion ($\sim$72 km s$^{-1}$) of the chrmospheric structure (or loops) rooted in P2 that merged with the new footpoint of the flux rope.
 
The comparison of the RHESSI X-ray sources location with the periodic brightening site (within ellipse) observed in the AIA 131 \AA~ channel (Figure \ref{stack2}(d)) suggests the occurrence of oscillatory reconnection (above the sunspots P1 and N1) during the interaction of the cool loops L1 and L2. AIA 304, 171, and 131 \AA~ composite movie shows the initial brightening between L1 and L2, followed by expansion of the rope and associated brightening toward eastern side. In addition, the outflows observed in the AIA 131 \AA~ channel are quasi-periodic as well and the period of oscillation ($\sim$0.5-1 min) is different from the preceding B6.0 flare.      
%%%%%%%%%%%%%%%%%%%%%%%%%%%%%%%%%%%%%%%%%%%%%%%%%%%%%%%%%%%%%%%%%%%%
\begin{figure*}
\centering{
\includegraphics[width=9.5cm]{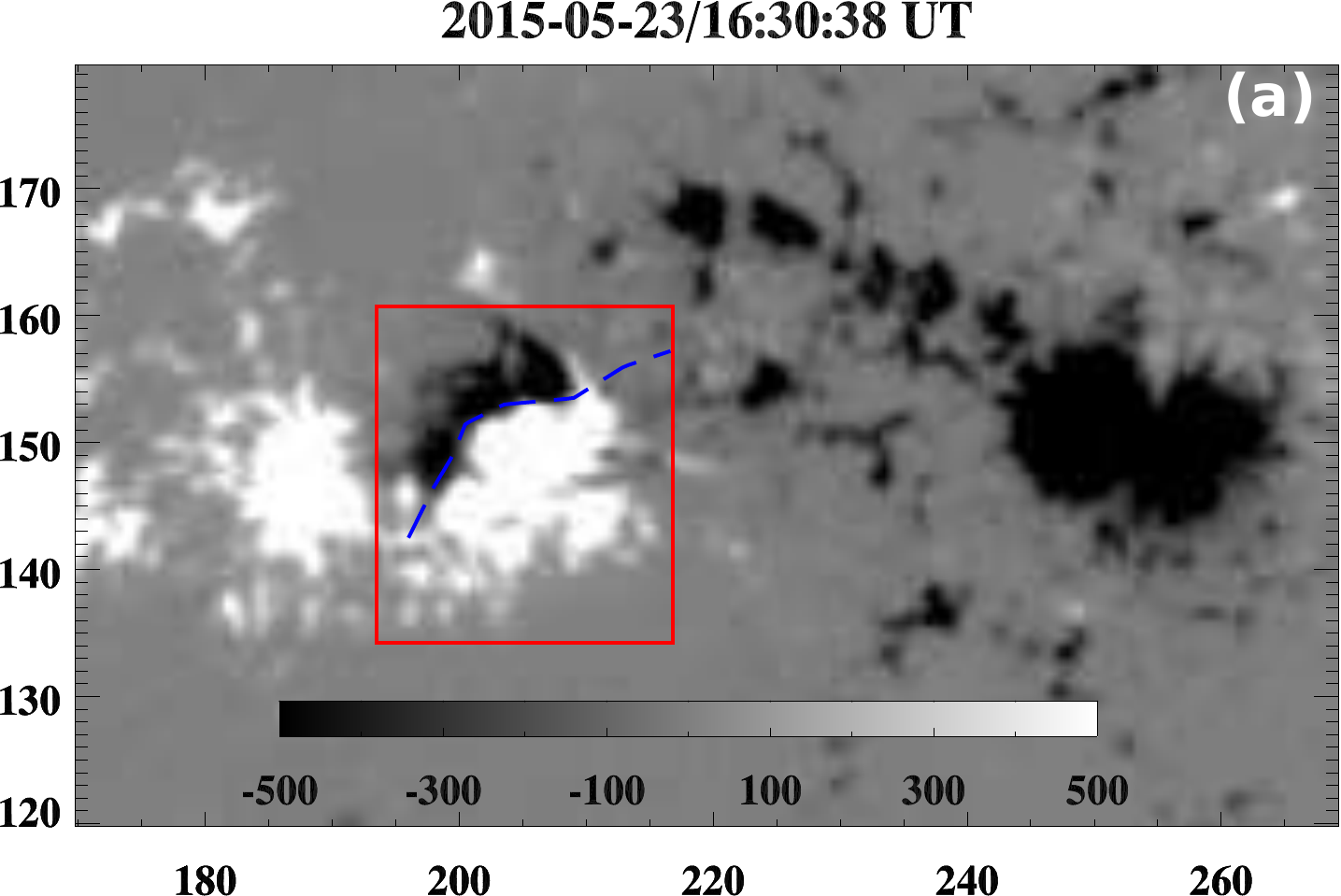}
\includegraphics[width=6.5cm]{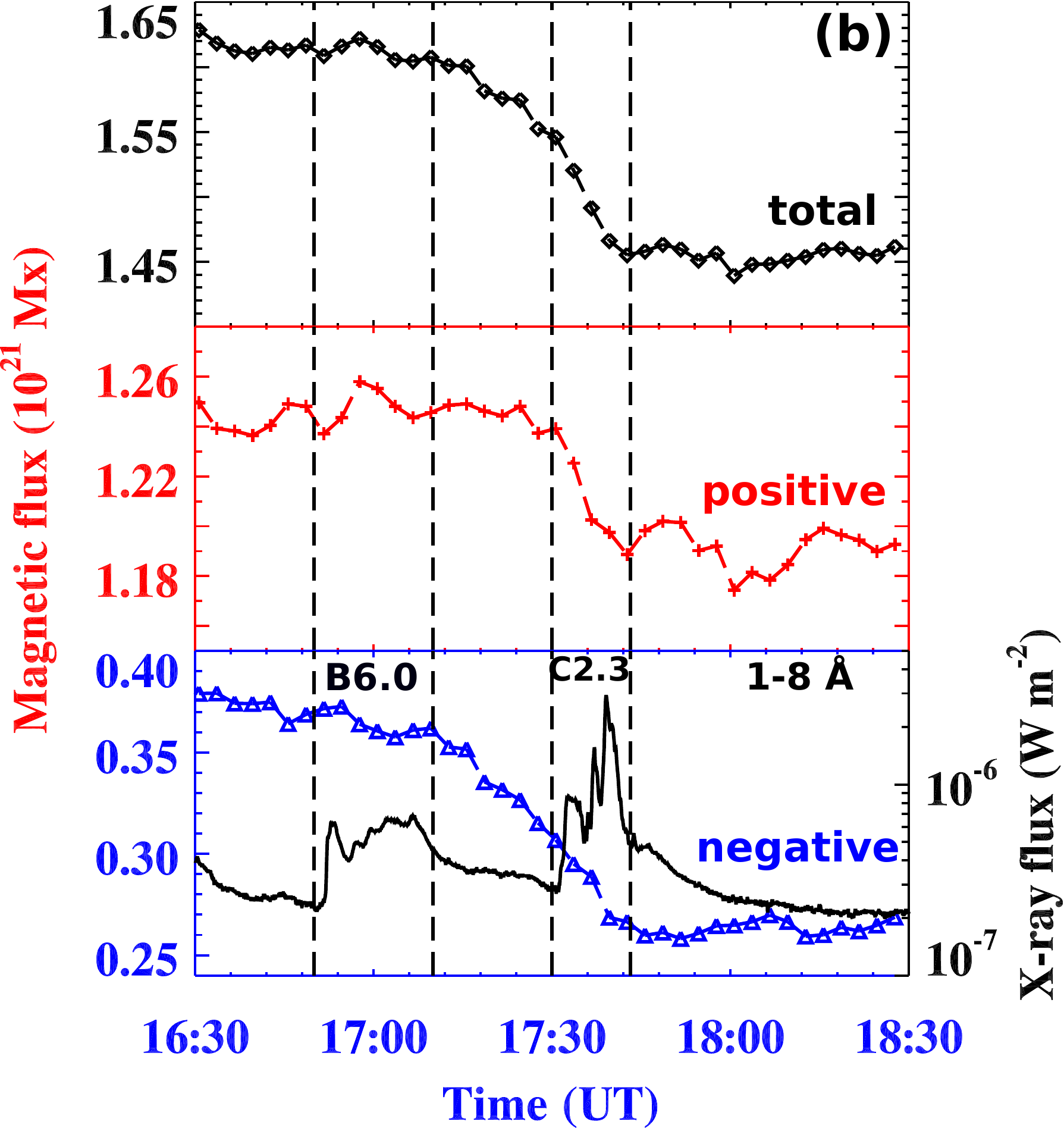}

\includegraphics[width=4cm]{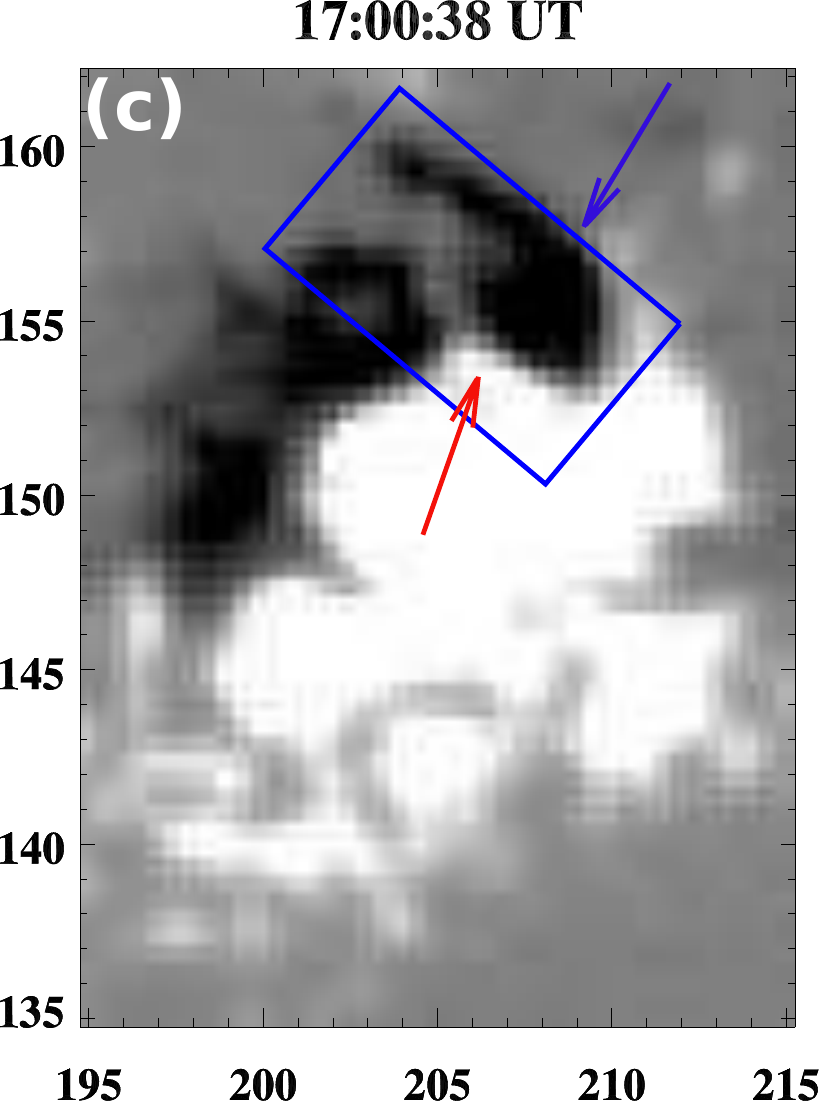}
\includegraphics[width=4cm]{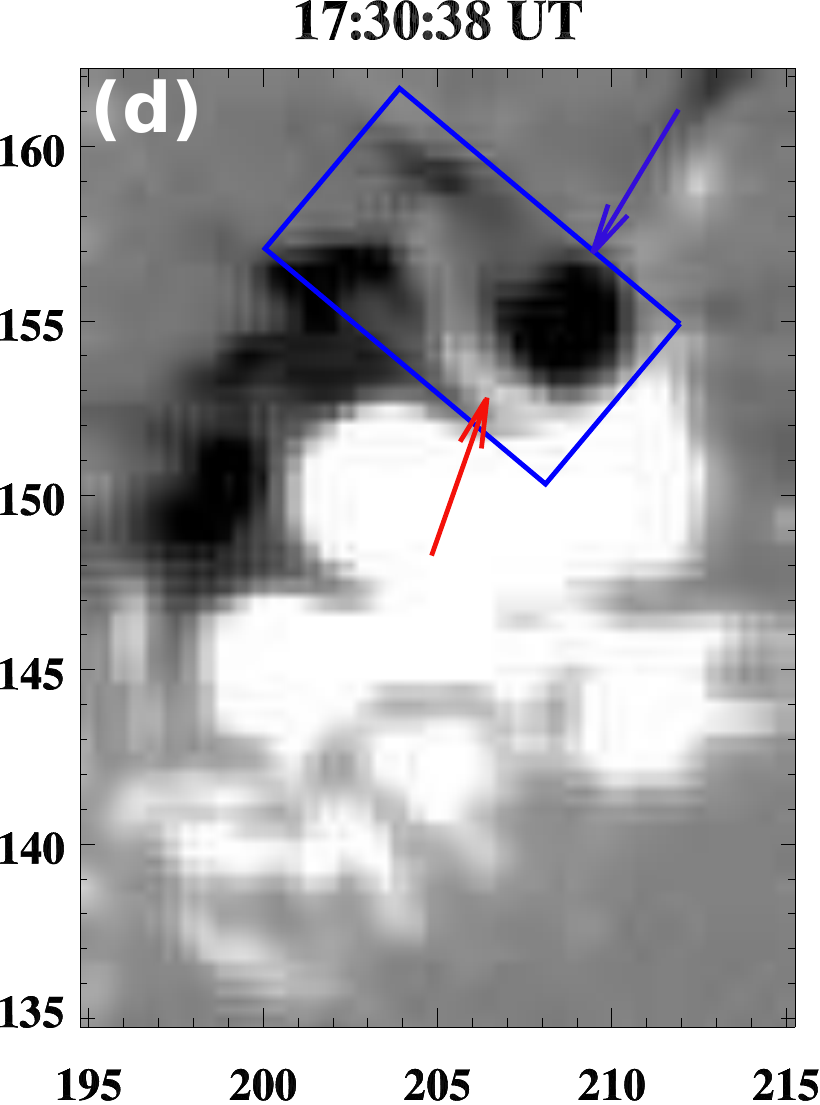}
\includegraphics[width=4cm]{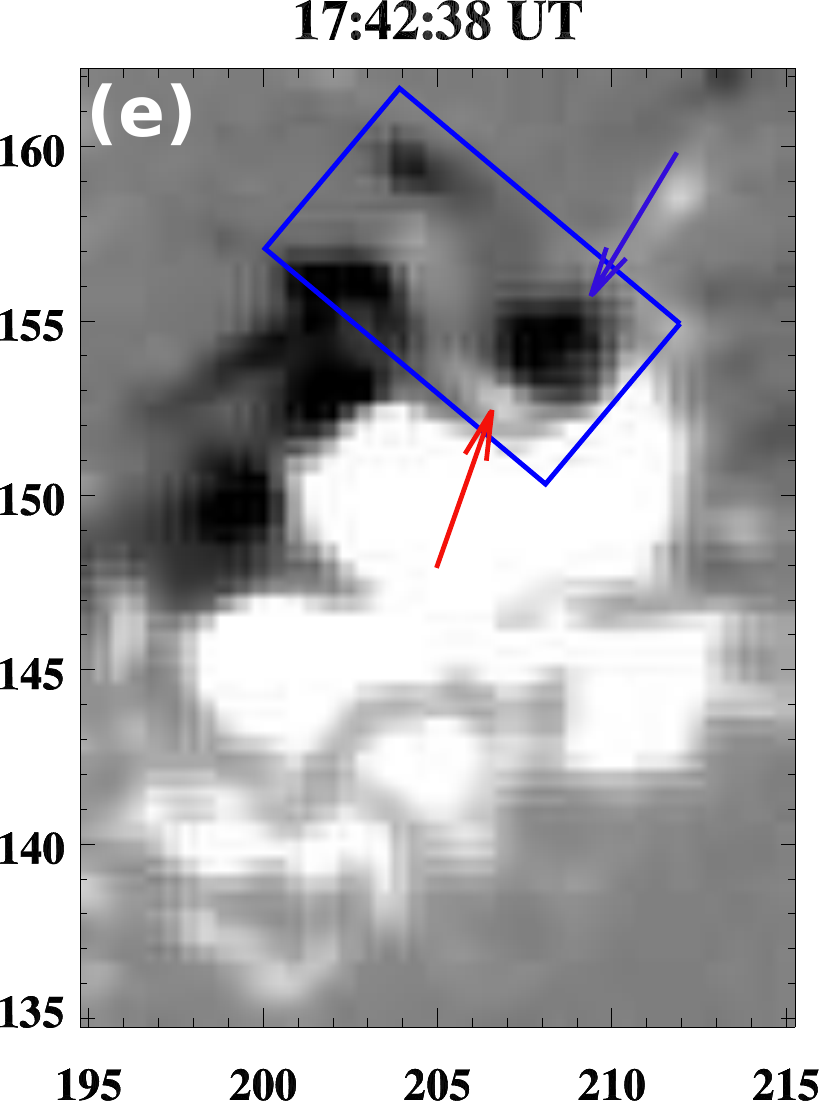}
\includegraphics[width=4cm]{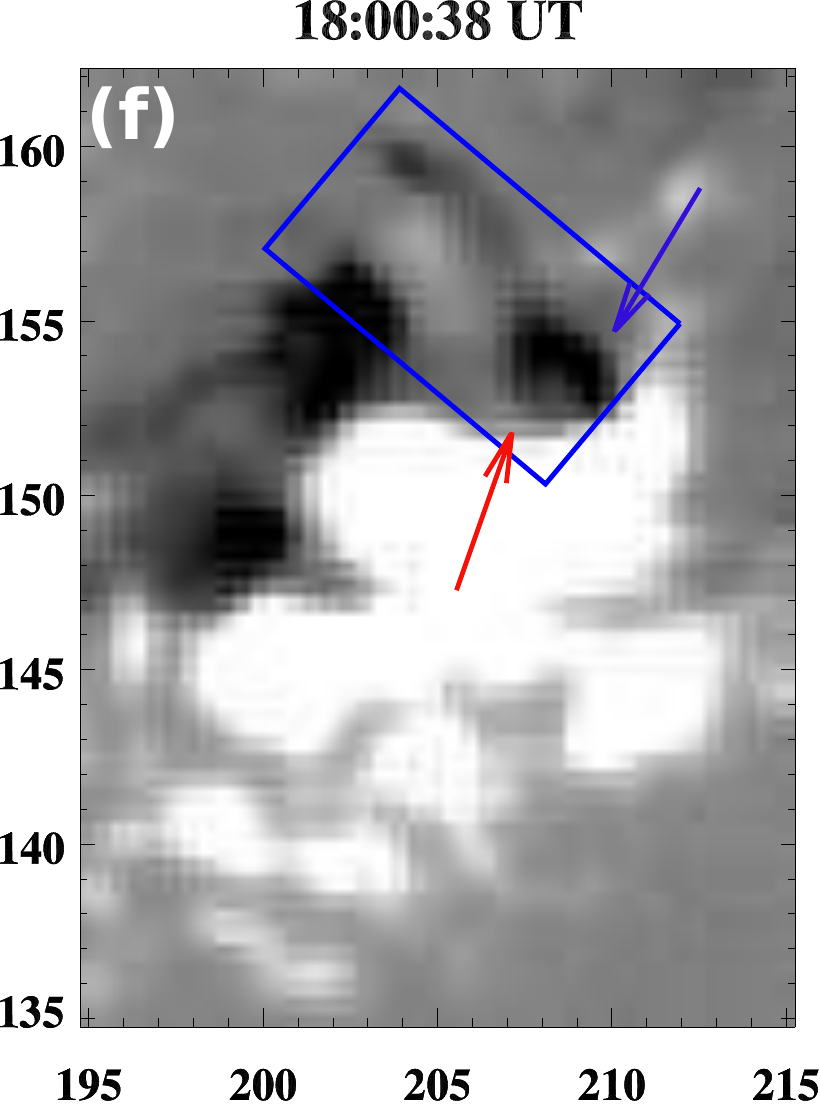}
}
\caption{(a) HMI magnetogram of the active region NOAA 12353. (b) Positive (red), absolute negative (blue), and total (black) flux profiles extracted from the rectangular box (red) region shown in the left panel. GOES soft X-ray flux profile in 1-8 \AA~ channel is also plotted with negative flux profile. The vertical dashed lines indicate the B6.0 and C2.3 flares duration. (c-f) Selected HMI magnetograms of the flare site showing flux cancellation (within blue rectangular box). The x- and y axes are labeled in arcsecs. The shear motion and rapid flux cancellation between N1 and P1 is shown in the HMI magnetograms movie (bottom panels) available in the online edition.}
\label{hmi}
\end{figure*}

%%%%%%%%%%%%%%%%%%%%%%%%%%%%%%%%%%%%%%%%%%%%%%%%%%%%%%%%%%%%%%%%%%%%%%%%%%%%%%%%%%%%%%%%%%%
%------------------------------------------------------------------------------------ 
\begin{figure*}
\centering{
\includegraphics[width=5cm]{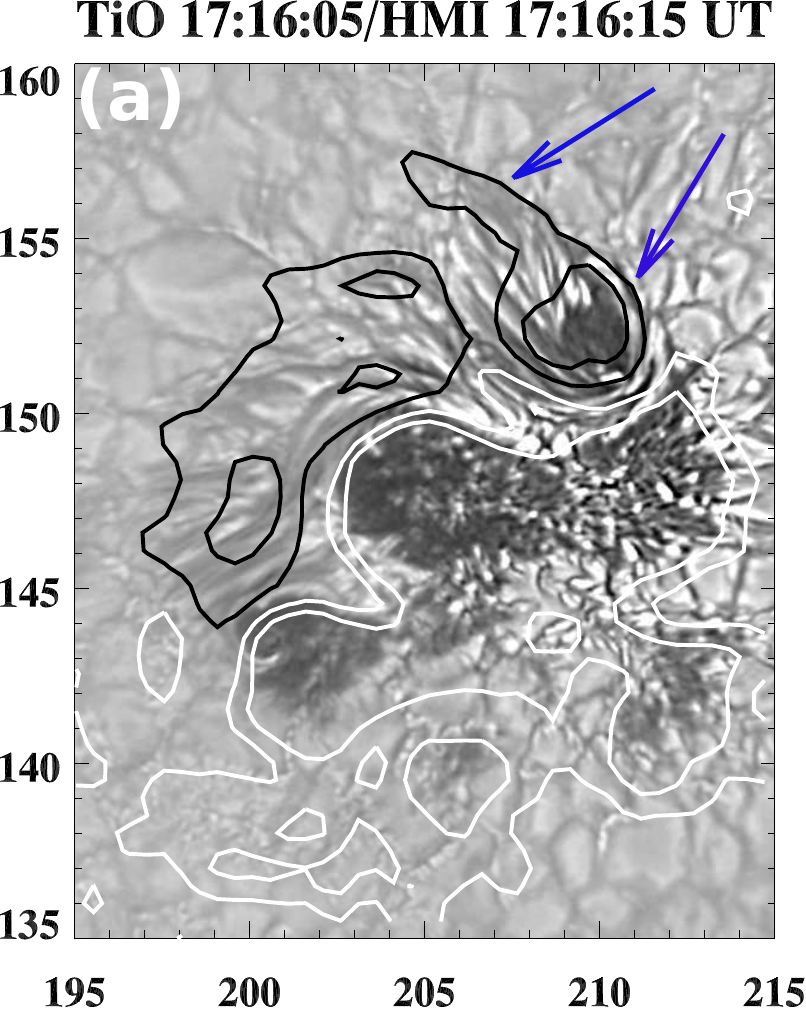}
\includegraphics[width=5cm]{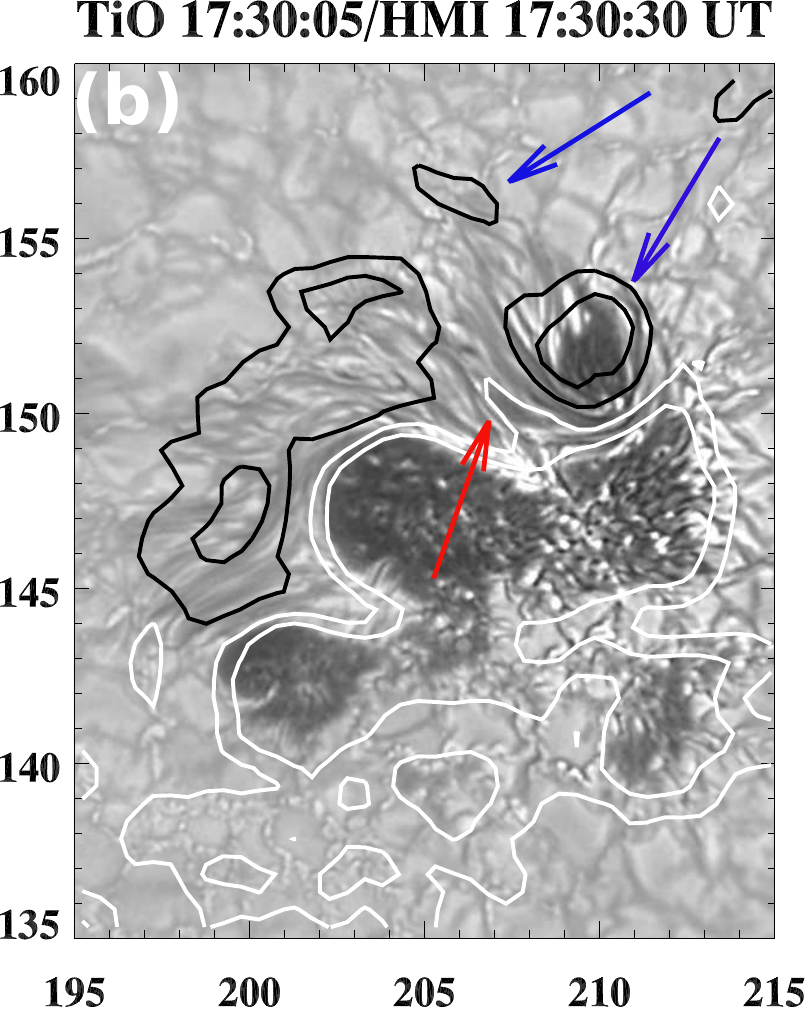}
\includegraphics[width=5cm]{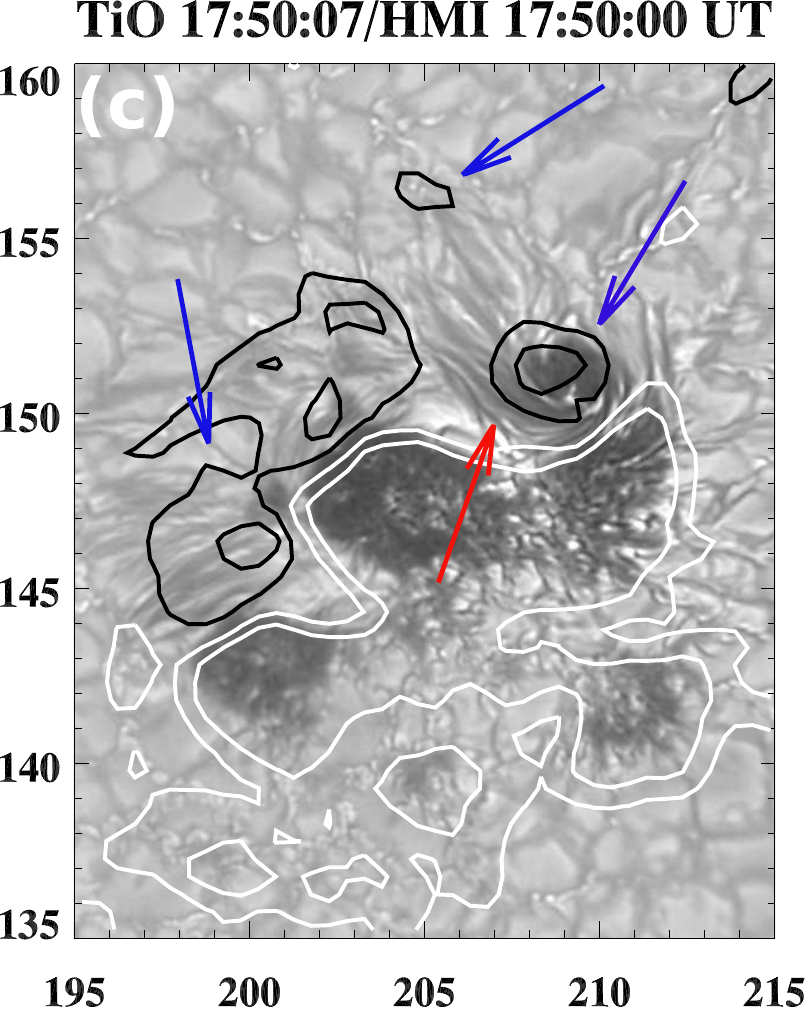}
}
\caption{NST TiO (7057 \AA~) images overlaid by HMI magnetogram contours of positive (white) and negative (black) polarities. The contour levels are $\pm$200 and $\pm$500 Gauss.  Flux cancellation and shrinkage of negative polarity sunspot are marked by arrows. The x- and y axes are labeled in arcsecs.
}
\label{tio}
\end{figure*}

%%%%%%%%%%%%%%%%%%%%%%%%%%%%%%%%%%%%%%%%%%%%%%%%%%%%%%%%%%%%%%%%%%%%%%%%%%%%%%%%%%%%%%%%%%%
%------------------------------------------------------------------------------------ 
\begin{figure*}
\centering{
\includegraphics[width=9.5cm,angle=90]{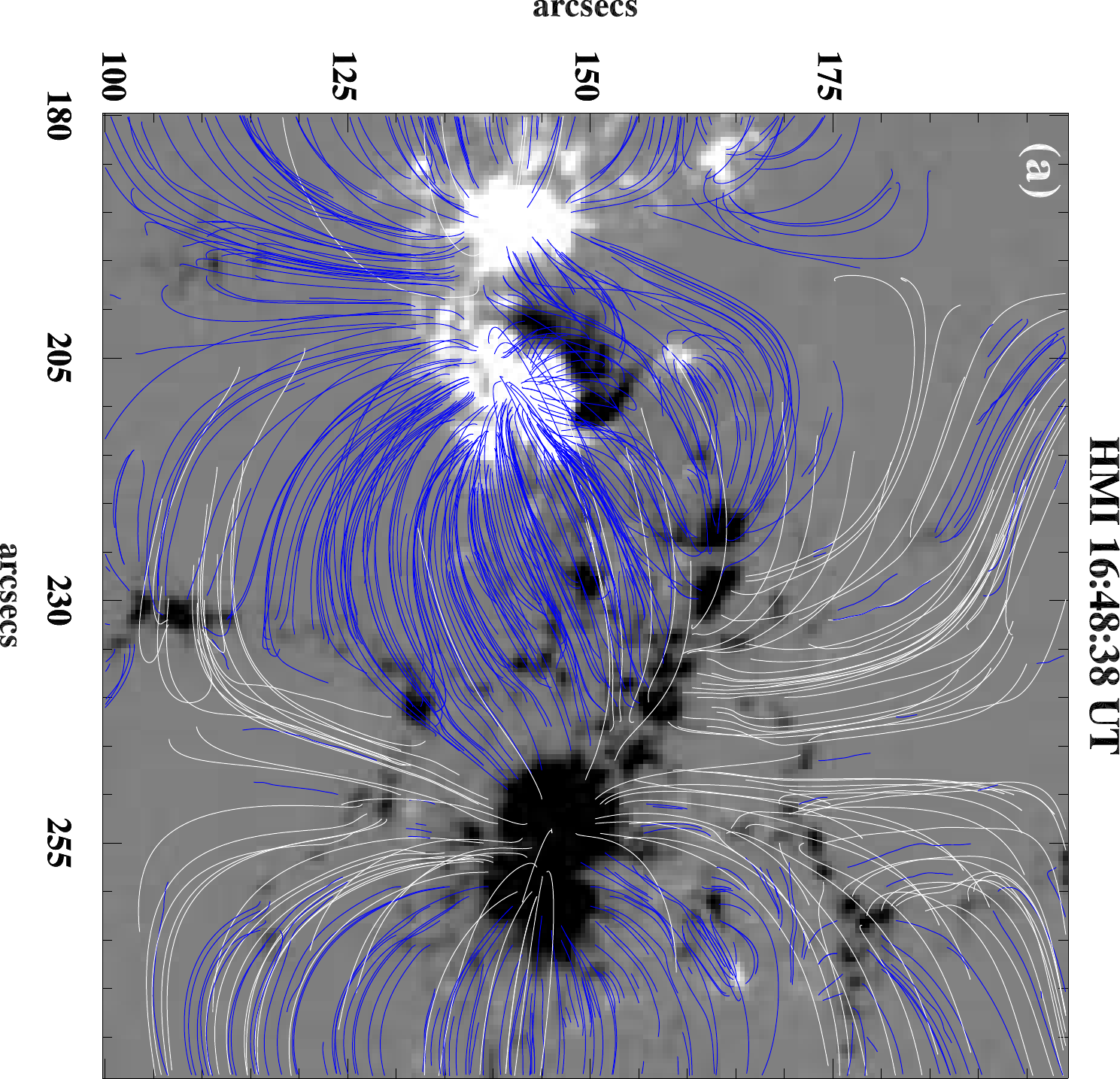}
\includegraphics[width=7.0cm]{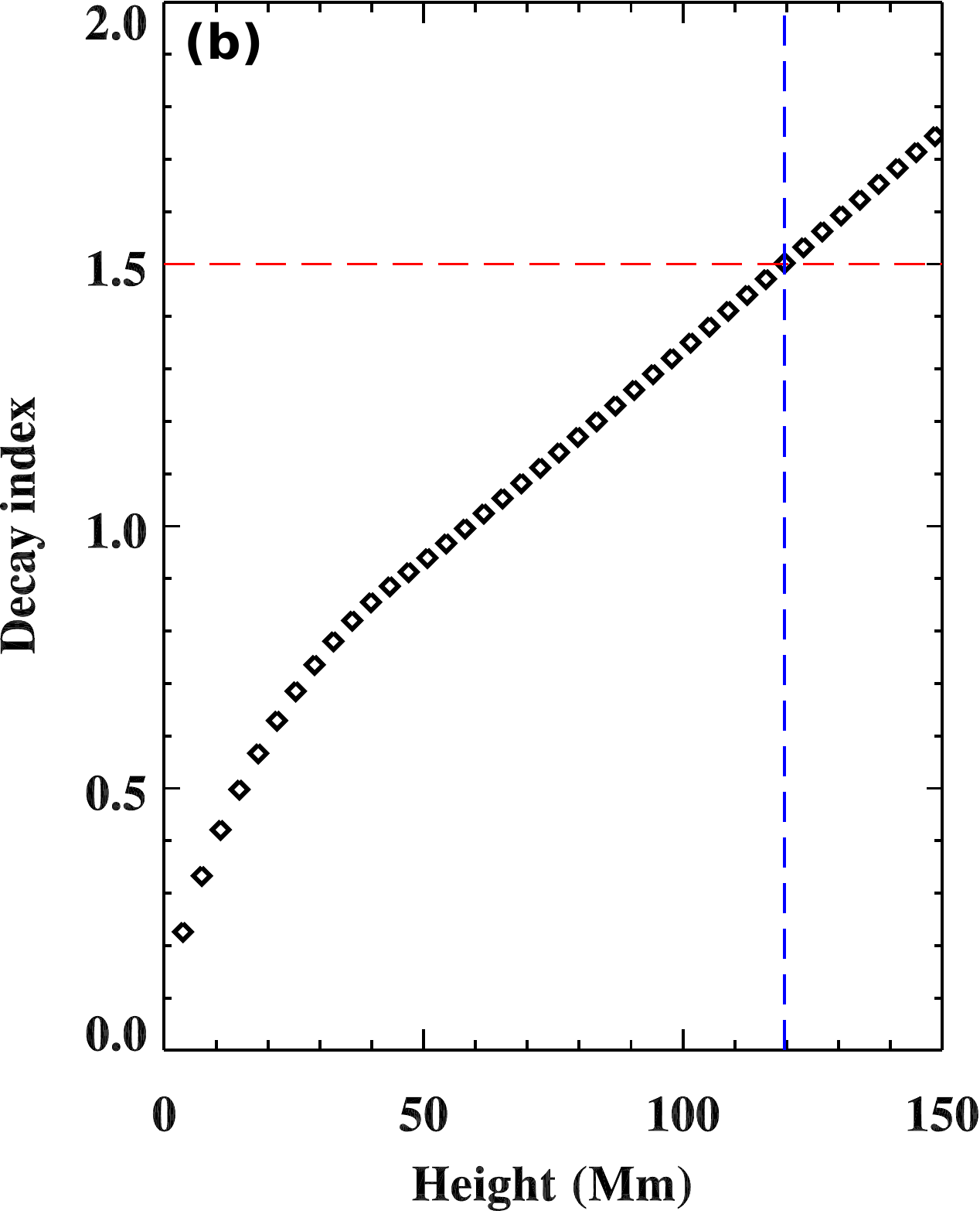}
}
\caption{(a) Potential field extrapolation of the AR using HMI magnetogram at 16:48:38 UT. The closed and open field lines are shown in blue and white colors, respectively. (b) Variation of the decay index with respect to height (Mm) above the PIL. The red horizontal dashed line indicates the threshold value of the decay index (1.5) at the height of $\sim$120 Mm (blue vertical dashed line).
}
\label{extra}
\end{figure*}

%%%%%%%%%%%%%%%%%%%%%%%%%%%%%%%%%%%%%%%%%%%%%%%%%%%%%%%%%%%%%%%%%%%%%%%%%%%%%%%%%%%%%%%%%%%
\subsection{Magnetic field evolution}
Figure \ref{hmi}(a) displays the complete view of the active region NOAA 12353 and the first brightening starts within the box region (red) between N1 and P1. To check the flux emergence or cancellation during the flare, we analysed HMI magnetograms for the interval of 16:30 UT to 18:30 UT. The HMI movie shows the shearing motions and rotation of the negative polarity sunspot (N1) that continuously moved southward and pushed the positive polarity region (P1). Interestingly, the HMI movie also shows rapid cancellation of both negative and positive flux during the loop's coalescence.

In order to quantify the flux cancellation, we extracted positive, negative, and total magnetic fluxes within a rectangular box region that encloses the flare site. Figure \ref{hmi}(b) shows the positive (red), absolute negative (blue), and total (black) magnetic flux profiles. 
GOES soft X-ray flux profile in the 1-8 \AA~ channel is also included (bottom panel) to compare the timing of the flares. The rapid flux cancellation of both polarities started during the B6.0 flare and continued until the end of the C2.3 flare. The rapid cancellation of negative/total flux proceeded even when there was no flares in progress ($\sim$17:08-17:28 UT), when loops (marked by 1, 2, 3) merged in the NST H$\alpha$ images (Figure \ref{nst1}). The positive flux cancellation was not as large as the negative flux cancellation. The negative flux changed from $\sim$3.7$\times$10$^{20}$ Mx to $\sim$2.7$\times$10$^{20}$ Mx ($\sim$27$\%$ decrease). The positive flux decreased only by 5\% from $\sim$1.25$\times$10$^{21}$ Mx to  $\sim$1.20$\times$10$^{21}$ Mx. The total unsigned flux changed from $\sim$1.62$\times$10$^{21}$ Mx to $\sim$1.47$\times$10$^{21}$ Mx. After the flare, the magnetic flux changes were nearly negligible.
To estimate the flux cancellation rate, we performed linear fit to the time profiles of magnetic fluxes between 17:18 UT and 17:42 UT and calculated the slopes. The estimated total unsigned, positive, and negative flux cancellation rates are comparable and equal to 3.44$\times$10$^{20}$ Mx h$^{-1}$,  1.57$\times$10$^{20}$ Mx h$^{-1}$, and  1.86$\times$10$^{20}$ Mx h$^{-1}$, respectively.

The evolution of the magnetic field before, during, and after the flare is shown in selected magnetograms (Figure \ref{hmi}(c-f)). The detachment of the tail of the tadpole-like negative sunspot N1 (blue rectangle, marked by arrow) is a result of flux cancellation during the flare. The motion of the negative polarity sunspot N1 and its cancellation with the ambient positive polarity region P1 (red arrow) is clearly observed.

To investigate the details of the morphological changes in the photospheric magnetic structures during the flares, we used NST TiO (7055 \AA) images. Figure \ref{tio} displays the co-temporal HMI magnetograms contour over the TiO Images at the start time (17:30:05 UT) and after the flare (17:50:07 UT). The cancellation of N1 and P1 is seen clearly. Initially N1 had a tail structure and we see the disappearance of the tail during the flare. Interestingly, we also observed the shrinkage of N1 (in the TiO movie) during the flare. The magnetogram contours of N1 over TiO images also show the reduced umbral area (i.e., collapse) of the spot N1. Furthermore, we see cancellation of an elongated structure in P1 which is marked by the red arrow.

%\citep{nakagawa1972}
Figure \ref{extra}(a) displays the potential field extrapolation of the active region using the HMI magnetogram (at 16:48:38 UT) before the flare trigger. The closed field lines are shown in blue color whereas the open field lines in white color. We can see the bunches of overlying field lines above the flare site. The connectivity of the field lines between P1, N3 and P2, N1 is very similar to the loops as observed in the H$\alpha$ and EUV images. 

To check the variation of the horizontal component of the overlying magnetic field with respect to height, we estimated the decay index n=-d(logB$_h$)/d(logH), where B$_h$ is the horizontal magnetic field strength and H is the height above the solar surface. We calculated the mean value of B$_h$ above the PIL between N1 and P1 (marked in Figure \ref{hmi}(a)). Figure \ref{extra}(b) displays the decay index vs. height above the PIL. According to the torus instability criteria \citep{kliem2006,olmedo2010}, the threshold value of the decay index should be $\ge$1.5. We can see that the threshold value of the decay index (1.5) can be achieved at the height of $\sim$120 Mm. Therefore, the flux rope should attain this height for a successful eruption. 
In addition, the AIA 304 \AA~ stack plot (Figure \ref{stack2}) reveals that the flux rope attains a height (projected) of about 25-30 Mm. The flux rope is most likely stopped by the strong overlying field of the active region and could not satisfied the condition for the torus instability, which result in the failed eruption. 

%%%%%%%%%%%%%%%%%%%%%%%%%%%%%%%%%%%
\section{SUMMARY AND DISCUSSION}
We presented high resolution multiwavelength observations of two homologous flares, accompanied by an oscillatory energy release and formation/disruption of a flux rope. The main results of this study are summarized below: 

We observed interaction and reconnection of cool chromospheric H$\alpha$ loops. The loop's interaction (merging of the loops) was associated with the formation of a left-handed flux rope along with a rapid photospheric flux cancellation observed by the HMI. The flux rope showed rotation in the counterclockwise direction and it failed to erupt due to the presence of strong overlying fields. We present the following  evidence of reconnection at the footoint F1: (i) clear interaction between the loops L1 and L2 seen in the high-resolution NST images. L2 starts to activate/appear during the reconnection between L1 and L2. We noticed a H$\alpha$ brightening between L1 and L2. (ii) Location of the X-ray sources suggests the particle acceleration/precipitation site above and between N1 and P1. (iii) A rapid photospheric flux cancellation between P1 and N1 during the interaction is a strong evidence of reconnection between the chromospheric loops.

The rotation in the counterclockwise direction indicates a left-handed flux rope. The strands in the flux rope confirm the left-handed twist.
 The clockwise rotation of N1 increases the right-handed twist, whereas the counterclockwise rotation of P2 adds the left-handed twist in the coronal flux rooted in these spots. The H$\alpha$ images indicate left-handed flux in the formed flux rope, which suggests a more dominant role of the rotation of P2. Moreover, P2 contains more magnetic flux than N1. 
%In addition, reconnection of L1 and L2 will generate a forward S-shaped sigmoid loop in the chromosphere which also suggests the left-handed twist of the rope \citep{green2007}.

\citet{vasyl2001} reported one hour long flux cancellation in the sunspot moat proceeding at a rate of 3$\times$10$^{19}$ Mx h$^{-1}$. They interpreted the cancellation to be due to photospheric reconnection between small flux elements and noted its possible connection to flares. Here we observed flux cancellation between two small sunspots not only during the flare but also 15-20 min before the C2.3 flare. In our case, the total absolute flux cancellation rate is $\sim$3.44$\times$10$^{20}$ Mx h$^{-1}$, which is almost 10 times larger than that reported by \citet{vasyl2001}. \citet{green2011} reported cancellation of total unsigned flux of $\sim$1.17$\times$10$^{21}$ Mx) at PIL in a decaying AR during $\sim$2.5 days preceding a  CME and suggested that the major flux cancellation was responsible for the formation of a S-shaped sigmoid/flux rope (prior to the eruption). However, here we observed rapid flux cancellation (total unsigned flux$\sim$1.7$\times$10$^{20}$ Mx) within only $\sim$20 min which is much faster (but smaller) than that observed by \citet{green2011}. Therefore, we relate the rapid/major flux cancellation with the formation of the flux rope. This much rapid flux cancellation is hardly observed before. In addition, the magnetic flux in interplanetary magnetic clouds lies between 10$^{20} $ Mx to 10$^{22}$ Mx \citep{qiu2007}. Here we observed a small flux rope (not a large-scale flux rope). Therefore, the magnetic flux associated with the rope may be of the order of 10$^{20}$ Mx (lower limit).
 
Without the use of high-resolution NST images, AIA observations could have been interpreted as activation of a pre-existing flux rope which is heated during the first B6.0 flare and trying to erupt during the second C2.3 flare. However, we do not see any pre-existing filament/flux rope before the onset of the first B6.0 flare. We only observe a system of cool chromospheric loops (connecting P1 and N3) in the AIA 304, 171 \AA, and labeled L2 in the NST H$\alpha$ images.
We are able to see clear interaction of cool loops at the PIL and the development of a twisted flux rope associated with a rapid flux cancellation and therefore we suggests that the flux rope was most likely developed/formed via magnetic reconnection of the loops associated with N1 and P1 sunspots. This result supports the previous report of the formation of a S-shaped flux rope via chromospheric reconnection \citep{kumar2015fr}. However, we did not observe such fast flux cancellation in our previous studies. In the tether cutting reconnection model \citep{moore2001}, the reconnection between the sheared arcade loops (in the corona) results in the formation of a coronal flux rope. Here we report the reconnection between the cool H$\alpha$ loops in the choromosphere that forms a flux rope.
 
The dynamic radio spectrum (25-180 MHz) observed at Sagamore Hill\footnote[1]{http://www.ngdc.noaa.gov/stp/space-weather/solar-data/solar-features/solar-radio/} radio station does not show any radio signatures during both flares. No type III radio burst was observed during both flares which  suggests the absence of particle acceleration into the interplanetary medium along the open field lines. Most of the accelerated particles were confined along the closed loops within the AR. A part of accelerated electrons most likely followed the closed field lines that led to their precipitation at the opposite footpoint of the reconnecting twisted loop. This resulted in chromospheric evaporation and heating of the loops up to $\sim$10 MK temperature.

Since the height (projected) of the overlying arcades in 171 \AA~ channel is more than the flux rope height ($\sim$25-30 Mm) during the C2.3 flare. In addition, the threshold value of the decay index above the PIL (1.5) could be achieved at a height of $\sim$120 Mm. Therefore, the flux rope did not attain the height of torus instability. In \citet{kumar2014}, a kinked small filament reached a height of $\sim$60 Mm, and was unable to erupt due to the interaction with overlying arcade loops. In addition, if the flux rope is very small, the total energy of the rope may not be sufficient to overcome/open the overlying field and could show a counterclockwise rotation within the overlying arcade loops. After the twist release, the plasma is drained back to the solar surface. 
 
The RHESSI X-ray sources (6-12 keV and 12-25 keV) do not show any significant motion during the flare, i.e., particle acceleration or energy release site remains stationary (between P1 and N1) during the flare. 
In addition, the RHESSI X-ray source (6-12 keV) was located above the sunspots P1 and N1, which is consistent with the intensity oscillation site observed in the EUV channels (1600 and 304 \AA). During the second C2.3 flare, we observed the cancellation of elongated penumbral field (see the TiO images) between N1 and P1. The footpoint of the loop L1 was rooted in the penumbral field of positive polarity P1. Both flares reveal the oscillatory behaviour in the X-ray/EUV flux profiles that most likely indicate the occurrence of repeated reconnection by the loop's coalescence between sunspot P1 and N1.

Previous observations of reconnection between loops in the chromosphere showed the standard flare reconnection in a current sheet below erupting flux \citep{yang2015,xue2016}, while here the process of tether-cutting and flux rope formation was observed for the first time in the chromosphere.

In conclusion, we have reported the direct observations of reconnection between chromospheric loops associated with rapid flux cancellation and flux rope formation with counterclockwise rotation. The hot flux ropes (observed in the AIA 131 and 94 \AA) are generally formed by reconnection among the coronal loops (e.g, \citealt{kumar2014}). Here, we reported the formation of a cool flux rope (H$\alpha$, AIA 304 and 171 \AA) by multiple reconnection between the chromospheric loops. These observations support the model of flux rope formation proposed by \citet{van1989}. Such a clear loop's coalescence associated with flux rope formation in a short interval ($\sim$30 minutes) is not reported before. Future multiwavelength studies with high-resolution observations (NST along with IRIS and SDO) will shed more light on the issues related to the flux rope formation/eruption.

%%%%%%%%%%%%%%%%%%%%%%%%%%%%%%%%%%%%%%%%%%%%%%%%%%%%%%%%%%%%%%%%
\begin{acknowledgements}
We thank the referee for his/her constructive comments/suggestions that improved the manuscript considerably.
SDO is a mission for NASA’s Living With a Star (LWS) program. The SDO data were (partly) provided by the Korean Data Center (KDC) for SDO in cooperation with NASA and SDO/HMI team. RHESSI
is a NASA Small Explorer. BBSO operation is supported by NJIT, US NSF AGS-1250818 and NASA NNX13AG14G, and NST operation is partly supported by the Korea Astronomy and Space Science Institute and Seoul National University, and by strategic priority research program of CAS with Grant No. XDB09000000. 
%Wavelet software was provided by C. Torrence and G. Compo, and is available at URL: http://atoc.colorado.edu/research/wavelets/.
 HW is supported by US NSF under grants AGS 1348513 and 1408703,  and NASA under grants NNX13AG13G/NNX16AF72G . K.-S. Cho acknowledges support by a grant from the US Air Force Research Laboratory, under agreement number FA 2386-14-1-4078 and by the ``Planetary system research for space exploration" from KASI. This work was supported by the ``Operation of Korea Space Weather Center" of KASI and the KASI basic research funds. This work was conducted as part of the effort of NASA's Living with a Star Focused Science Team ``Jets" NASA LWS NNX11AO73G grant.  VYu acknowledges support from NSF AGS-1146896, AFOSR-FA9550-15-1-0322 grants and Korea Astronomy and Space Science Institute. PK thanks Prof. Valery M. Nakariakov for several helpful discussions.
\end{acknowledgements}
%%%%%%%%%%%%%%%%%%%%%%%%%%%%%%%%%%%%%%%%%%%%%%%%%%%%%%%%
\bibliographystyle{aa}
\bibliography{reference}

\begin{thebibliography}{58}
\expandafter\ifx\csname natexlab\endcsname\relax\def\natexlab#1{#1}\fi

\bibitem[{{Antiochos}(1998)}]{antiochos1998}
{Antiochos}, S.~K. 1998, \apjl, 502, L181

\bibitem[{{Aschwanden} {et~al.}(2013){Aschwanden}, {Boerner}, {Schrijver}, \&
  {Malanushenko}}]{asc2013}
{Aschwanden}, M.~J., {Boerner}, P., {Schrijver}, C.~J., \& {Malanushenko}, A.
  2013, \solphys, 283, 5

\bibitem[{{Aulanier} {et~al.}(2010){Aulanier}, {T{\"o}r{\"o}k}, {D{\'e}moulin},
  \& {DeLuca}}]{aulanier2010}
{Aulanier}, G., {T{\"o}r{\"o}k}, T., {D{\'e}moulin}, P., \& {DeLuca}, E.~E.
  2010, \apj, 708, 314

\bibitem[{{B{\c a}k-St{\c e}{\'s}licka} {et~al.}(2013){B{\c a}k-St{\c
  e}{\'s}licka}, {Gibson}, {Fan}, {Bethge}, {Forland}, \&
  {Rachmeler}}]{bak2013}
{B{\c a}k-St{\c e}{\'s}licka}, U., {Gibson}, S.~E., {Fan}, Y., {et~al.} 2013,
  \apjl, 770, L28

\bibitem[{{Bothmer} \& {Schwenn}(1998)}]{bothmer1998}
{Bothmer}, V. \& {Schwenn}, R. 1998, Annales Geophysicae, 16, 1

\bibitem[{{Burlaga} {et~al.}(1982){Burlaga}, {Klein}, {Sheeley}, {Michels},
  {Howard}, {Koomen}, {Schwenn}, \& {Rosenbauer}}]{burlaga1982}
{Burlaga}, L.~F., {Klein}, L., {Sheeley}, Jr., N.~R., {et~al.} 1982, \grl, 9,
  1317

\bibitem[{{Canfield} {et~al.}(1999){Canfield}, {Hudson}, \&
  {McKenzie}}]{canfield1999}
{Canfield}, R.~C., {Hudson}, H.~S., \& {McKenzie}, D.~E. 1999, \grl, 26, 627

\bibitem[{{Chen}(2011)}]{chen2011}
{Chen}, P.~F. 2011, Living Reviews in Solar Physics, 8, 1

\bibitem[{{Chen} \& {Shibata}(2000)}]{chen2000}
{Chen}, P.~F. \& {Shibata}, K. 2000, \apj, 545, 524

\bibitem[{{Cheng} {et~al.}(2011){Cheng}, {Zhang}, {Liu}, \& {Ding}}]{cheng2011}
{Cheng}, X., {Zhang}, J., {Liu}, Y., \& {Ding}, M.~D. 2011, \apjl, 732, L25

\bibitem[{{Einaudi} \& {van Hoven}(1983)}]{einaudi1983}
{Einaudi}, G. \& {van Hoven}, G. 1983, \solphys, 88, 163

\bibitem[{{Fan} \& {Gibson}(2004)}]{fan2004}
{Fan}, Y. \& {Gibson}, S.~E. 2004, \apj, 609, 1123

\bibitem[{{Gary} \& {Moore}(2004)}]{gary2004}
{Gary}, G.~A. \& {Moore}, R.~L. 2004, \apj, 611, 545

\bibitem[{{Gibson} {et~al.}(2006){Gibson}, {Fan}, {T{\"o}r{\"o}k}, \&
  {Kliem}}]{gibson2006}
{Gibson}, S.~E., {Fan}, Y., {T{\"o}r{\"o}k}, T., \& {Kliem}, B. 2006, \ssr,
  124, 131

\bibitem[{{Gopalswamy} {et~al.}(2005){Gopalswamy}, {Yashiro}, {Michalek},
  {Xie}, {Lepping}, \& {Howard}}]{gopalswamy2005}
{Gopalswamy}, N., {Yashiro}, S., {Michalek}, G., {et~al.} 2005, \grl, 32, 12

\bibitem[{{Gosling} {et~al.}(1995){Gosling}, {Birn}, \& {Hesse}}]{gosling1995}
{Gosling}, J.~T., {Birn}, J., \& {Hesse}, M. 1995, \grl, 22, 869

\bibitem[{{Green} \& {Kliem}(2009)}]{green2009}
{Green}, L.~M. \& {Kliem}, B. 2009, \apjl, 700, L83

\bibitem[{{Green} {et~al.}(2011){Green}, {Kliem}, \& {Wallace}}]{green2011}
{Green}, L.~M., {Kliem}, B., \& {Wallace}, A.~J. 2011, \aap, 526, A2

\bibitem[{{Hood} \& {Priest}(1981)}]{hood1981}
{Hood}, A.~W. \& {Priest}, E.~R. 1981, Geophysical and Astrophysical Fluid
  Dynamics, 17, 297

\bibitem[{{Karpen} {et~al.}(2012){Karpen}, {Antiochos}, \&
  {DeVore}}]{karpen2012}
{Karpen}, J.~T., {Antiochos}, S.~K., \& {DeVore}, C.~R. 2012, \apj, 760, 81

\bibitem[{{Kliem} {et~al.}(2004){Kliem}, {Titov}, \&
  {T{\"o}r{\"o}k}}]{kliem2004}
{Kliem}, B., {Titov}, V.~S., \& {T{\"o}r{\"o}k}, T. 2004, \aap, 413, L23

\bibitem[{{Kliem} \& {T{\"o}r{\"o}k}(2006)}]{kliem2006}
{Kliem}, B. \& {T{\"o}r{\"o}k}, T. 2006, Physical Review Letters, 96, 255002

\bibitem[{{Kumar} \& {Cho}(2014)}]{kumar2014}
{Kumar}, P. \& {Cho}, K.-S. 2014, \aap, 572, A83

\bibitem[{{Kumar} {et~al.}(2012){Kumar}, {Cho}, {Bong}, {Park}, \&
  {Kim}}]{kumar2012}
{Kumar}, P., {Cho}, K.-S., {Bong}, S.-C., {Park}, S.-H., \& {Kim}, Y.~H. 2012,
  \apj, 746, 67

\bibitem[{{Kumar} \& {Innes}(2013)}]{kumar2013b}
{Kumar}, P. \& {Innes}, D.~E. 2013, \solphys, 288, 255

\bibitem[{{Kumar} {et~al.}(2011{\natexlab{a}}){Kumar}, {Manoharan}, \&
  {Uddin}}]{kumar2011}
{Kumar}, P., {Manoharan}, P.~K., \& {Uddin}, W. 2011{\natexlab{a}}, \solphys,
  271, 149

\bibitem[{{Kumar} {et~al.}(2011{\natexlab{b}}){Kumar}, {Srivastava},
  {Filippov}, {Erd{\'e}lyi}, \& {Uddin}}]{Kumar2011a}
{Kumar}, P., {Srivastava}, A.~K., {Filippov}, B., {Erd{\'e}lyi}, R., \&
  {Uddin}, W. 2011{\natexlab{b}}, \solphys, 272, 301

\bibitem[{{Kumar} {et~al.}(2010){Kumar}, {Srivastava}, {Filippov}, \&
  {Uddin}}]{kumar2010b}
{Kumar}, P., {Srivastava}, A.~K., {Filippov}, B., \& {Uddin}, W. 2010,
  \solphys, 266, 39

\bibitem[{{Kumar} {et~al.}(2015){Kumar}, {Yurchyshyn}, {Wang}, \&
  {Cho}}]{kumar2015fr}
{Kumar}, P., {Yurchyshyn}, V., {Wang}, H., \& {Cho}, K.-S. 2015, \apj, 809, 83

\bibitem[{{Lemen} {et~al.}(2012){Lemen}, {Title}, {Akin}, {Boerner}, {Chou},
  {Drake}, {Duncan}, {Edwards}, {Friedlaender}, {Heyman}, {Hurlburt}, {Katz},
  {Kushner}, {Levay}, {Lindgren}, {Mathur}, {McFeaters}, {Mitchell}, {Rehse},
  {Schrijver}, {Springer}, {Stern}, {Tarbell}, {Wuelser}, {Wolfson}, {Yanari},
  {Bookbinder}, {Cheimets}, {Caldwell}, {Deluca}, {Gates}, {Golub}, {Park},
  {Podgorski}, {Bush}, {Scherrer}, {Gummin}, {Smith}, {Auker}, {Jerram},
  {Pool}, {Soufli}, {Windt}, {Beardsley}, {Clapp}, {Lang}, \&
  {Waltham}}]{lemen2012}
{Lemen}, J.~R., {Title}, A.~M., {Akin}, D.~J., {et~al.} 2012, \solphys, 275, 17

\bibitem[{{Lin} {et~al.}(2002){Lin}, {Dennis}, {Hurford}, {Smith}, {Zehnder},
  {Harvey}, {Curtis}, {Pankow}, {Turin}, {Bester}, {Csillaghy}, {Lewis},
  {Madden}, {van Beek}, {Appleby}, {Raudorf}, {McTiernan}, {Ramaty}, {Schmahl},
  {Schwartz}, {Krucker}, {Abiad}, {Quinn}, {Berg}, {Hashii}, {Sterling},
  {Jackson}, {Pratt}, {Campbell}, {Malone}, {Landis}, {Barrington-Leigh},
  {Slassi-Sennou}, {Cork}, {Clark}, {Amato}, {Orwig}, {Boyle}, {Banks},
  {Shirey}, {Tolbert}, {Zarro}, {Snow}, {Thomsen}, {Henneck}, {McHedlishvili},
  {Ming}, {Fivian}, {Jordan}, {Wanner}, {Crubb}, {Preble}, {Matranga}, {Benz},
  {Hudson}, {Canfield}, {Holman}, {Crannell}, {Kosugi}, {Emslie}, {Vilmer},
  {Brown}, {Johns-Krull}, {Aschwanden}, {Metcalf}, \& {Conway}}]{lin2002}
{Lin}, R.~P., {Dennis}, B.~R., {Hurford}, G.~J., {et~al.} 2002, \solphys, 210,
  3

\bibitem[{{Liu} {et~al.}(2010){Liu}, {Liu}, {Wang}, {Deng}, \&
  {Wang}}]{rui2010}
{Liu}, R., {Liu}, C., {Wang}, S., {Deng}, N., \& {Wang}, H. 2010, \apjl, 725,
  L84

\bibitem[{{Longcope} \& {Beveridge}(2007)}]{longcope2007}
{Longcope}, D.~W. \& {Beveridge}, C. 2007, \apj, 669, 621

\bibitem[{{Manoharan}(2010)}]{manoharan2010}
{Manoharan}, P.~K. 2010, \solphys, 265, 137

\bibitem[{{Marubashi}(1986)}]{marubashi1986}
{Marubashi}, K. 1986, Advances in Space Research, 6, 335

\bibitem[{{McKaig}(2001)}]{mckaig2001}
{McKaig}, I. 2001, \aap, 371, 328

\bibitem[{{Meegan} {et~al.}(2009){Meegan}, {Lichti}, {Bhat}, {Bissaldi},
  {Briggs}, {Connaughton}, {Diehl}, {Fishman}, {Greiner}, {Hoover}, {van der
  Horst}, {von Kienlin}, {Kippen}, {Kouveliotou}, {McBreen}, {Paciesas},
  {Preece}, {Steinle}, {Wallace}, {Wilson}, \& {Wilson-Hodge}}]{meegan2009}
{Meegan}, C., {Lichti}, G., {Bhat}, P.~N., {et~al.} 2009, \apj, 702, 791

\bibitem[{{Moore} {et~al.}(2001){Moore}, {Sterling}, {Hudson}, \&
  {Lemen}}]{moore2001}
{Moore}, R.~L., {Sterling}, A.~C., {Hudson}, H.~S., \& {Lemen}, J.~R. 2001,
  \apj, 552, 833

\bibitem[{{Okamoto} {et~al.}(2008){Okamoto}, {Tsuneta}, {Lites}, {Kubo},
  {Yokoyama}, {Berger}, {Ichimoto}, {Katsukawa}, {Nagata}, {Shibata},
  {Shimizu}, {Shine}, {Suematsu}, {Tarbell}, \& {Title}}]{Okamoto2008}
{Okamoto}, T.~J., {Tsuneta}, S., {Lites}, B.~W., {et~al.} 2008, \apjl, 673,
  L215

\bibitem[{{Olmedo} \& {Zhang}(2010)}]{olmedo2010}
{Olmedo}, O. \& {Zhang}, J. 2010, \apj, 718, 433

\bibitem[{{Pesnell} {et~al.}(2012){Pesnell}, {Thompson}, \&
  {Chamberlin}}]{pesnell2012}
{Pesnell}, W.~D., {Thompson}, B.~J., \& {Chamberlin}, P.~C. 2012, \solphys,
  275, 3

\bibitem[{{Priest} {et~al.}(1989){Priest}, {Hood}, \& {Anzer}}]{priest1989}
{Priest}, E.~R., {Hood}, A.~W., \& {Anzer}, U. 1989, \apj, 344, 1010

\bibitem[{{Qiu} {et~al.}(2007){Qiu}, {Hu}, {Howard}, \& {Yurchyshyn}}]{qiu2007}
{Qiu}, J., {Hu}, Q., {Howard}, T.~A., \& {Yurchyshyn}, V.~B. 2007, \apj, 659,
  758

\bibitem[{Rust \& Kumar(1996)}]{rust1996}
Rust, D. \& Kumar, A. 1996, The Astrophysical Journal Letters, 464, L199

\bibitem[{{Rust} \& {Kumar}(1994)}]{rust1994}
{Rust}, D.~M. \& {Kumar}, A. 1994, \solphys, 155, 69

\bibitem[{{Schou} {et~al.}(2012){Schou}, {Scherrer}, {Bush}, {Wachter},
  {Couvidat}, {Rabello-Soares}, {Bogart}, {Hoeksema}, {Liu}, {Duvall}, {Akin},
  {Allard}, {Miles}, {Rairden}, {Shine}, {Tarbell}, {Title}, {Wolfson},
  {Elmore}, {Norton}, \& {Tomczyk}}]{schou2012}
{Schou}, J., {Scherrer}, P.~H., {Bush}, R.~I., {et~al.} 2012, \solphys, 275,
  229

\bibitem[{{Srivastava} {et~al.}(2010){Srivastava}, {Zaqarashvili}, {Kumar}, \&
  {Khodachenko}}]{srivastava2010}
{Srivastava}, A.~K., {Zaqarashvili}, T.~V., {Kumar}, P., \& {Khodachenko},
  M.~L. 2010, \apj, 715, 292

\bibitem[{{Srivastava} \& {Venkatakrishnan}(2004)}]{srivastava2004}
{Srivastava}, N. \& {Venkatakrishnan}, P. 2004, Journal of Geophysical Research
  (Space Physics), 109, 10103

\bibitem[{{T{\"o}r{\"o}k} \& {Kliem}(2005)}]{torok2005}
{T{\"o}r{\"o}k}, T. \& {Kliem}, B. 2005, \apjl, 630, L97

\bibitem[{{van Ballegooijen} \& {Martens}(1989)}]{van1989}
{van Ballegooijen}, A.~A. \& {Martens}, P.~C.~H. 1989, \apj, 343, 971

\bibitem[{{Vourlidas}(2014)}]{vourlidas2014}
{Vourlidas}, A. 2014, Plasma Physics and Controlled Fusion, 56, 064001

\bibitem[{{Vourlidas} {et~al.}(2013){Vourlidas}, {Lynch}, {Howard}, \&
  {Li}}]{vourlidas2013}
{Vourlidas}, A., {Lynch}, B.~J., {Howard}, R.~A., \& {Li}, Y. 2013, \solphys,
  284, 179

\bibitem[{{Wang} {et~al.}(2015){Wang}, {Cao}, {Chang}, {Yan}, {Liu}, {Zheng},
  {Chae}, \& {Ji}}]{wang2015}
{Wang}, H., {Cao}, W., {Chang}, L., {et~al.} 2015, Nature Communication, 6,
  7008

\bibitem[{{Xue} {et~al.}(2016){Xue}, {Yan}, {Cheng}, {Yang}, {Su}, {Kliem},
  {Zhang}, {Liu}, {Bi}, {Xiang}, {Yang}, \& {Zhao}}]{xue2016}
{Xue}, Z., {Yan}, X., {Cheng}, X., {et~al.} 2016, Nature Communications, 7,
  11837

\bibitem[{{Yang} {et~al.}(2015){Yang}, {Zhang}, \& {Xiang}}]{yang2015}
{Yang}, S., {Zhang}, J., \& {Xiang}, Y. 2015, \apjl, 798, L11

\bibitem[{{Yurchyshyn} {et~al.}(2015){Yurchyshyn}, {Kumar}, {Cho}, {Lim}, \&
  {Abramenko}}]{vasyl2015}
{Yurchyshyn}, V., {Kumar}, P., {Cho}, K.-S., {Lim}, E.-K., \& {Abramenko},
  V.~I. 2015, \apj, 812, 172

\bibitem[{{Yurchyshyn} {et~al.}(2001){Yurchyshyn}, {Wang}, {Goode}, \&
  {Deng}}]{vasyl2001}
{Yurchyshyn}, V.~B., {Wang}, H., {Goode}, P.~R., \& {Deng}, Y. 2001, \apj, 563,
  381

\bibitem[{{Zhang} {et~al.}(2012){Zhang}, {Cheng}, \& {Ding}}]{zhang2012}
{Zhang}, J., {Cheng}, X., \& {Ding}, M.-D. 2012, Nature Communications, 3, 747

\end{thebibliography}
%%%%%%%%%%%%%%%%%%%%%%%%%%%%%%%%%%%%%%%%%%%%%%%%%%%%%%%%

\end{document}